\documentclass[acmtog, nonacm]{acmart} 
\AtBeginDocument{%
  }

\copyrightyear{2025}
\acmYear{2025}
\setcopyright{cc}
\setcctype{by}
\acmConference[SA Conference Papers '25]{SIGGRAPH Asia 2025 Conference Papers}{December 15--18, 2025}{Hong Kong, Hong Kong}
\acmBooktitle{SIGGRAPH Asia 2025 Conference Papers (SA Conference Papers '25), December 15--18, 2025, Hong Kong, Hong Kong}
\acmDOI{10.1145/3757377.3764002}
\acmISBN{979-8-4007-2137-3/25/12} 

\acmSubmissionID{papers\_2447}

\usepackage{subcaption}

\usepackage{xcolor} 

\usepackage{float}
\usepackage{multirow}
\usepackage{cuted}

\definecolor{ys}{rgb}{0,0,0}
\newcommand{\ys}[1]{\textcolor{ys}{#1}}

\begin{document}

\title{\ys{FreeMusco: Motion-Free Learning of Latent Control for Morphology-Adaptive Locomotion in Musculoskeletal Characters}}
 
\author{Minkwan Kim}
\affiliation{%
  \institution{Hanyang University}
  \city{Seoul}
  \country{Republic of Korea}} 
\email{palkan21@hanyang.ac.kr}

\author{Yoonsang Lee}
\authornote{Corresponding author.}
\affiliation{%
  \institution{Hanyang University}
  \city{Seoul}
  \country{Republic of Korea}} 
\email{yoonsanglee@hanyang.ac.kr}

\begin{abstract}

\ys{We propose FreeMusco, a motion-free framework that jointly learns latent representations and control policies for musculoskeletal characters. By leveraging the musculoskeletal model as a strong prior, our method enables energy-aware and morphology-adaptive locomotion to emerge without motion data. The framework generalizes across human, non-human, and synthetic morphologies, where distinct energy-efficient strategies naturally appear—for example, quadrupedal gaits in Chimanoid versus bipedal gaits in Humanoid.
The latent space and corresponding control policy are constructed from scratch, without demonstration, and enable downstream tasks such as goal navigation and path following—representing, to our knowledge, the first motion-free method to provide such capabilities.
FreeMusco learns diverse and physically plausible locomotion behaviors through model-based reinforcement learning, guided by the locomotion objective that combines control, balancing, and biomechanical terms. To better capture the periodic structure of natural gait, we introduce the temporally averaged loss formulation, which compares simulated and target states over a time window rather than on a per-frame basis.
We further encourage behavioral diversity by randomizing target poses and energy levels during training, enabling locomotion to be flexibly modulated in both form and intensity at runtime. Together, these results demonstrate that versatile and adaptive locomotion control can emerge without motion capture, offering a new direction for simulating movement in characters where data collection is impractical or impossible.}

\end{abstract}



\begin{teaserfigure}
    \centering
  \includegraphics[trim=900 560 650 1.5, clip, width=0.245\textwidth]{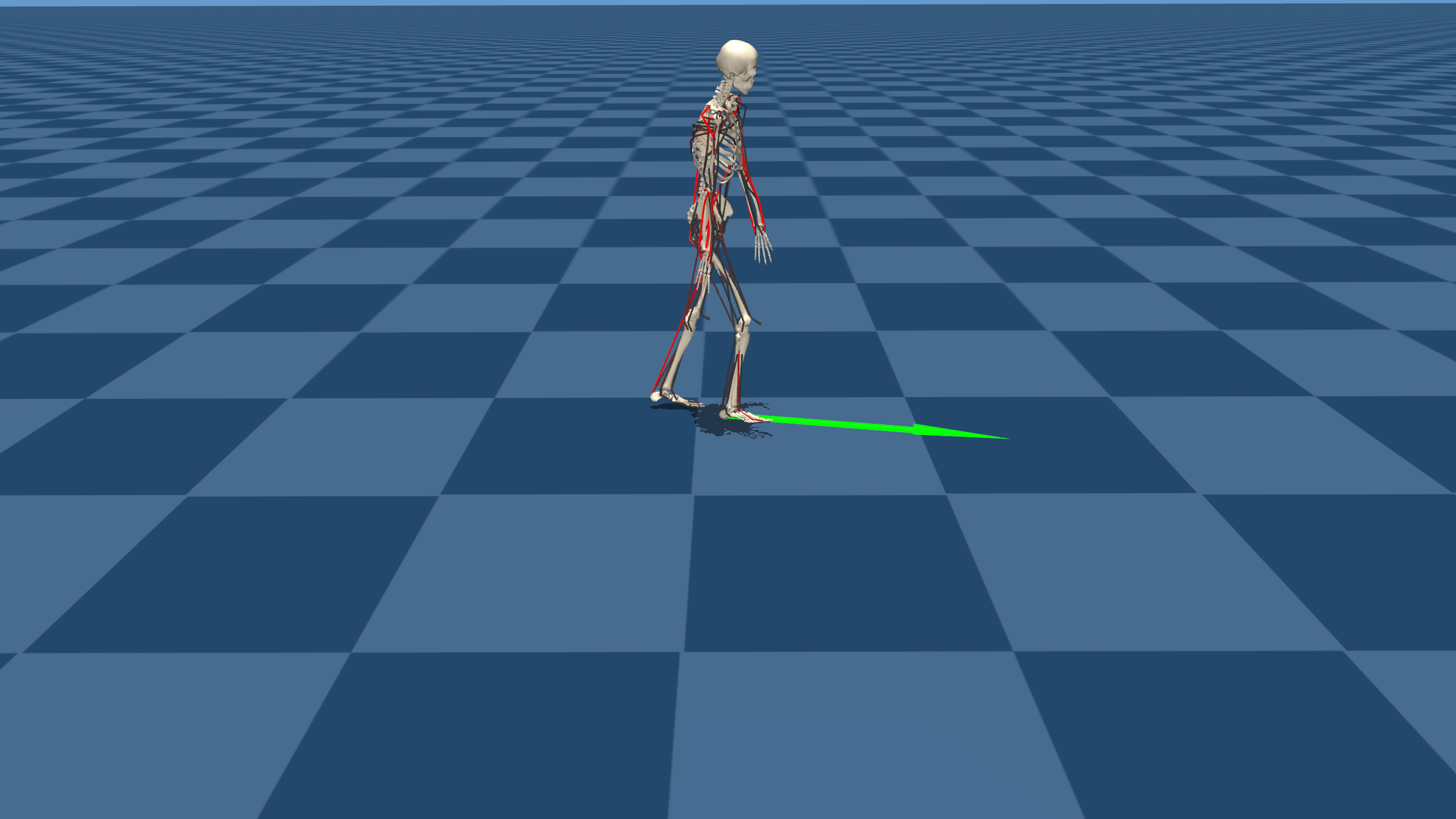}
  \includegraphics[trim=500 40 450 0, clip, width=0.245\textwidth]{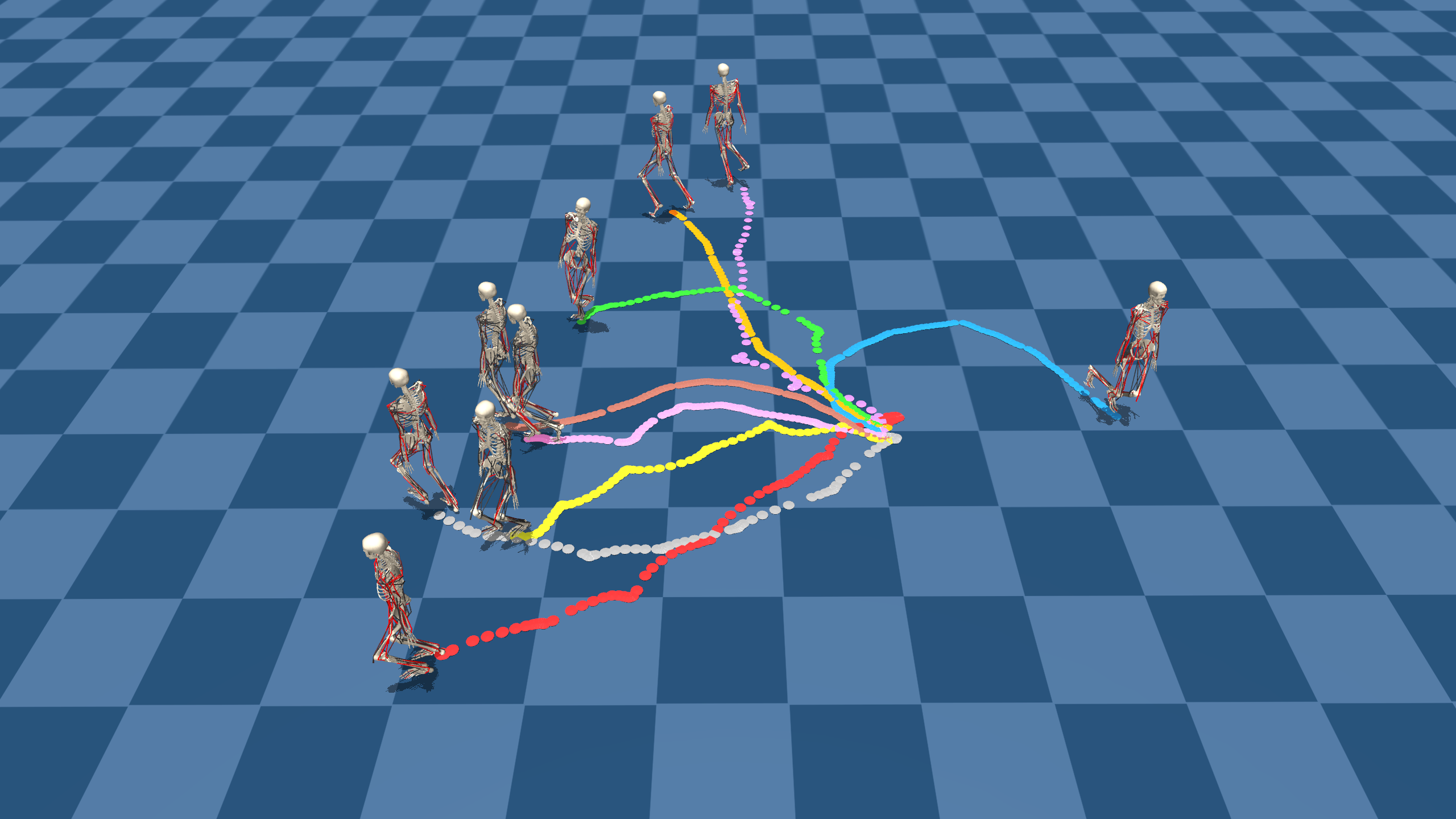}
  \includegraphics[trim=600 240 580 0, clip, width=0.245\textwidth]{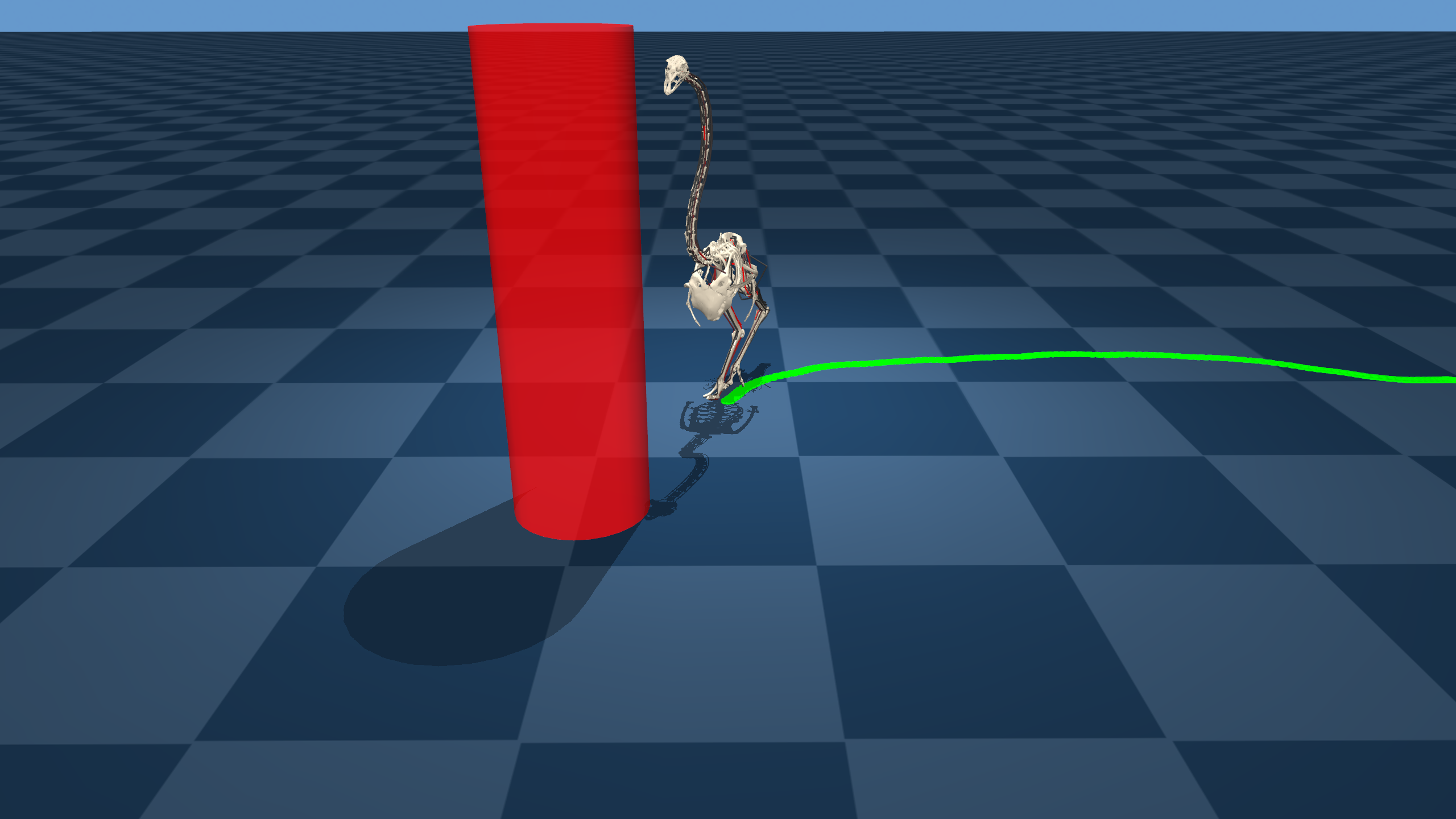}
  \includegraphics[trim=810 410 910 299.5, clip, width=0.245\textwidth]{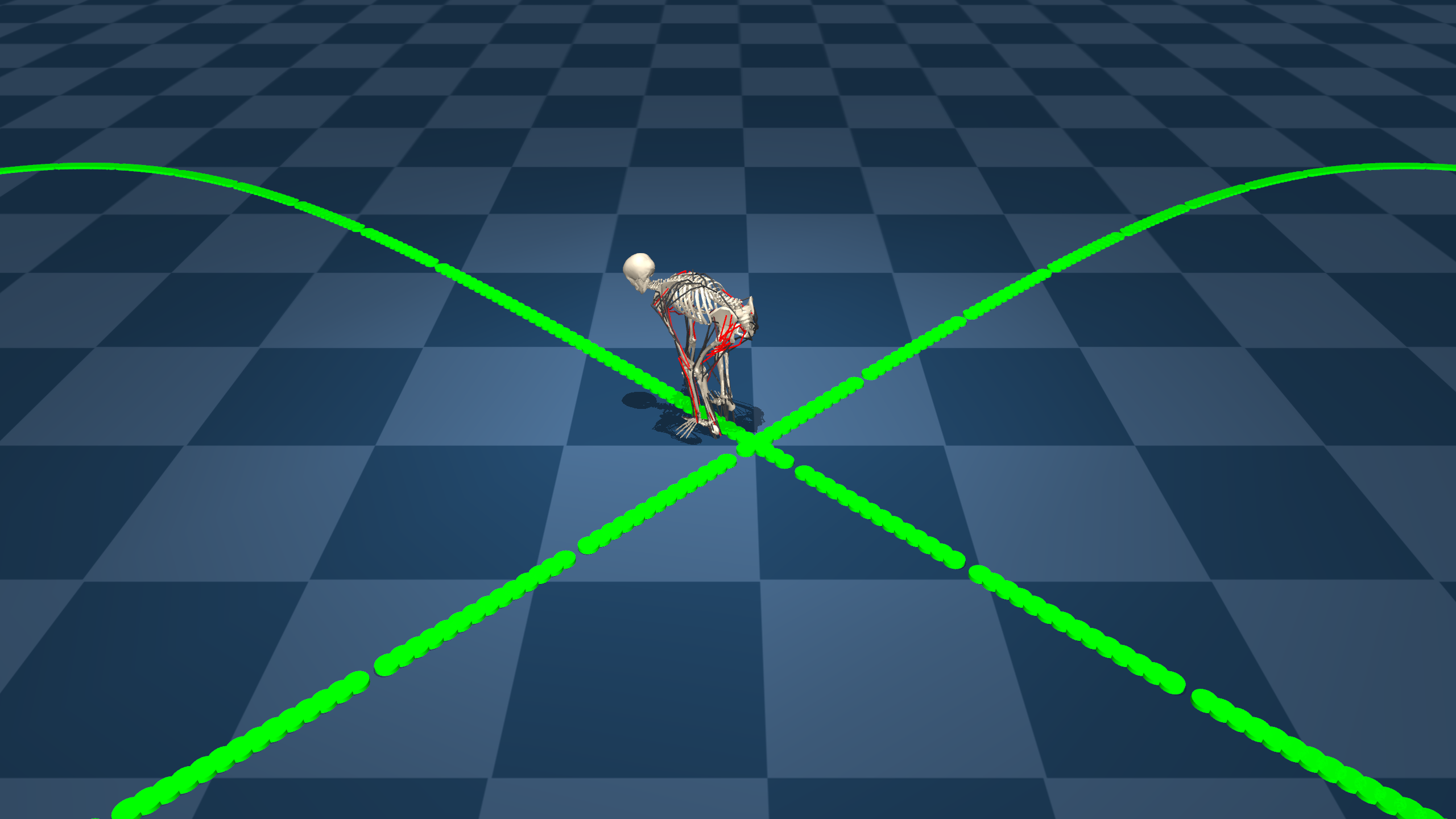}
  \caption{
FreeMusco is a motion-free framework that learns a latent representation of morphology-adaptive locomotion from musculoskeletal simulation, without motion data. The learned latent space enables high-level control for downstream tasks such as goal navigation and path following. The figure shows: (1) Humanoid locomotion control, (2) diverse motions sampled from the latent space, (3) goal navigation with Ostrich, and (4) path following with Chimanoid.
  }
  \label{fig:teaser}
\end{teaserfigure}

\maketitle

\section{Introduction}

Most existing research on controlling physically simulated characters relies on motion capture data to train models that reflect the motion distribution in the dataset.
In particular, a growing body of work has explored latent representation-based approaches, 
which embed character motions and transitions from motion datasets into low-dimensional latent spaces using generative models such as VAEs~\cite{yao_controlvae_2022}, GANs~\cite{CALM}, and VQ-VAEs~\cite{zhu_neural_2023}. 
These latent spaces are then used to train control policies capable of performing various downstream tasks.
While such \textit{motion-driven} approaches are highly effective in imitating diverse behaviors and solving complex control problems, they are inherently limited by the distribution of the training data. As a result, the generated motions tend to reflect the styles and ranges present in the motion dataset and often inherit performer-specific characteristics.

In contrast, the \textit{motion-free} approach—which learns control policies without relying on motion data—holds the potential to generate more generalizable behaviors that are not biased toward a particular dataset. This paradigm is particularly useful for simulating characters for which motion capture is impractical, such as animals or fictional creatures. However, generating natural and stable locomotion without motion data often requires incorporating prior knowledge about gait, such as careful tuning of joint power ratios and torque limits~\cite{wang_optimizing_2009,yu_learningsymmetric_2018}, control heuristics based on gait cycles~\cite{wang_optimizing_2012}, or curriculum learning strategies~\cite{yu_learningsymmetric_2018} that gradually increase task difficulty.
Moreover, while motion-driven methods have been widely extended to a range of downstream tasks, motion-free approaches have predominantly focused on learning control policies for a single behavior, such as walking or running~\cite{yu_learningsymmetric_2018,schumacher_emergence_2025}. This limitation stems from the structural challenge of lacking explicit guidance during training.

\ys{To this end, we propose FreeMusco, a motion-free framework that jointly learns latent representations and control policies for musculoskeletal characters.
Leveraging the musculoskeletal model as a strong prior, our method enables energy-aware and morphology-adaptive locomotion to emerge without motion references.
The approach relies solely on each character’s morphology and biomechanics, and we demonstrate its generalization across human, non-human, and synthetic morphologies.
Moreover, the learned latent space supports high-level control for downstream tasks such as goal navigation and path following—representing, to our knowledge, the first motion-free method to provide such capabilities.
Interestingly, different morphologies reveal distinct energy-efficient strategies in our framework: for example, Chimanoid tends to adopt quadrupedal gaits, whereas Humanoid consistently remains bipedal, indicating that energy-efficient gaits depend on morphology—and that our motion-free method allows such strategies to emerge.}

These results are enabled by our training framework and loss formulation.
We design the learning pipeline to generate diverse and physically plausible locomotion behaviors through model-based reinforcement learning (RL) under varying goal conditions.
A central component of this approach is the \textit{locomotion objective loss}, which integrates control, balancing, and biomechanical terms to guide \ys{characters} toward stable and goal-directed \ys{movement}. To better model the natural rhythm of animal locomotion, we apply a \textit{temporally averaged loss} formulation to selected \ys{terms, comparing} average simulated and target states over a time window rather than \ys{frame by frame}. In contrast to imitation-based settings—where reference motions inherently exhibit rhythmic oscillatory patterns—our motion-free setting uses randomly assigned goals that lack such patterns. The temporally averaged formulation therefore better encourages oscillatory behavior around the target, capturing the periodic nature of gait. 
\ys{Our training framework and loss design encourage diverse behaviors, while the musculoskeletal simulation acts as a prior that guides motion toward physiologically plausible ranges; together they support the emergence of natural locomotion.}

We further enhance behavioral diversity by randomizing the target pose and energy level during training.
This encourages the control policy to learn a wider range of postural and energetic behaviors, leading the latent space to reflect diverse movement styles.
At runtime, this allows flexible modulation of both form and intensity—even under the same target speed.

Our main contributions are summarized as follows:

\begin{itemize}
\item We propose a \textit{motion-free} framework that learns a latent space of diverse locomotion and a control policy end-to-end from musculoskeletal models, without motion capture data. The learned space supports high-level tasks such as goal navigation and path following.

\item \ys{Our framework generalizes to non-humanoid and synthetic characters, where the musculoskeletal prior promotes physiologically plausible and morphology-adaptive behaviors, e.g., quadrupedal gaits in Chimanoid versus bipedal in Humanoid.}

\item To guide motion generation without imitation, we introduce a \textit{locomotion objective loss} that \ys{integrates} control, balancing, and biomechanical terms, \ys{combined} with a \textit{temporally averaged loss} to encourage the emergence of \ys{oscillatory,} periodic gait patterns.

\item \ys{We enhance behavioral diversity by randomizing target poses and energy levels during training, which allows flexible modulation of locomotion form and intensity at runtime.}

\end{itemize}

\section{Related Work}

\subsection{Motion-Driven Physics-based Character Control} 

Early advances in motion-driven physics-based character control focused on constructing feedback controllers from motion capture data, often relying on handcrafted or optimized rules~\cite{lee_data-driven_2010,liu_terrain_2012}.

The advent of deep reinforcement learning (DRL) brought rapid progress to this paradigm, enabling more scalable and flexible policy learning directly from demonstration data.
DeepMimic~\cite{peng2018deepmimic} established imitation learning as a viable approach for producing high-quality motion tracking, inspiring a wave of research that sought to move beyond simple imitation.
Subsequent work aimed to broaden motion repertoires and task coverage by tracking synthesized motions from large datasets\ys{~\cite{bergamin_drecon_2019,Park2019,won_scalable_2020,wang2020unicon}}, exploring variations around reference clips~\cite{Lee:2021:Parameterized,chimeras22}, leveraging simplified physics models for generalization~\cite{kwon2020fast,kwon_adaptive_2023}, and decomposing control into modular components~\cite{bae_pmp_2023,xu_composite_2023}.

Another major direction \ys{has} focused on learning latent motion representations~\cite{DBLP:conf/iclr/MerelHGAPWTH19,ASE}, which enable skill reuse and goal-directed control. These representations have been further enhanced through task conditioning~\cite{CALM,dou_case_2023} and integration with differentiable world models~\cite{yao_controlvae_2022,won_physics-based_2022}. More recently, discrete latent spaces have been introduced to stabilize training and support unified multi-task control\ys{~\cite{yao_moconvq_2024,zhu_neural_2023,bae_versatile_2025}.
These latent representations have been applied beyond basic velocity control, to downstream tasks including
goal navigation~\cite{won_physics-based_2022,CALM,bae_versatile_2025}, path following~\cite{won_physics-based_2022,dou_case_2023}, skill control~\cite{yao_controlvae_2022,dou_case_2023}, text-to-motion~\cite{yao_moconvq_2024}, domain-specific applications such as sports~\cite{zhu_neural_2023,kim_physicsfc_2025,wang_strategy_2024,zhang_learning_2023}.
}

Beyond structured latent learning, recent generative approaches have focused on directly modeling the distribution of motion trajectories. 
Diffusion-based methods improve diversity and realism through denoising-based generation~\cite{serifi_robot_2024,truong_pdp_2024}, 
while transformer-based models such as MaskedMimic~\cite{tessler_maskedmimic_2024} learn action priors from masked partial sequences.

While prior work has leveraged motion data to build expressive and task-capable controllers, our approach explores a motion-free alternative. This formulation supports the generation of plausible, morphology-consistent locomotion without relying on demonstration data.

\subsection{Motion-Free Physics-based Character Control}

Motion-free approaches to physics-based character control emerged earlier than motion-driven methods.
A seminal work by Hodgins et al.~\shortcite{hodgins_animating_1995} showed that dynamic simulation with handcrafted control could generate realistic behaviors like running, bicycling, and vaulting.
This line of research continued with SIMBICON~\cite{yin2007simbicon}, a finite state machine (FSM)-based controller built with hand-tuned feedback rules for robust biped locomotion.  
Subsequent work expanded this paradigm by optimizing FSM controller parameters \cite{wang_optimizing_2009,wang_optimizing_2010} or integrating predictive models and trajectory generators \cite{coros2010generalized}.

With the advent of RL, more flexible motion-free training frameworks have emerged. 
Heess et al.~\shortcite{heess_learning_2016} and Peng et al.~\shortcite{peng2015dynamic,peng2016terrain} introduced RL-based systems that learn locomotion skills from scratch, relying solely on reward signals without any reference trajectories. 
To improve learning efficiency and behavior quality, techniques such as curriculum learning~\cite{allsteps20}, symmetry regularization, and energy-aware objectives~\cite{yu_learningsymmetric_2018} have been adopted. These strategies guide exploration and help shape motion patterns without relying on motion priors.

Although motion-free approaches were more common in the early stages of physics-based character control, the rise of DRL has made motion-driven methods dominant, enabling effective learning from large-scale demonstrations. However, motion-free methods offer unique advantages, such as improved generalization to novel morphologies and the ability to learn without reference motions.
Motivated by these strengths, we revisit motion-free learning and propose an advanced framework that jointly learns a latent representation and control policy, enabling diverse and morphology-adaptive locomotion without demonstration data.

\subsection{Musculoskeletal Character Control}

Musculoskeletal simulation enables physically grounded character control by modeling muscle dynamics and anatomical structure. Early work demonstrated that plausible locomotion could be achieved without motion capture using biologically inspired actuators and objectives \cite{wang_optimizing_2012,thomas_flexiblemuscle_2013}

Scalable frameworks have been proposed that combine trajectory optimization \cite{lee_locomotion_2014} and hierarchical control \cite{lee_scalablemuscle_2019} to coordinate hundreds of muscle units, guided by motion data. Building on these models, gait prediction networks have been introduced that generate personalized gait cycles conditioned on anatomical parameters, trained across a large number of anatomical conditions \cite{park_generative_2022}. 
Variational approaches have been applied to learn muscle-space skill representations from demonstration data, incorporating fatigue modeling for efficient skill transfer \cite{feng_musclevae_2023}.
Scalable frameworks have also been proposed to generate physiologically plausible muscle activations across diverse human movements using large-scale datasets~\cite{park_magnet_2025}.

In parallel, DRL has been applied to muscle-level control without demonstrations.

Recent methods explore synergy-based actuation, energy-based rewards, and exploration strategies to achieve robust and generalizable behaviors~\cite{kim_learning_2023,schumacher2023:deprl,schumacher_emergence_2025,berg_sar_2023}.

These studies demonstrate the potential of learning-based musculoskeletal control. Our work follows this direction by proposing a motion-free framework that learns diverse locomotion skills through model-based RL.

\section{FreeMusco Framework}

FreeMusco is a framework designed to control locomotion and learn its latent representation without relying on motion-capture data. By combining musculoskeletal physics simulation with model-based RL, the framework aims to achieve both biomechanical plausibility and expressive motion generation.
Unlike prior approaches that typically rely on imitating reference motions, FreeMusco explores the motion space based on varying goal conditions,
\ys{leveraging the musculoskeletal simulation that guides motion toward physiologically plausible ranges,}
enabling meaningful representations to emerge without imitation.
This section describes the overall training framework of FreeMusco, including its latent control architecture, training objectives, and techniques to promote diverse and controllable locomotion.

\paragraph{Muscle Dynamics}
To simulate biomechanically realistic movement, we use the Hill-type muscle-tendon actuator model implemented in the MuJoCo physics engine \cite{todorov_mujoco_2012}. This model computes muscle forces based on muscle length and contraction velocity, incorporating both active forces driven by muscle activation and passive elastic components. 
\ys{Muscle paths vary with the character’s posture and are automatically determined by line-segment-based routing from path points with predefined fixed offsets relative to each body segment. Along these paths, all forces act in a pulling direction and are translated into joint torques.}
In our system, the control input corresponds to muscle activation, allowing movements to emerge within physiologically plausible limits.
The detailed equations and modeling assumptions are provided in Section~A of the supplementary material.

\paragraph{Musculoskeletal Characters}

\begin{figure}[h] 
    \centering
    \subfloat[Humanoid]{   
        \includegraphics[trim=150 100 50 0, clip, width=0.32\linewidth]{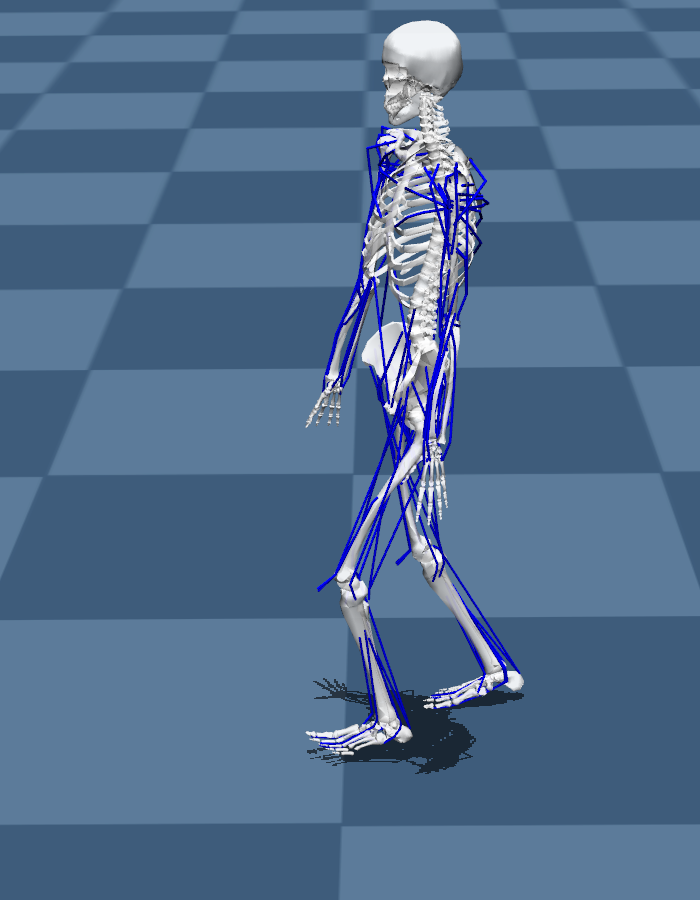}
    }%
    \subfloat[Ostrich]{         
        \includegraphics[trim=150 100 50 0, clip, width=0.32\linewidth]{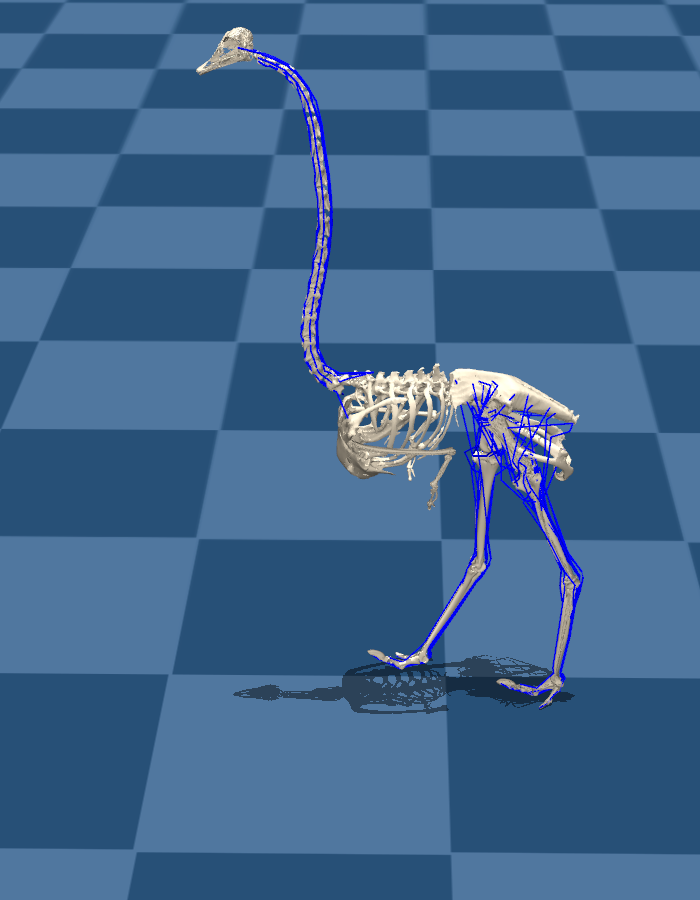}
    }%
    \subfloat[Chimanoid]{   
        \includegraphics[trim=50 0 150 100, clip, width=0.32\linewidth]{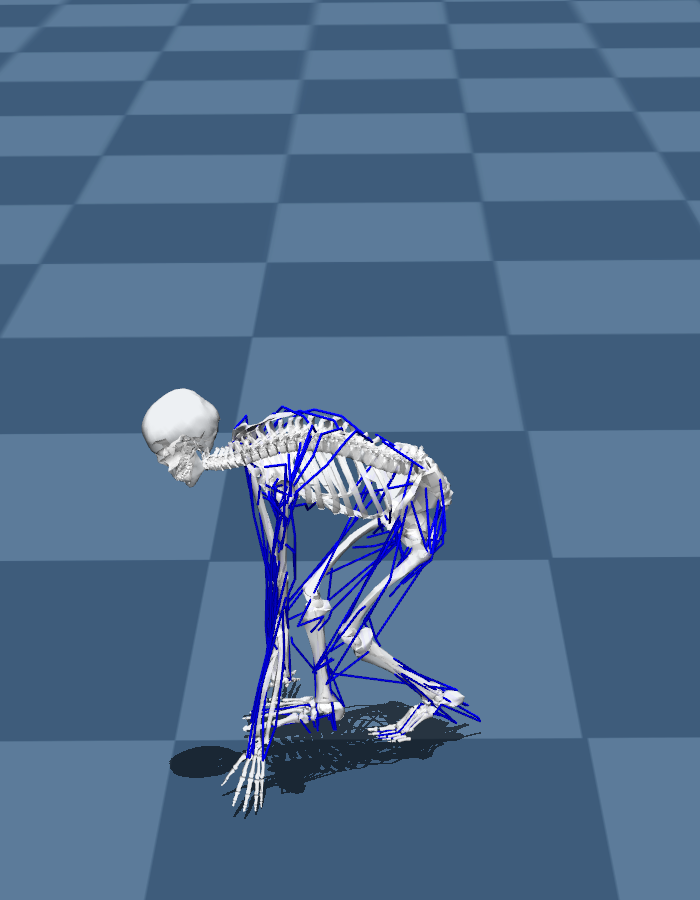}
    }%
    \caption{Musculoskeletal characters used for training and simulation.}
    \label{fig:characters}
\end{figure}

We demonstrate our method on three musculoskeletal characters, shown in Figure~\ref{fig:characters}.

\begin{description}
    \item[Humanoid] A human musculoskeletal model with 120 muscles from \cite{lee_locomotion_2014}.

    \item[Ostrich] A 120-muscle ostrich model from \cite{barbera_ostrichrl_2021}.

    \item[Chimanoid] A fictional 120-muscle character created by modifying the Humanoid model, with changes such as elongated arms and shortened legs.

\end{description}

For details on the musculoskeletal characters, please see Section~B of the supplementary material.

\subsection{Latent Control Architecture}
\label{sec:architecture}

\ys{Model-based RL} with a world model enables end-to-end training by directly propagating gradients through differentiable dynamics, and has shown strong performance across various tasks, including single-clip imitation~\cite{fussel_supertrack_2021}, latent space embedding from motion datasets~\cite{yao_controlvae_2022}, and fatigue-aware muscle control~\cite{feng_musclevae_2023}.
Our architecture builds on ControlVAE~\cite{yao_controlvae_2022}, which incorporates a posterior encoder, prior encoder, decoder (policy), and a differentiable world model. While ControlVAE was originally developed for motion imitation, we extend this structure to a motion-free setting, enabling the learning of the latent representation and policy directly from muscle-actuated dynamics. We also adopt its three-stage training procedure, adapted to operate without reference motions.

\begin{figure}
    \centering
    
    \includegraphics[trim=0 60 0 20, clip, width=0.45\textwidth]{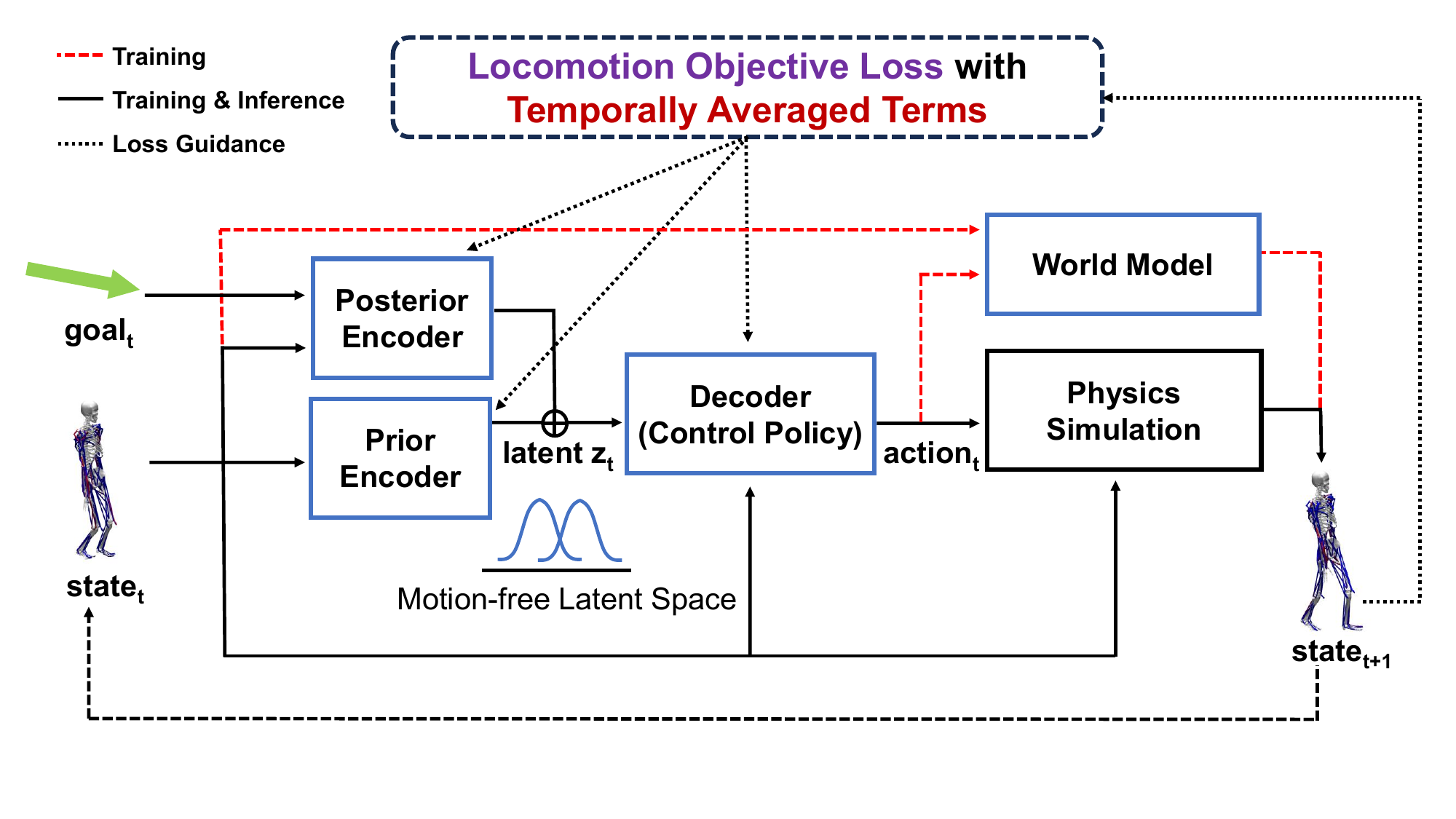}
    \caption{Overview of the FreeMusco framework.}
    \label{fig:overview}
\end{figure}

The posterior encoder learns the posterior distribution $q(\mathbf{z}_t | \mathbf{s}_t, \mathbf{g}_t)$, which models the distribution over latent variables $\mathbf{z}_t$ that achieve the goal $\mathbf{g}_t$ given the current character state $\mathbf{s}_t$ (Figure~\ref{fig:overview}).  
The prior encoder learns the state-conditioned prior distribution $p(\mathbf{z}_t | \mathbf{s}_t)$, representing the distribution of feasible latent variables based solely on the current state.
The decoder (policy) $\pi(\mathbf{a}_t | \mathbf{s}_t, \mathbf{z}_t)$ takes the state $\mathbf{s}_t$ and a latent vector $\mathbf{z}_t$ as input, and outputs muscle activations $\mathbf{a}_t$ for the character. During training, $\mathbf{z}_t$ is formed by summing the outputs from both the prior and posterior encoders.
The world model $\omega(\mathbf{s}_{t+1}, \mathbf{e}_t | \mathbf{s}_t, \mathbf{a}_t)$ predicts the next state $\mathbf{s}_{t+1}$ and the metabolic energy expenditure $\mathbf{e}_t$ for each muscle, given the current state and the applied activations.
Each of the four components is modeled as a normal distribution with a fixed diagonal covariance $\sigma^2 \mathbf I$ and a learnable mean $\mathbf \mu_q$, $\mathbf \mu_p$, $\mathbf \mu_\pi$, and $\mathbf \mu_\omega$, predicted by the corresponding neural networks.  
The prior and posterior distributions are regularized using a KL divergence loss. 
\ys{In practice, the posterior encoder predicts a residual over the prior, and the posterior distribution is defined by adding this residual to the prior encoder output, from which the latent variable $\mathbf{z}_t$ is sampled.
The KL divergence loss thus regularizes the posterior encoder's residual output, ensuring that the residual remains small and only captures goal-relevant variations.}

Unlike prior work such as~\cite{yao_controlvae_2022}, our method does not assume access to a future reference motion.  
Instead, the posterior encoder is conditioned on a goal signal $\mathbf{g}_t$, which provides behavior-guiding information.  
We explore multiple goal configurations, each specifying a different combination of target features such as movement velocity, facing direction, pose, and desired energy expenditure.  
All configurations are centered around a target velocity, with optional extensions that enable control over movement style and physical characteristics.
\ys{Another difference is that our world model also estimates per-muscle metabolic energy expenditure $\mathbf{e}_t$.}
Together, these components enable the FreeMusco model to learn diverse and physically plausible motions across varying goal conditions, forming an energy-aware latent representation without relying on imitation.

For details on this architecture, please see Section~C of the supplementary material.

\paragraph{Training Procedure}

The model is trained by repeating the following three stages in each iteration.

\textit{Stage 1:} A random goal \( \mathbf{g}_t \) is sampled and fed into the current model to produce actions $\mathbf{a}_t$, which are applied to the physics simulator to obtain transitions \( (\mathbf{s}_t, \mathbf{a}_t, \mathbf{s}_{t+1}) \). These transitions, the goal \( \mathbf{g}_t \), and the metabolic energy expenditure \( \mathbf{e}_t \) are stored in a trajectory buffer for training.
While this procedure is common in model-based RL settings with differentiable world models, our system differs in that it explores and collects training data without relying on any reference motion, instead using randomly sampled goals to induce diverse behaviors across the action space.

\textit{Stage 2:} We train the world model using the data stored in the buffer.
We minimize the $\ell_1$ error between predicted and ground-truth rollouts over $T_W$ steps, including both state $\mathbf s_{t+1}$ and metabolic energy expenditure $\mathbf e_t$.
The predicted energy is incorporated into a dedicated loss term in the next stage, which serves as a regularizer to promote energy-aware and physically stable motion.

\textit{Stage 3:} With the world model frozen, we generate $T_P$-step trajectories using the current encoder-decoder and world model, starting from an initial state $\mathbf s_0$ and goal $\mathbf{g}_t$ retrieved from a randomly sampled trajectory in the buffer, and update the model parameters by minimizing the loss $L_{\text{VAE}}$.
It combines the locomotion objective loss $L_{\text{objective}}$ and the KL divergence loss $L_{\text{KL}}$ between the latent prior and posterior:
\begin{align}
    L_{\text{VAE}} = L_{\text{objective}} + \beta \cdot L_{\text{KL}},
\end{align}
Here, $L_{\text{objective}}$ integrates control, balancing, and biomechanical terms to guide the policy toward stable, purposeful, and physically plausible motion without relying on imitation, as detailed in Sections \ref{sec:avg_loss} and \ref{sec:obj_loss}.
$\beta$ is a weighting factor scheduled during training.

The details on this training procedure are provided in the Section~D of the supplementary material.

\subsection{Temporally Averaged Loss}
\label{sec:avg_loss}

We introduce a \textit{temporally averaged loss} to promote biologically plausible locomotion by accounting for natural oscillations in movement. 
Instead of enforcing per-step consistency—as in conventional formulations—this loss compares averages of the simulated and target states over a short temporal window.

This design is motivated by the observation that, in real animal locomotion, attributes like velocity, posture, and up direction fluctuate rhythmically around a desired value. Enforcing frame-by-frame targets in such settings often leads to overly constrained and unnatural motion.
While per-step loss has proven effective in imitation-based frameworks~\cite{fussel_supertrack_2021,yao_controlvae_2022}, where reference trajectories naturally exhibit rhythmic variation, our motion-free setting relies on randomized low-level goals lacking such patterns. Thus, temporally averaged supervision better captures the inherent variability of gait.

Accordingly, we define main terms in the locomotion objective loss $L_{\text{objective}}$ using the following temporally averaged formulation:
\begin{align}
    \mathcal{L}_{\text{avg}}(\{\overline{\mathbf x}_t\}, \{\mathbf x_t\}) &= \left\lVert  \frac{1}{T_P} \sum_{t=0}^{T_P - 1} \gamma^t \overline{\mathbf x}_t - \frac{1}{T_P} \sum_{t=0}^{T_P - 1} \gamma^t \mathbf x_t \right\rVert_1,
    \label{eq:avg_loss}
\end{align}
where $\{\overline{\mathbf x}_t\}$ and $\{\mathbf x_t\}$ denote length-$T_P$ sequences of target and simulated values for a motion-related quantity, such as velocity, posture, height, or up direction.
Note that $\{\overline{\mathbf x}_t\}$ are retrieved from trajectories sampled from the buffer, while $\{\mathbf x_t\}$ are generated by rolling out the current model from the same initial state and goal as the sampled trajectories (see Stage 3 of the training procedure).
The discount factor $\gamma^t$ accounts for the fact that the synthetic trajectory is generated using an approximate world model and may accumulate errors over time, assigning lower weights to potentially inaccurate future states, as in ~\cite{yao_controlvae_2022}.

In contrast, the standard per-step formulation, which is used for some terms in $L_{\text{objective}}$, can be defined as follows:
\begin{align}
    \mathcal{L}_{\text{step}}(\{\overline{\mathbf x}_t\}, \{\mathbf x_t\}) &= \frac{1}{T_P} \sum_{t=0}^{T_P - 1} \gamma^t \cdot \left\lVert \overline{\mathbf x}_t - \mathbf x_t \right\rVert_1.
    \label{eq:step_loss}   
\end{align}
We used $T_P = 32$ steps and a discount factor of $\gamma = 0.99$ in our experiments.

\subsection{Locomotion Objective Loss}
\label{sec:obj_loss}

To train the encoders and decoder (policy) for goal-directed locomotion, we define the objective loss $L_{\text{objective}}$ as a combination of control, balancing, and biomechanical terms.
This loss encourages the character to move in a manner consistent with the goal $\mathbf{g}_t$, while maintaining balance and exhibiting stable and physically plausible motion without relying on imitation.
As described in Section~\ref{sec:architecture}, each target component in the goal $\mathbf{g}_t$ randomly sampled during Stage~1 of the training procedure.

\paragraph{Control Objective}
The control objective comprises $L_{\text{vel}}$ and $L_{\text{dir}}$, which guide the character to move in a manner aligned with the control targets specified in $\mathbf g_t$:
\begin{align}
    L_{\text{vel}} &= \mathcal{L}_{\text{avg}}(\{\overline{\mathbf v}_t\}, \{\mathbf v_t\}),\\
    L_{\text{dir}} &= \mathcal{L}_{\text{step}}(\{\overline{\mathbf d}_t\}, \{\mathbf d_t\}),
\end{align}
where $\{\overline{\mathbf v}_t\}$ and $\{\mathbf v_t\}$ denote the length-$T_P$ sequences of target and simulated horizontal root velocities, and $\{\overline{\mathbf d}_t\}$ and $\{\mathbf d_t\}$ denote the corresponding sequences of root facing directions.
$L_{\text{vel}}$ penalizes deviations from the target velocity in the local frame, while $L_{\text{dir}}$ aligns the root facing direction with the target direction.
We apply the per-step loss to $L_{\text{dir}}$, as we empirically found that temporal averaging reduces control responsiveness for directional alignment, which benefits from immediate per-frame feedback.
In goal configurations where $\{\overline{\mathbf d}_t\}$ is not explicitly included in $\mathbf{g}_t$, it is set to match the direction of the velocities in $\{\overline{\mathbf v}_t\}$, implying that the character faces its movement direction.

\paragraph{Balancing Objective}

The balancing objective includes \( L_{\text{height}} \) and \( L_{\text{up}} \), which help the character maintain an upright and stable posture during motion:
\begin{align}
    L_{\text{height}} &= \frac{1}{T_P} \sum_{t=0}^{T_P - 1} \gamma^t \cdot \left\lVert \max(0, \overline{h}_t - h_t) \right\rVert_1, \\
    L_{\text{up}} &= \mathcal{L}_{\text{avg}}(\{\overline{\mathbf u}_t\}, \{\mathbf u_t\}).
\end{align}

Here, \( L_{\text{height}} \) penalizes cases where the root height \( h_t \) (e.g., pelvis) falls below a predefined target threshold \( \overline{h}_t \), which is set slightly lower than the initial standing root height.
The loss is activated only when the character drops below the threshold, encouraging it to stay upright and avoid falling. Since it disregards values above the target, it naturally permits vertical oscillation during walking.
Because of this design, temporal averaging has little impact on the behavior of \( L_{\text{height}} \), and we therefore adopt a per-step formulation for simplicity. In practice, we observed no meaningful difference compared to its temporally averaged variant.

\( L_{\text{up}} \) aligns the character root’s local up vector sequence \( \{\mathbf{u}_t\} \) with the target up direction sequence \( \{\overline{\mathbf{u}}_t\} \), which is set to the global \( y \)-axis at every timestep.
This encourages the character to maintain an upright posture throughout motion, supporting stable and balanced locomotion.

\paragraph{Biomechanical Objective}
The biomechanical objective combines \( L_{\text{pose}} \) and \( L_{\text{energy}} \) to encourage biomechanically plausible and energy-efficient movement:
\begin{align}
    L_{\text{pose}} &= \mathcal{L}_{\text{avg}}(\{\overline{\mathbf p}_t\}, \{\mathbf p_t\}),
    \label{eq:L_pose}\\
    L_{\text{energy}} &= \mathcal{L}_{\text{step}}(\{\overline{e}_t\}, \{e_t\}),
    \label{eq:L_energy}
\end{align}
where \( \{\overline{\mathbf{p}}_t\} \) and \( \{\mathbf{p}_t\} \) denote the sequences of target and simulated poses, 
and \( \{\overline{e}_t\} \) and \( \{e_t\} \) represent the sequences of target and simulated total metabolic energy expenditure at each timestep.

Our motion-free setting does not assume access to demonstration-based priors. However, we found that learning failed when no pose regularization was applied, indicating that some form of guidance is necessary.
To this end, we set the target pose sequence \( \{\overline{\mathbf{p}}_t\} \) based on the character’s rest pose—i.e., the default standing posture, providing a minimal guide.
This regularization does not strictly enforce a specific posture, but instead allows for wide oscillations around the rest pose using the temporally averaged loss.
Notably, we observed that even with \ys{the upright standing pose,} quadrupedal characters naturally developed four-legged locomotion patterns, demonstrating that the guide does not hinder the emergence of morphology-appropriate behaviors.

The simulated total energy \( e_t \) is computed by summing the per-muscle metabolic energy values predicted by the world model at time \( t \). The target values \( \{\overline{e}_t\} \) are set to zero to guide the policy toward minimizing energy usage. \( L_{\text{energy}} \) penalizes excessive energy consumption at each frame using the per-step loss, promoting consistent energy efficiency throughout the motion. This term acts as a regularizer that reflects the biological tendency of animals to move efficiently.

\paragraph{Total Objective Loss}

The total loss \( L_{\text{objective}} \) is defined as:
\begin{align}
L_{\text{objective}} = &\ 
 w_v L_{\text{vel}} + w_d L_{\text{dir}}
 +\ w_h L_{\text{height}} + w_u L_{\text{up}} \nonumber \\
& +\ w_p L_{\text{pose}} + w_e L_{\text{energy}},
\end{align}
where \( w_v, w_d, w_h, w_u, w_p, \) and \( w_e \) are scalar weights.
\ys{The details on the locomotion objective loss, including the weights, are provided in Section~E of the supplementary material.}

\paragraph{Target Pose and Energy Randomization}

In our motion-free setting, we define the target pose and energy in \( L_{\text{pose}} \) (Equation~\eqref{eq:L_pose}) and \( L_{\text{energy}} \) (Equation~\eqref{eq:L_energy}) using the character’s \ys{default standing} pose and zero energy expenditure, respectively. While these settings
help stabilize early training, they may limit the diversity of learned behaviors by failing to capture sufficient variation in plausible postures and energy expenditure levels.

To address this, we randomly sample the target pose and energy from predefined ranges during Stage~1 of the training procedure, as a part of the goal \( \mathbf{g}_t \).
These sampled targets are also used in \( L_{\text{pose}} \) and \( L_{\text{energy}} \) as target sequences, allowing the model to learn across a broader range of postures and energy expenditure levels.
At runtime, the character can be guided toward higher-energy movement or stylized gaits with distinct postural characteristics.

\section{Experimental Result}

Our encoder, decoder, and world model operate at 33\,Hz. The physics simulator runs at a finer resolution, with a simulation timestep of \(1/(33 \times 15)\) seconds. At each control step, the muscle activation output is held constant for 15 simulation steps, and the world model predicts the state transition over a duration of \(1/33\) seconds.

FreeMusco model and it latent space was trained using an NVIDIA GeForce RTX 4070 GPU and four cores of an AMD Ryzen 7 7800X3D CPU.
\ys{Training converged within about one day across all morphologies and goal configurations.}
The high-level policy, used for goal navigation or path following tasks, was trained separately and required around 30 minutes. The results are best observed in the accompanying video. \ys{The code and data used in this study are available at https://github.com/palkan21/FreeMusco.}

\subsection{Locomotion Control}

\begin{figure}
    \centering
    
    \includegraphics[trim=50 30 50 30, clip,width=0.092\textwidth]{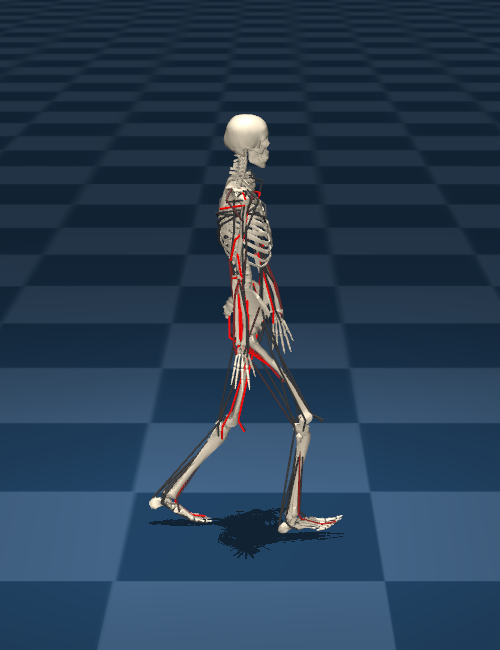}
    \includegraphics[trim=50 30 50 30, clip,width=0.092\textwidth]{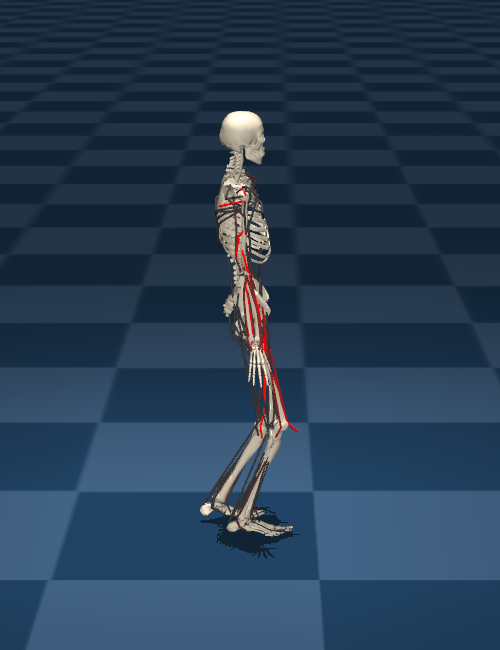}
    \includegraphics[trim=50 30 50 30, clip, width=0.092\textwidth]{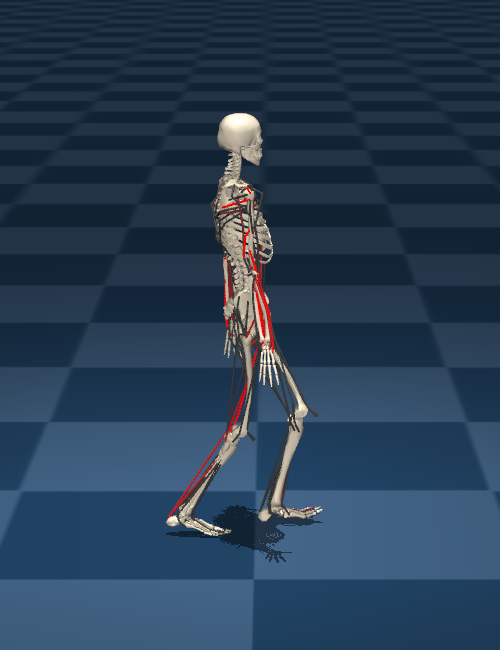}
    \includegraphics[trim=50 30 50 30, clip,width=0.092\textwidth]{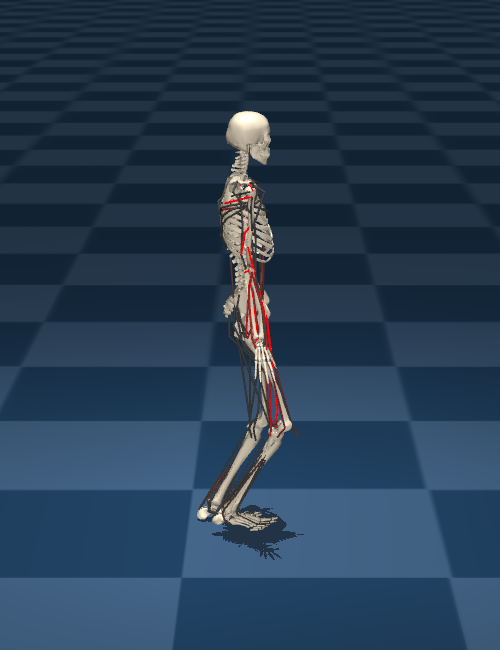}
    \includegraphics[trim=50 30 50 30, clip,width=0.092\textwidth]{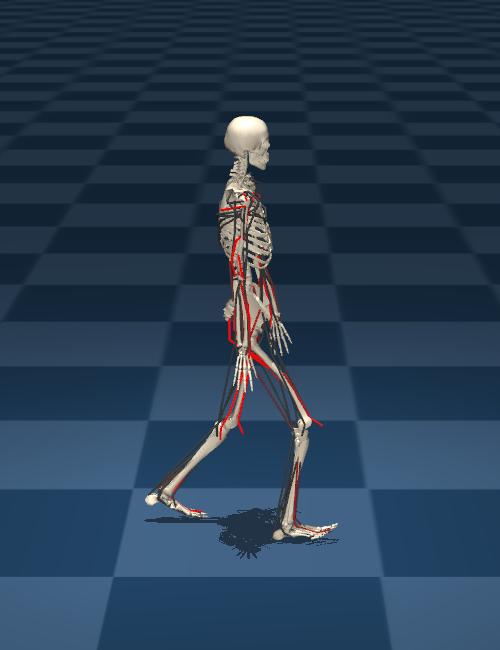}

    \includegraphics[width=0.092\textwidth]{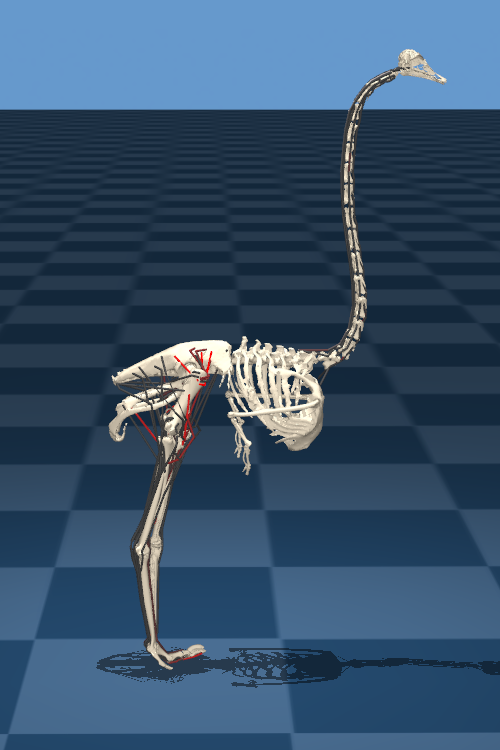}
    \includegraphics[width=0.092\textwidth]{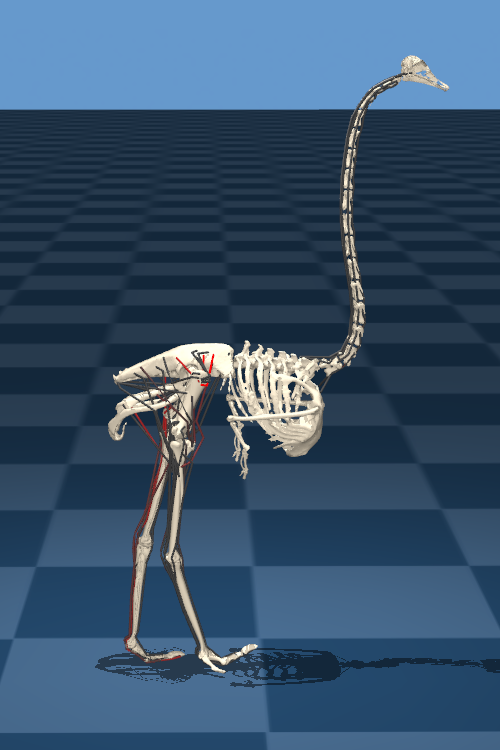}
    \includegraphics[width=0.092\textwidth]{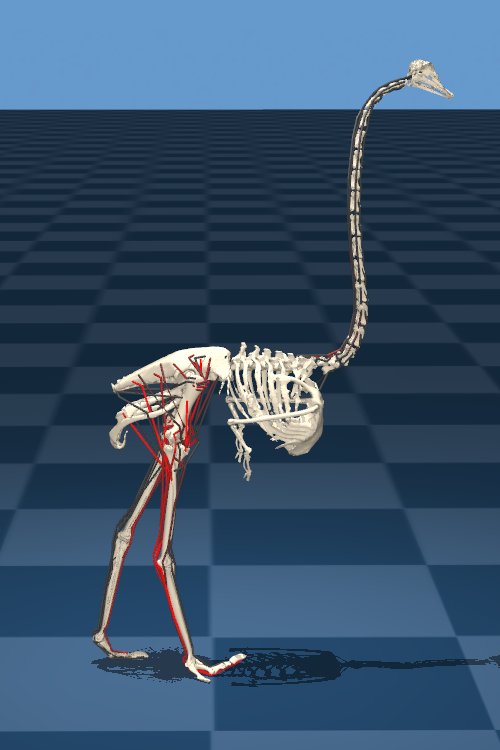}
    \includegraphics[width=0.092\textwidth]{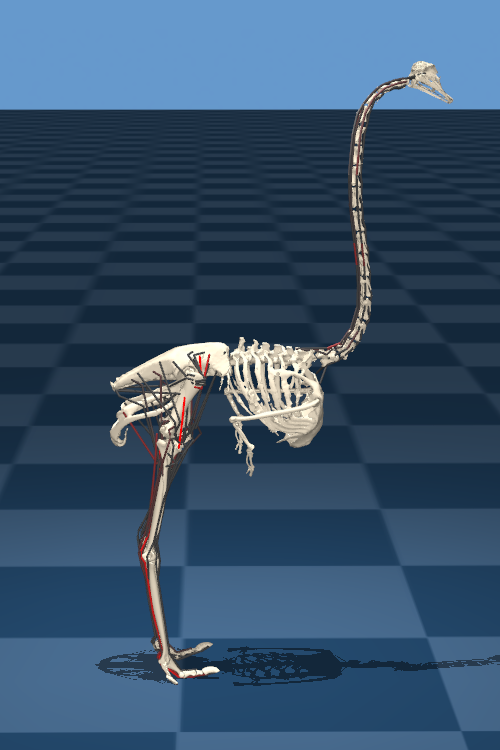}
    \includegraphics[width=0.092\textwidth]{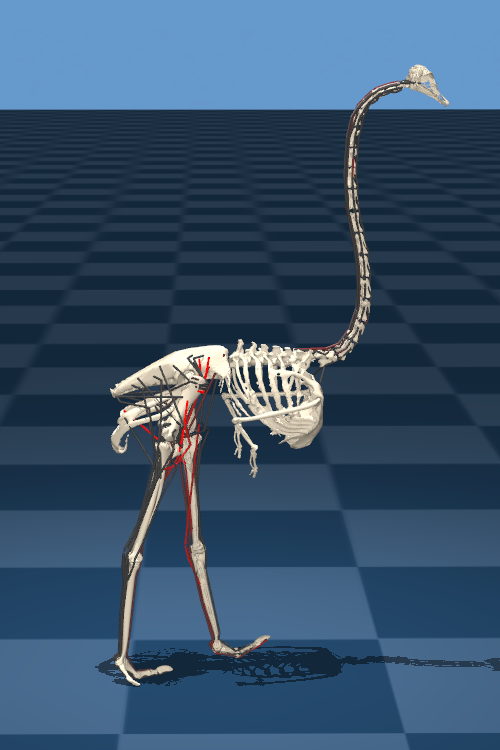}
    
    \caption{Locomotion sequences of Humanoid (top) and Ostrich (bottom).
    \ys{The policies are trained under the \textit{Velocity Only} configuration where the goal includes a target horizontal velocity.}}
    \label{fig:locomotion_human_ostrich}
\end{figure}

\begin{figure}
    \centering
    \includegraphics[trim=300 100 400 200, clip, width=0.325\linewidth]{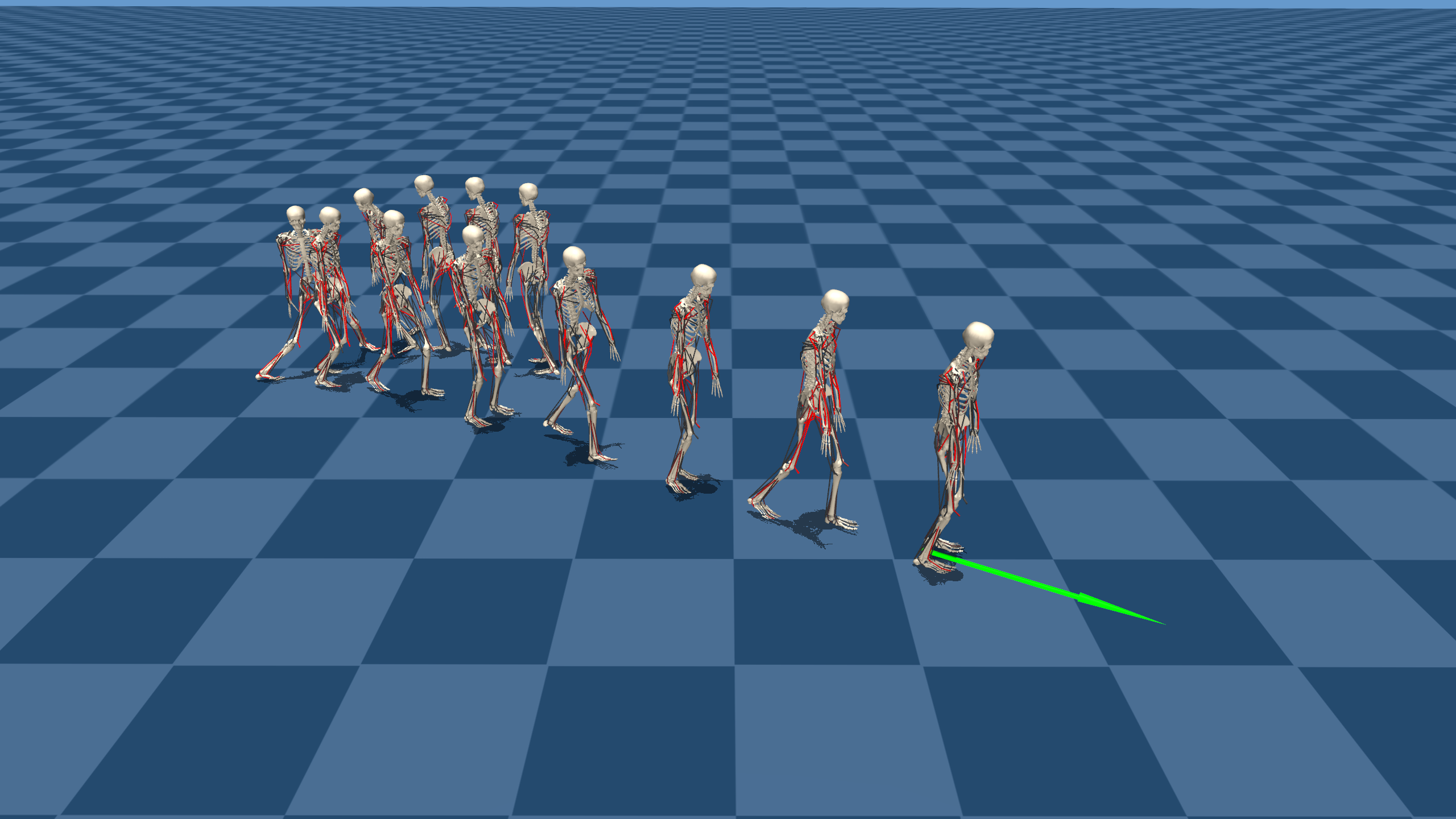}
    \includegraphics[trim=300 100 400 200, clip, width=0.325\linewidth]{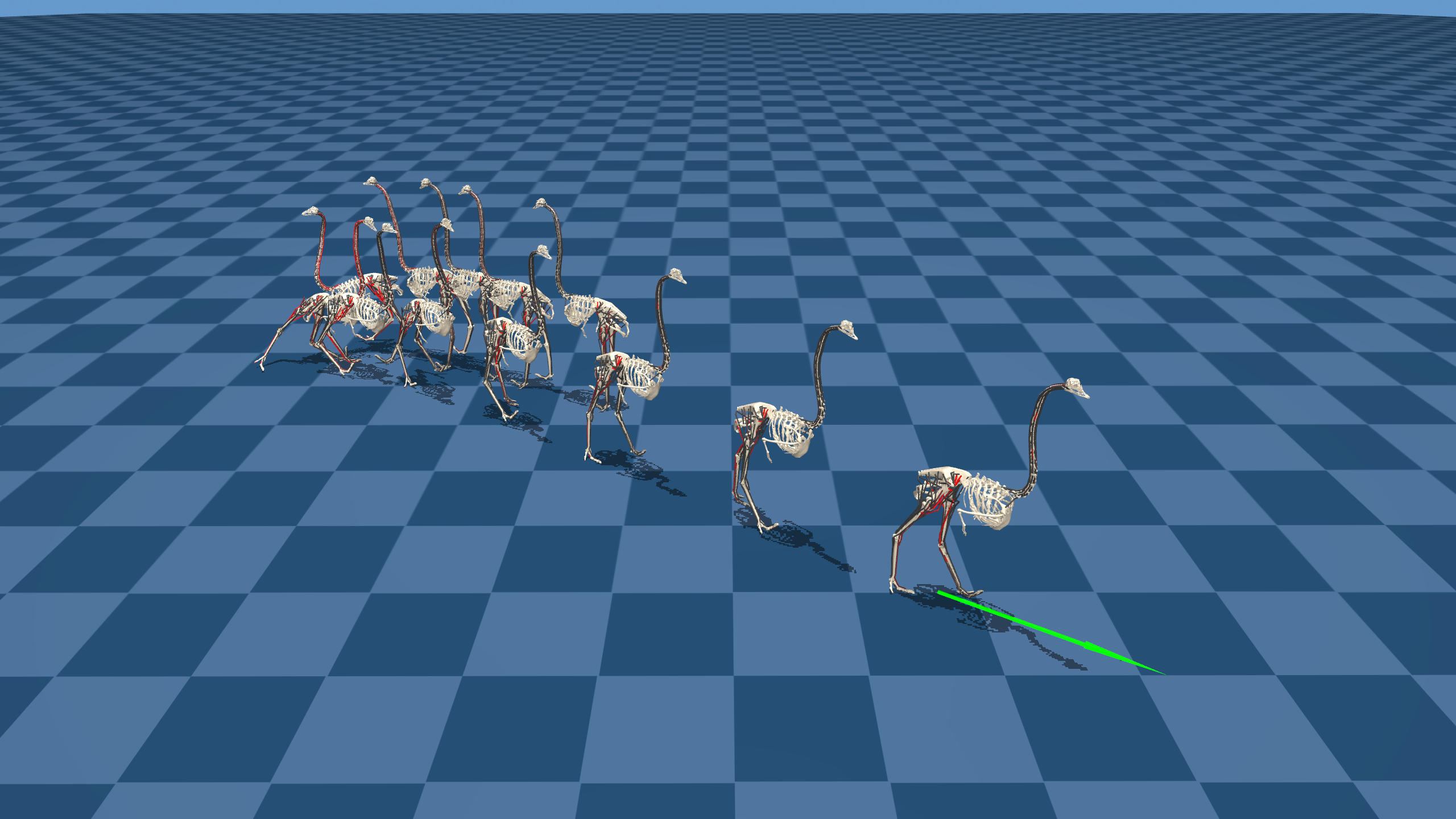}
    \includegraphics[trim=300 100 400 200, clip, width=0.325\linewidth]{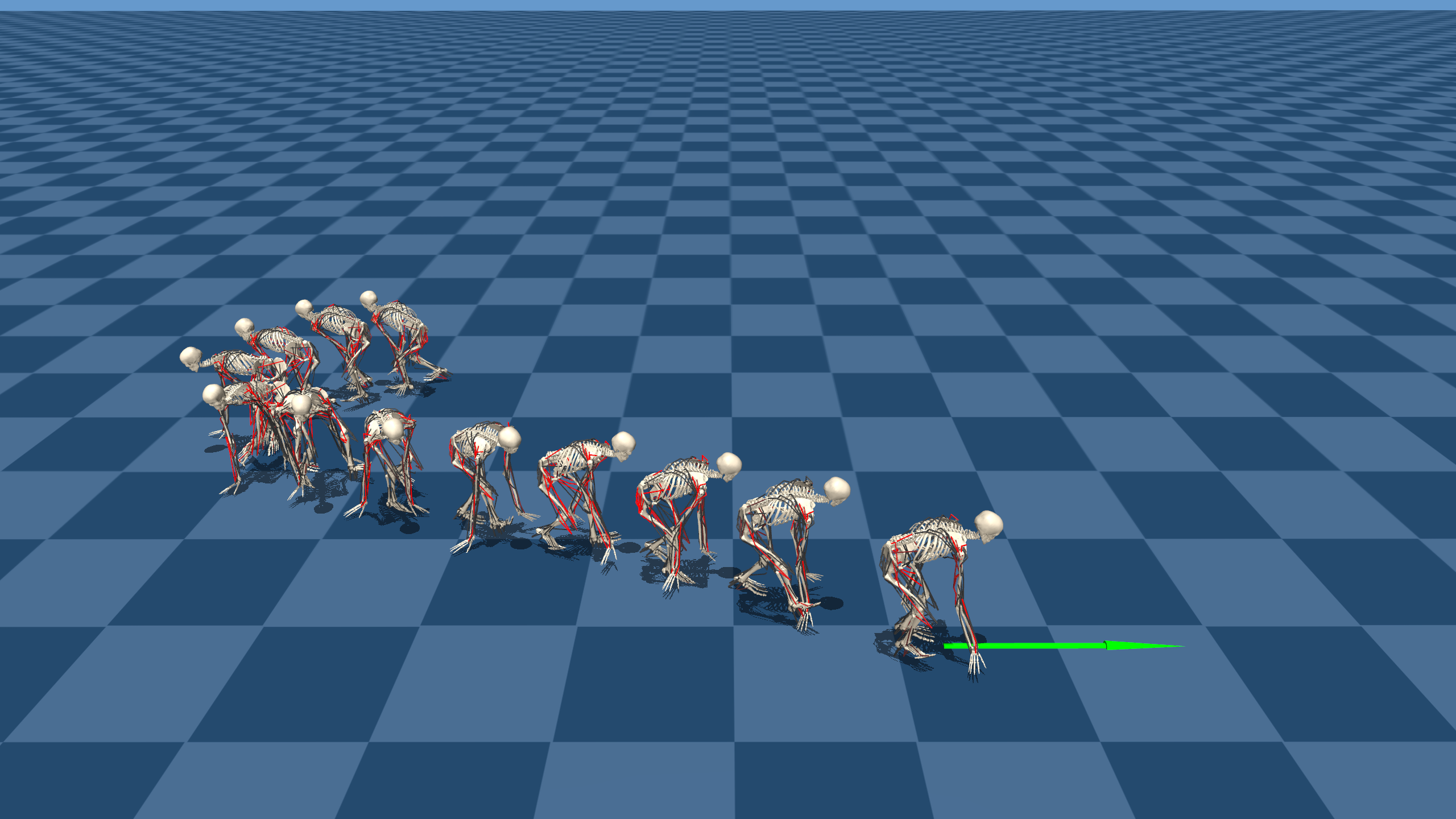}
    \caption{
    \ys{Locomotion sequences of Humanoid (left), Ostrich (center), and Chimanoid (right), trained under the \textit{Velocity Only} configuration.}
    }    
    \label{fig:locomotion_direction_last}
\end{figure}

\begin{figure}
    \centering
    
    \includegraphics[trim=50 30 50 30, clip, width=0.092\textwidth]{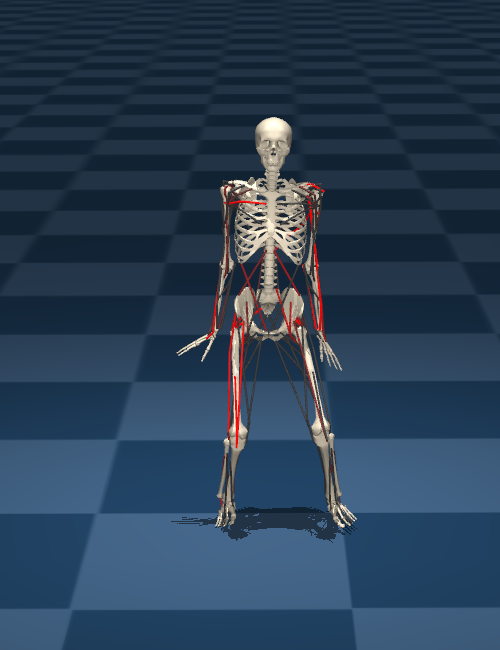}
    \includegraphics[trim=50 30 50 30, clip,width=0.092\textwidth]{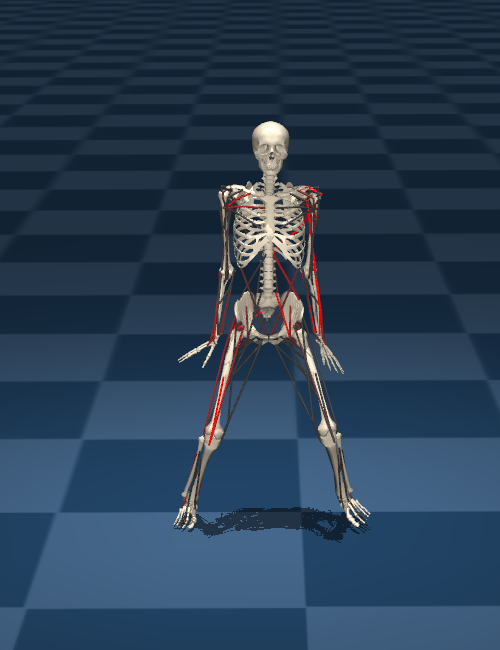}
    \includegraphics[trim=50 30 50 30, clip,width=0.092\textwidth]{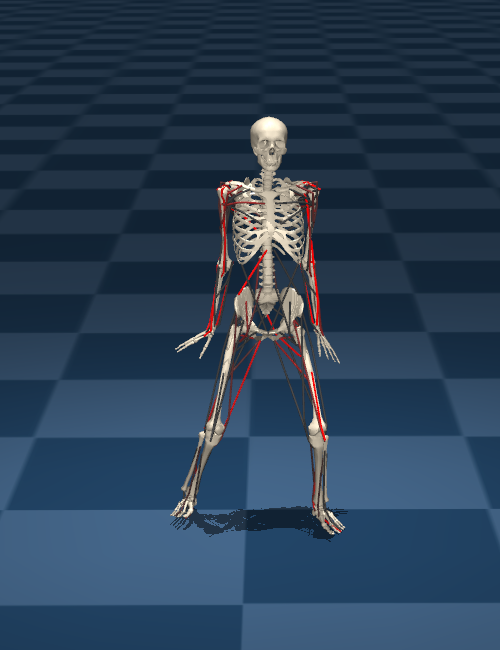}
    \includegraphics[trim=50 30 50 30, clip,width=0.092\textwidth]{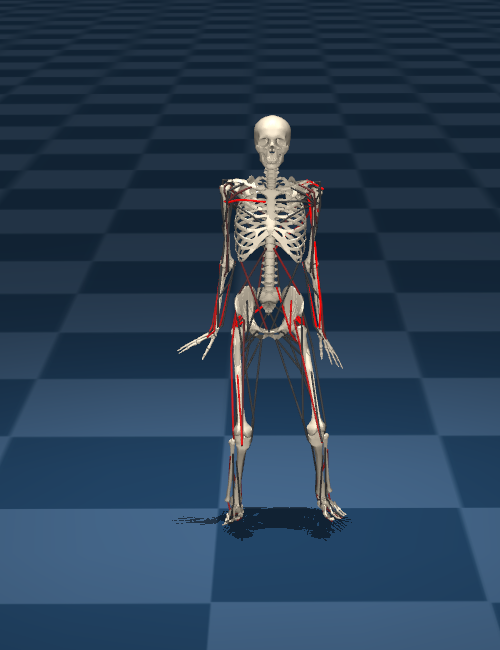}
    \includegraphics[trim=50 30 50 30, clip,width=0.092\textwidth]{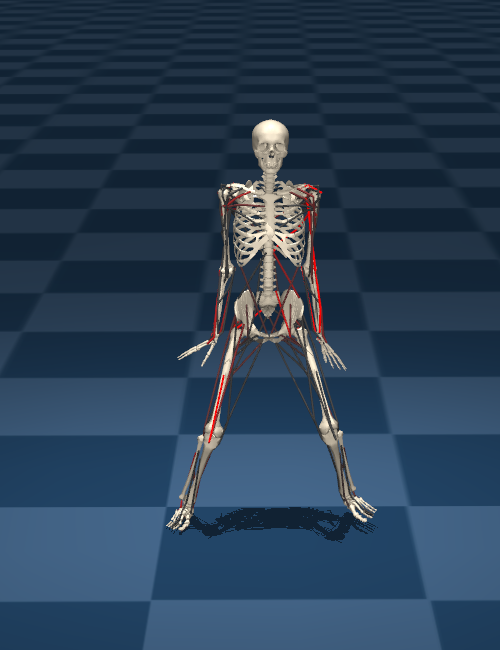}
    \includegraphics[trim=50 30 50 30, clip,width=0.092\textwidth]{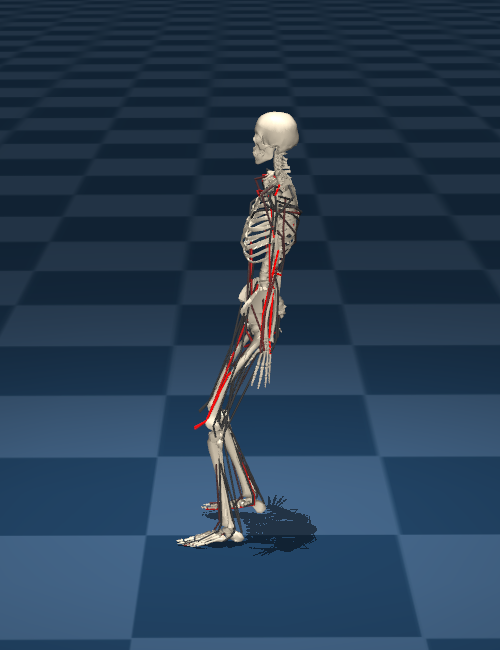}
    \includegraphics[trim=50 30 50 30, clip,width=0.092\textwidth]{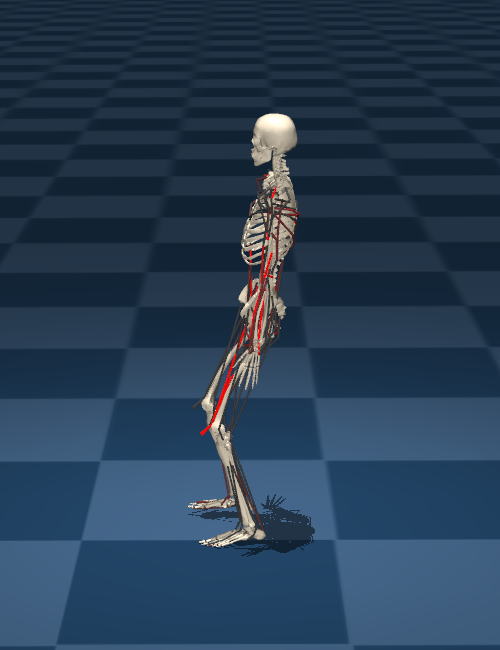}
    \includegraphics[trim=50 30 50 30, clip,width=0.092\textwidth]{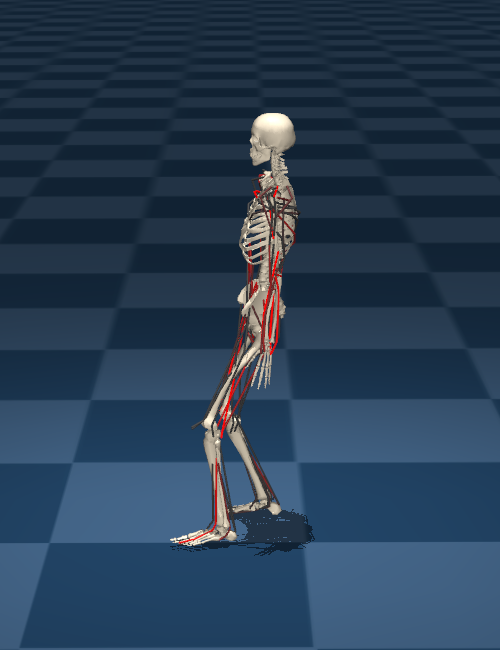}
    \includegraphics[trim=50 30 50 30, clip,width=0.092\textwidth]{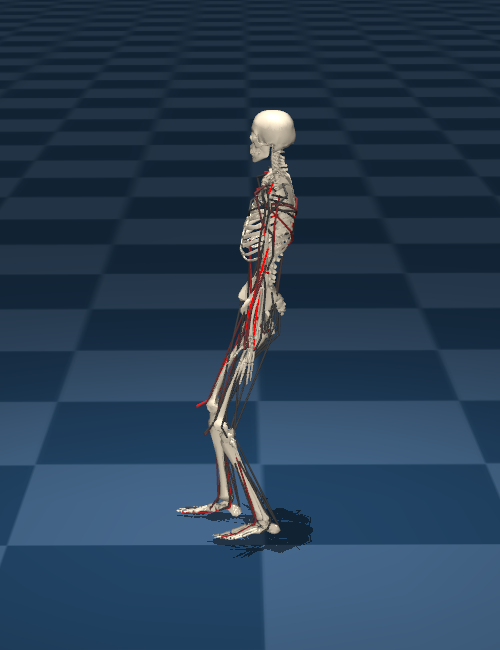}
    \includegraphics[trim=50 30 50 30, clip,width=0.092\textwidth]{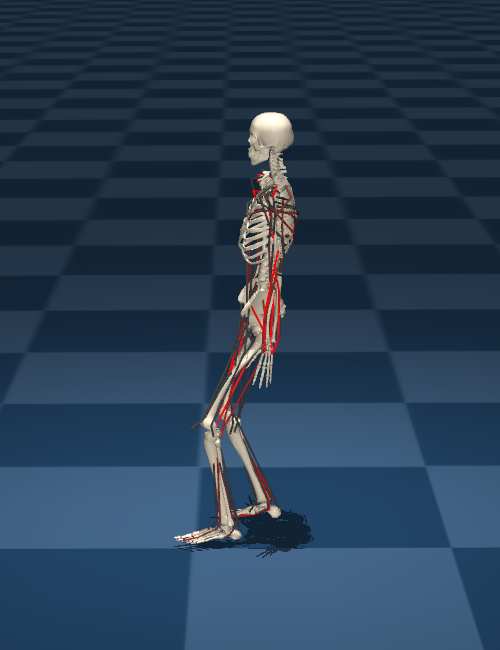}

    \caption{
    \ys{Sidewise walk (top) and backward walk (bottom) sequences of Humanoid.
    The policies are trained under the \textit{Velocity + Direction} configuration where the goal additionally includes a target facing direction.}
    }    
    \label{fig:direction_humanoid_last}
\end{figure}

\begin{figure}
    \centering
    
    \includegraphics[trim=60 30 30 130,  clip,width=0.192\columnwidth]{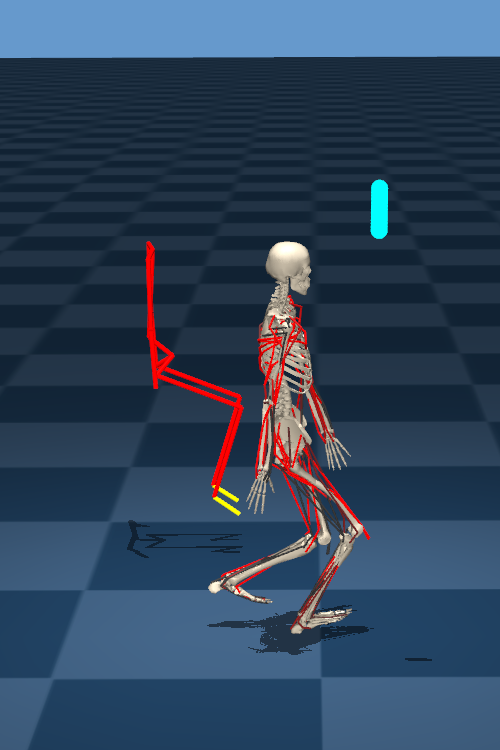}
    \includegraphics[trim=60 30 30 130, clip,width=0.192\columnwidth]{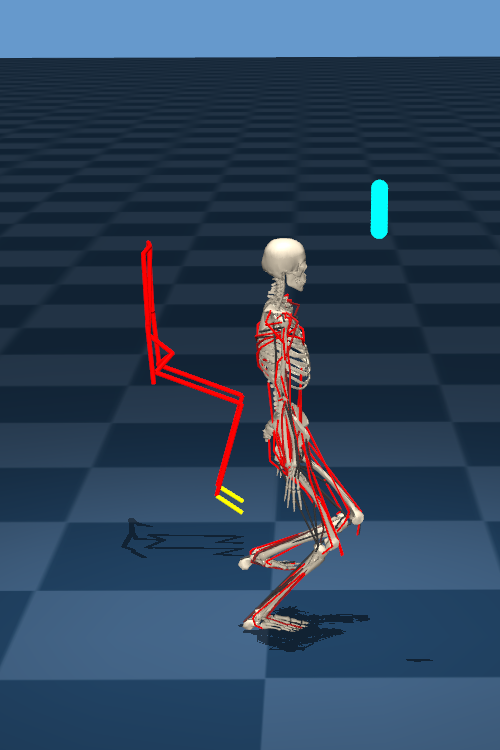}
    \includegraphics[trim=60 30 30 130,  clip,width=0.192\columnwidth]{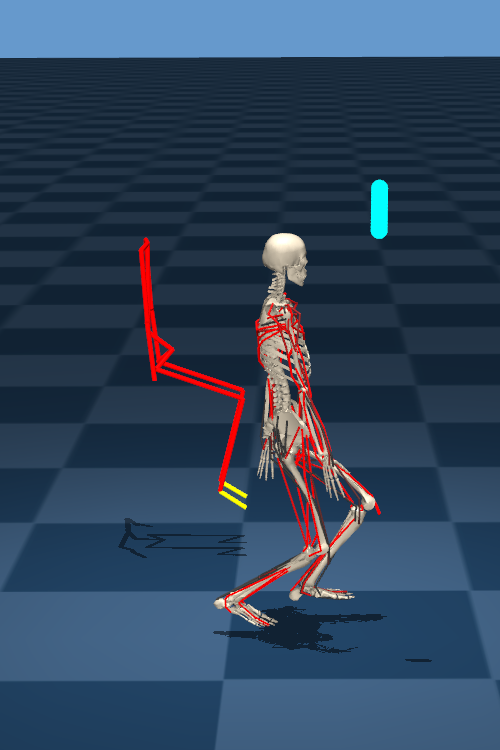}
    \includegraphics[trim=60 30 30 130, clip,width=0.192\columnwidth]{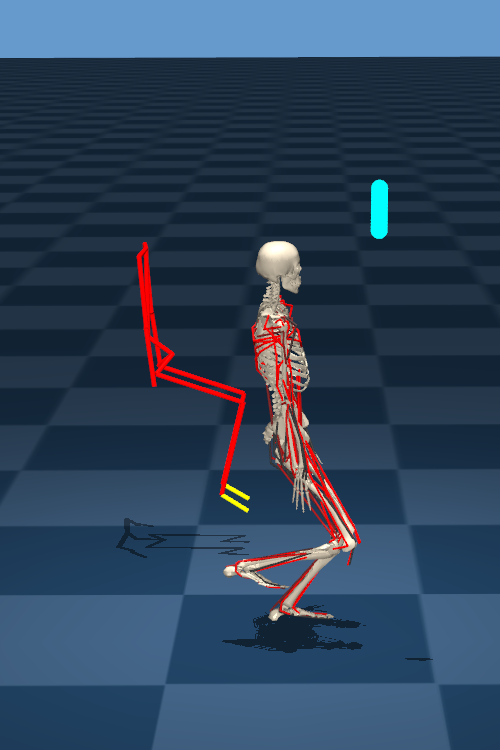}
    \includegraphics[trim=60 30 30 130,  clip,width=0.192\columnwidth]{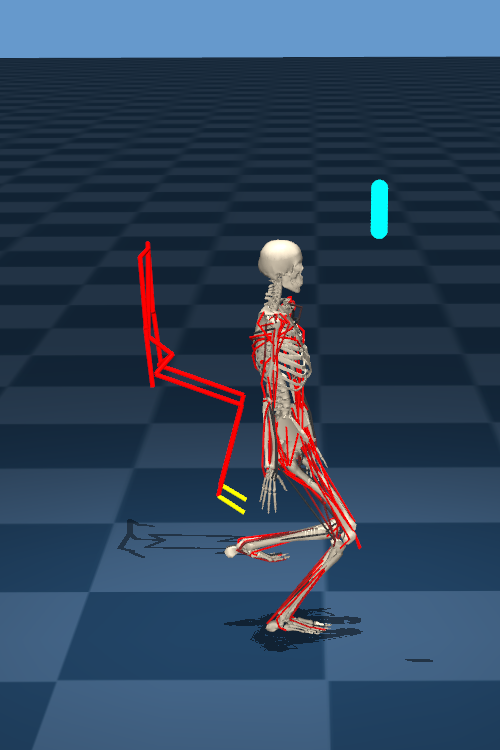}

    \includegraphics[trim=60 30 30 130, clip,width=0.192\columnwidth]{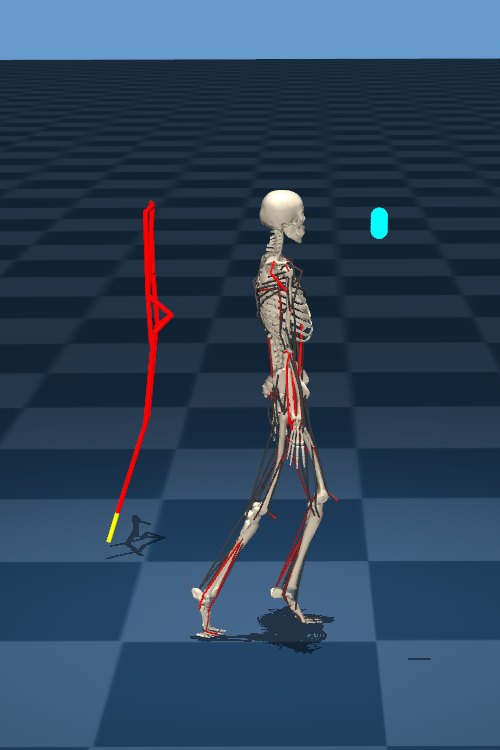}
    \includegraphics[trim=60 30 30 130,  clip,width=0.192\columnwidth]{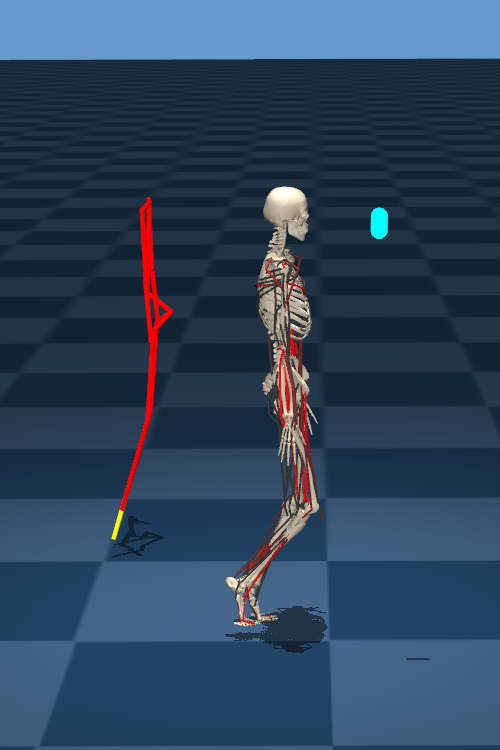}
    \includegraphics[trim=60 30 30 130, clip,width=0.192\columnwidth]{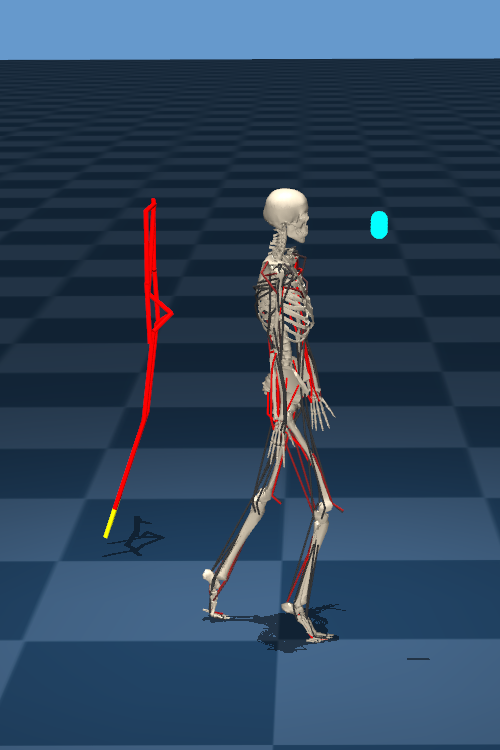}
    \includegraphics[trim=60 30 30 130,  clip,width=0.192\columnwidth]{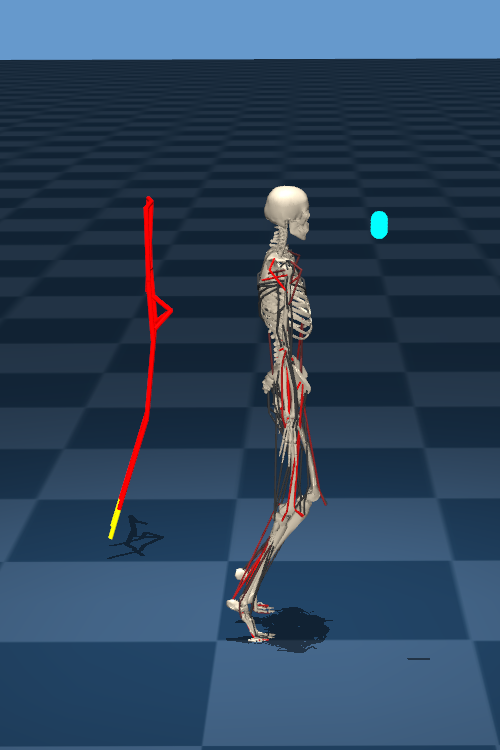}
    \includegraphics[trim=60 30 30 130,  clip,width=0.192\columnwidth]{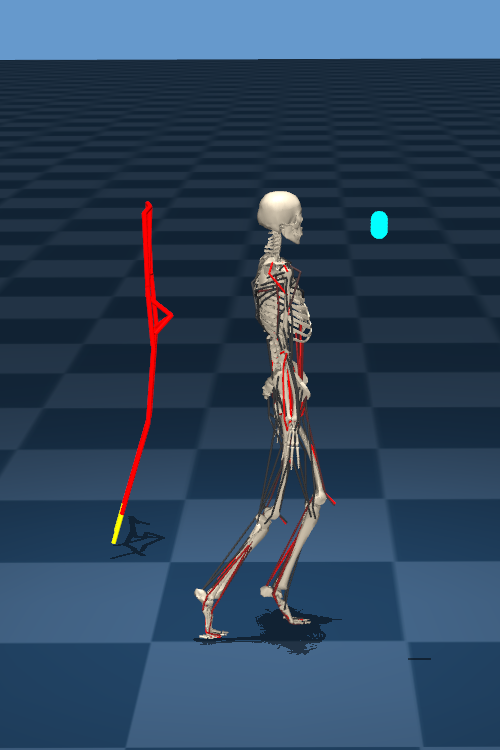}

    \includegraphics[trim=60 30 30 130, clip,width=0.192\columnwidth]{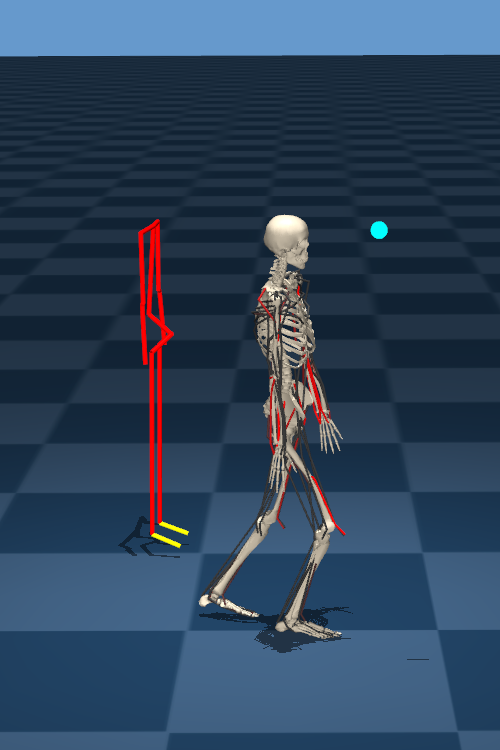}
    \includegraphics[trim=60 30 30 130,  clip,width=0.192\columnwidth]{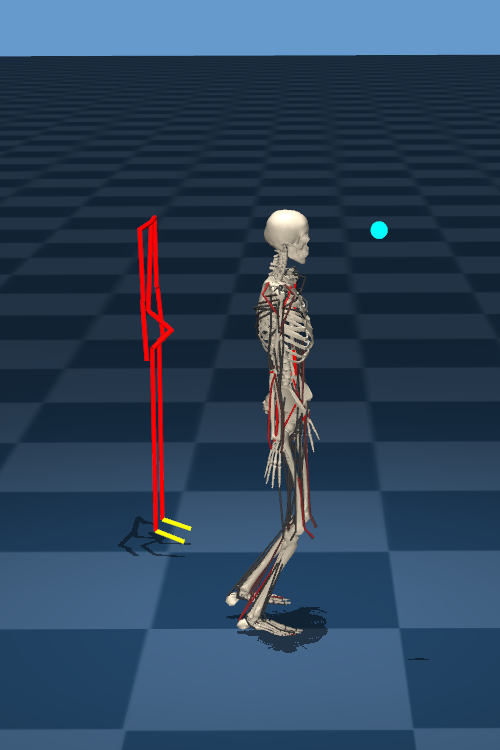}
    \includegraphics[trim=60 30 30 130,  clip,width=0.192\columnwidth]{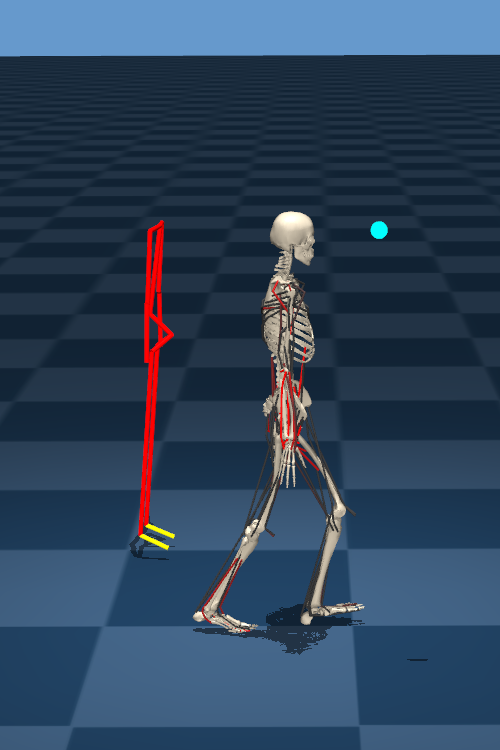}
    \includegraphics[trim=60 30 30 130,  clip,width=0.192\columnwidth]{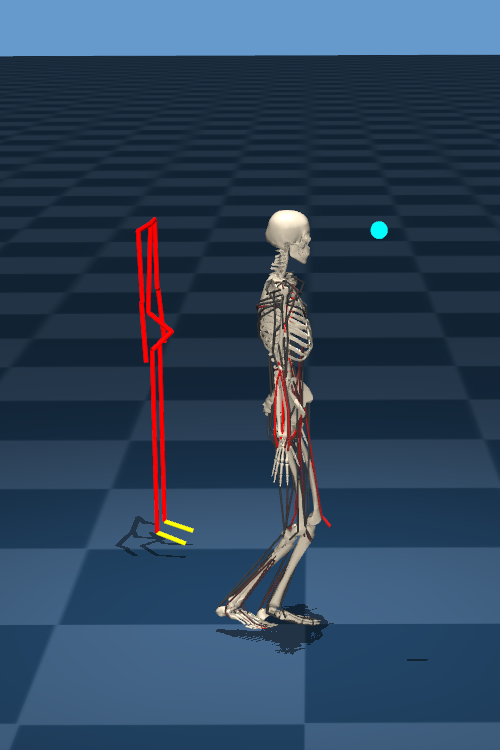}
    \includegraphics[trim=60 30 30 130, clip,width=0.192\columnwidth]{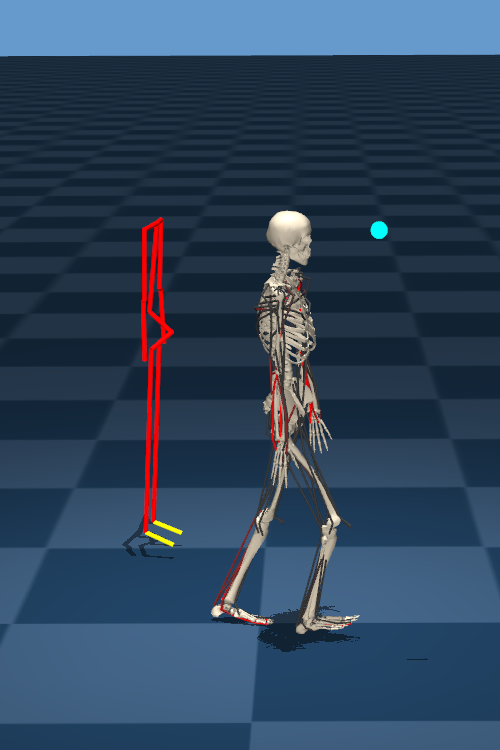}

    \includegraphics[trim=160 180 30 80, clip,width=0.192\columnwidth]{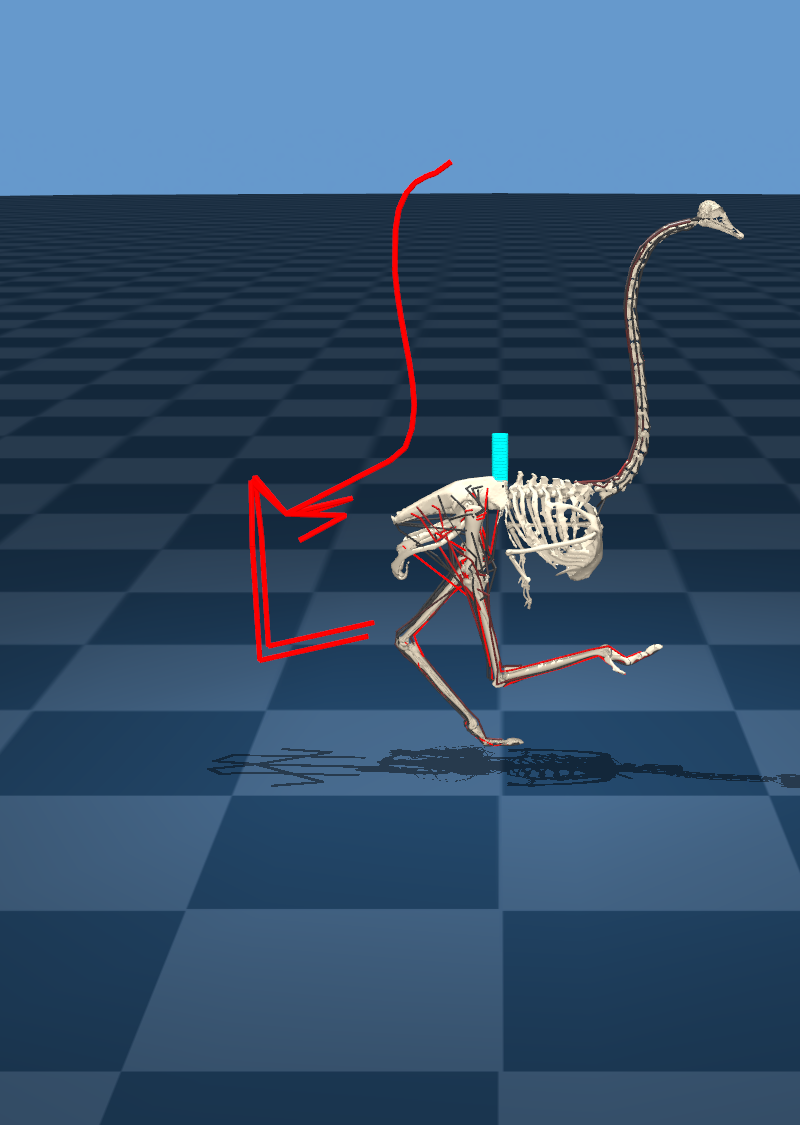}
    \includegraphics[trim=160 180 30 80, clip,width=0.192\columnwidth]{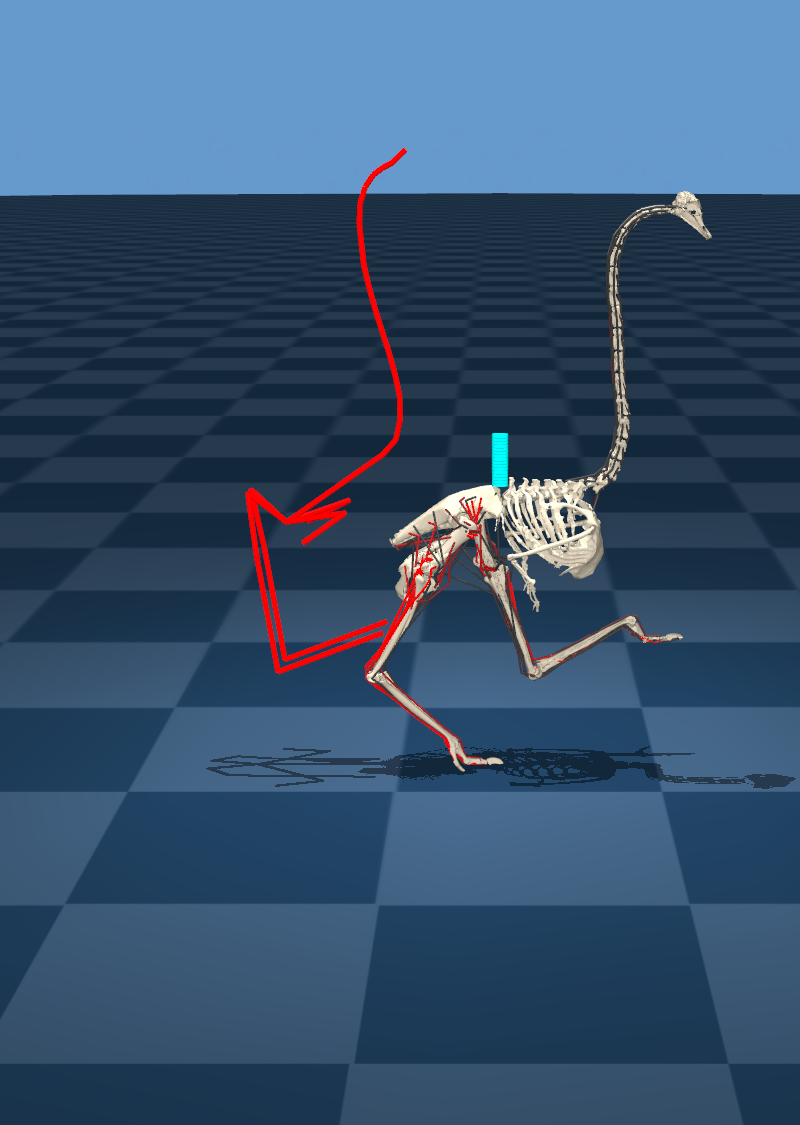}
    \includegraphics[trim=160 180 30 80, clip,width=0.192\columnwidth]{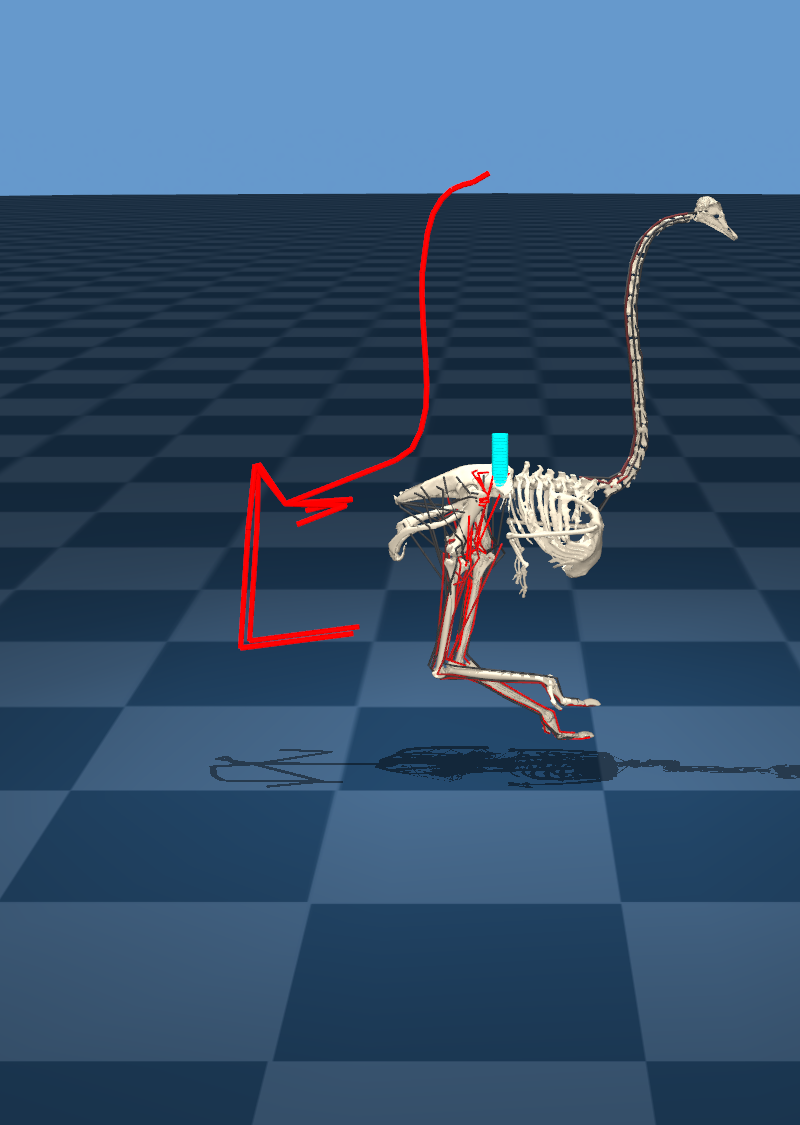}
    \includegraphics[trim=160 180 30 80, clip,width=0.192\columnwidth]{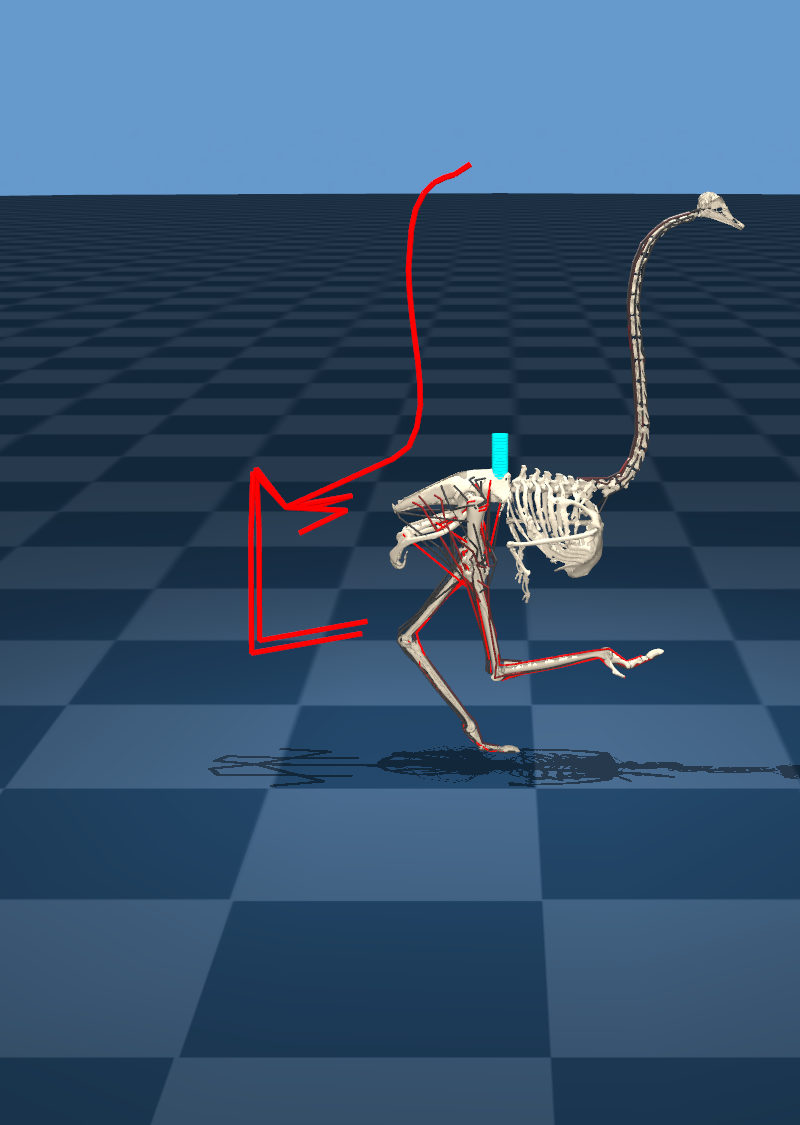}
    \includegraphics[trim=160 180 30 80, clip,width=0.192\columnwidth]{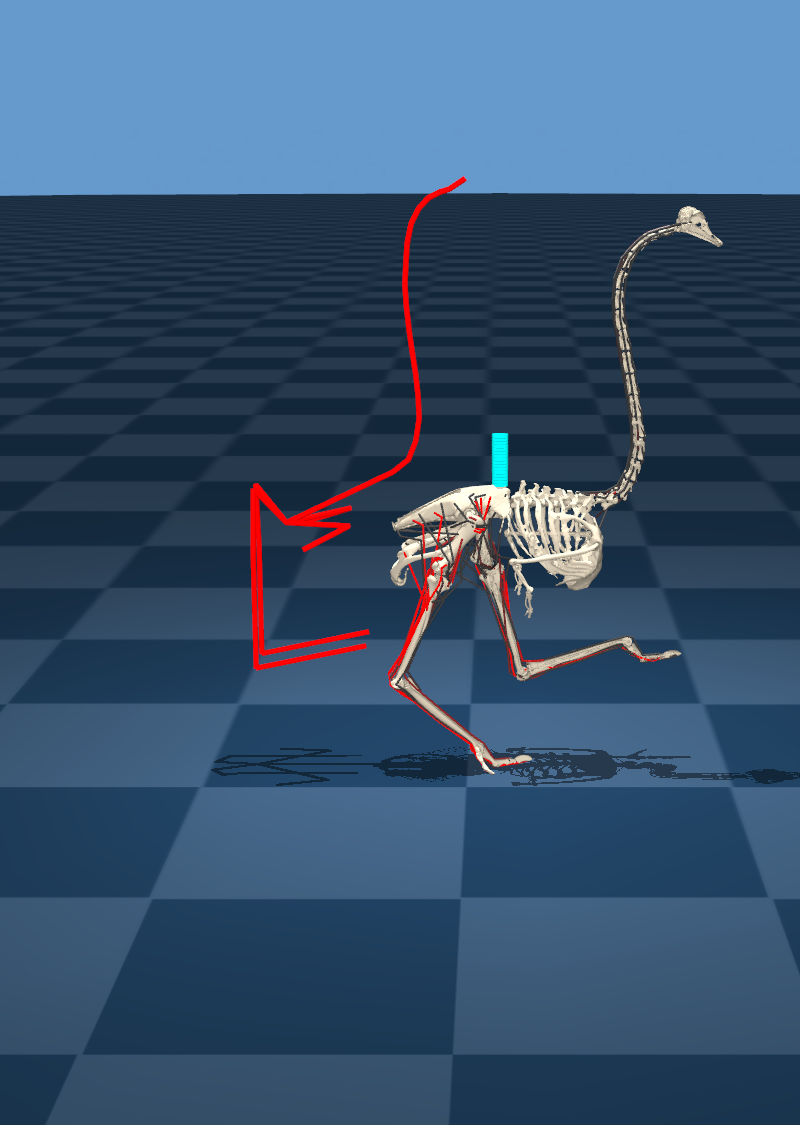}

    \includegraphics[trim=160 180 30 80, clip,width=0.192\columnwidth]{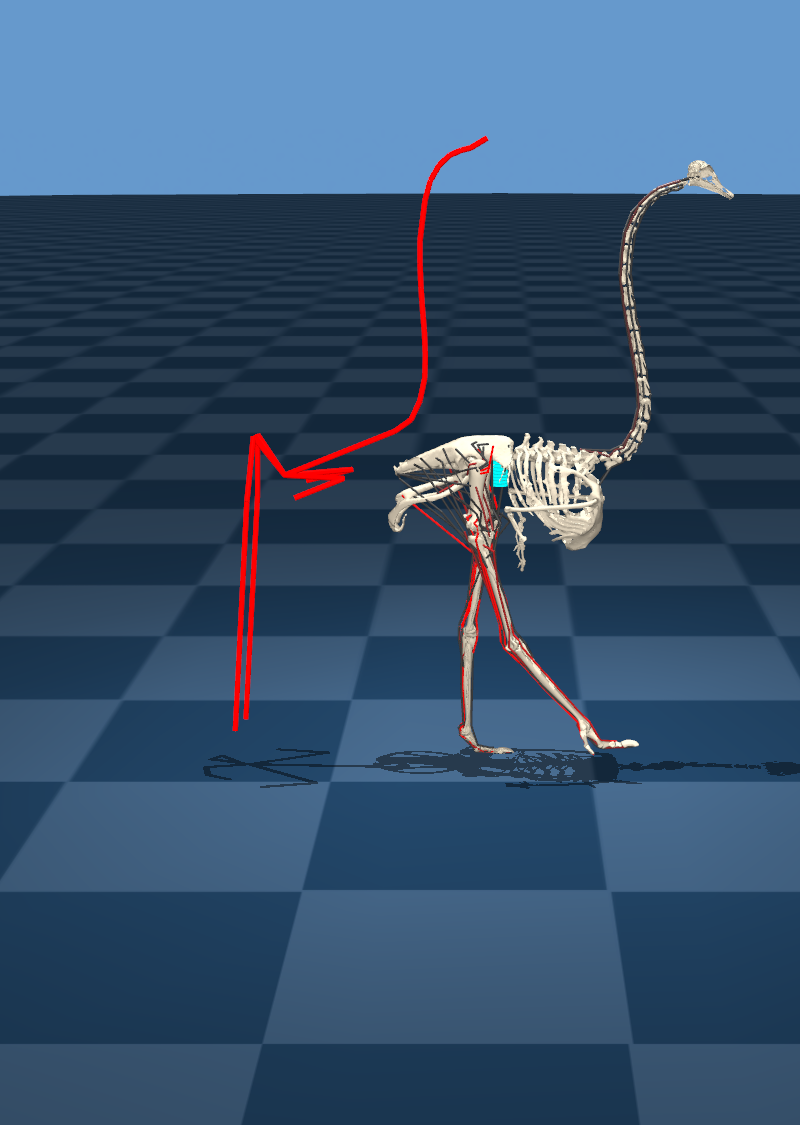}
    \includegraphics[trim=160 180 30 80, clip,width=0.192\columnwidth]{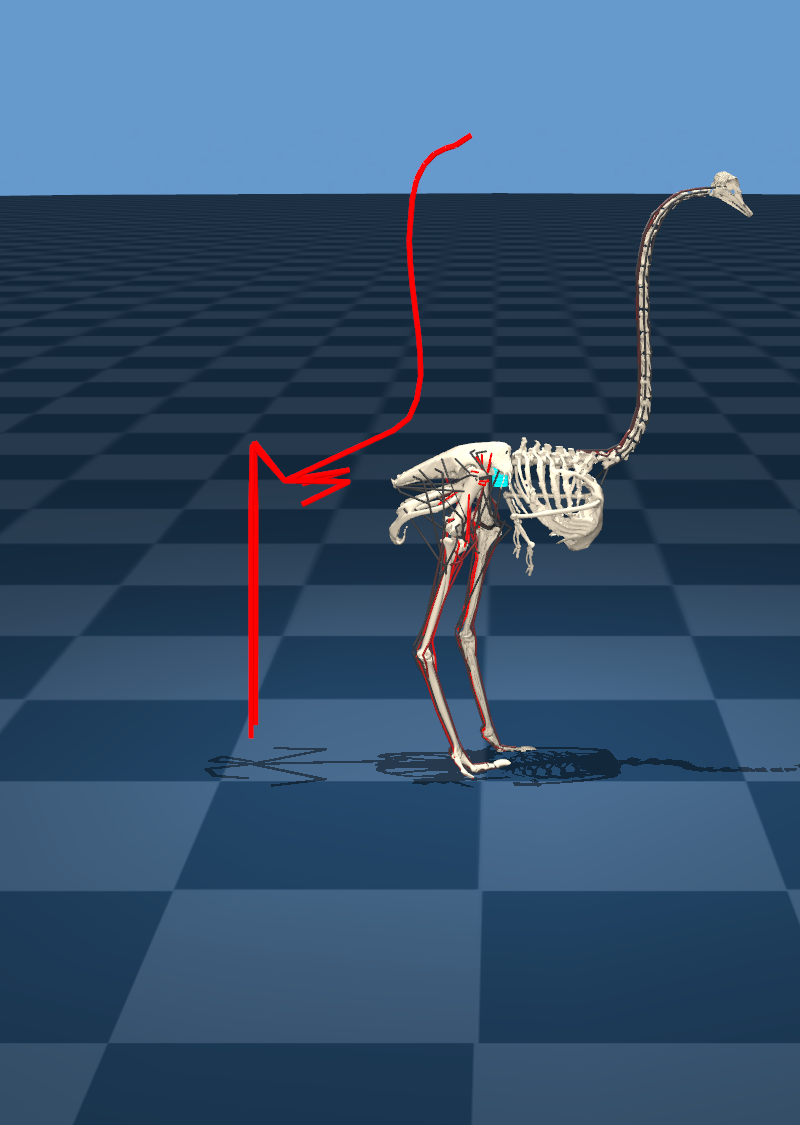}
    \includegraphics[trim=160 180 30 80, clip,width=0.192\columnwidth]{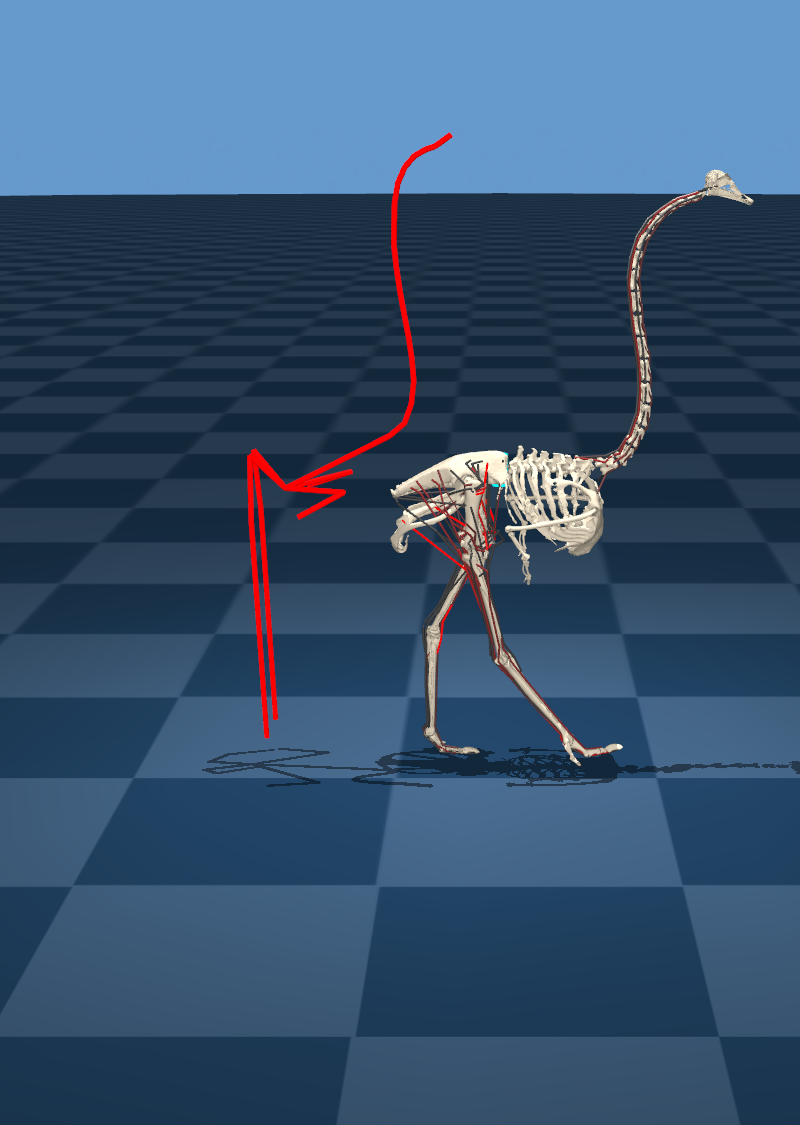}
    \includegraphics[trim=160 180 30 80, clip,width=0.192\columnwidth]{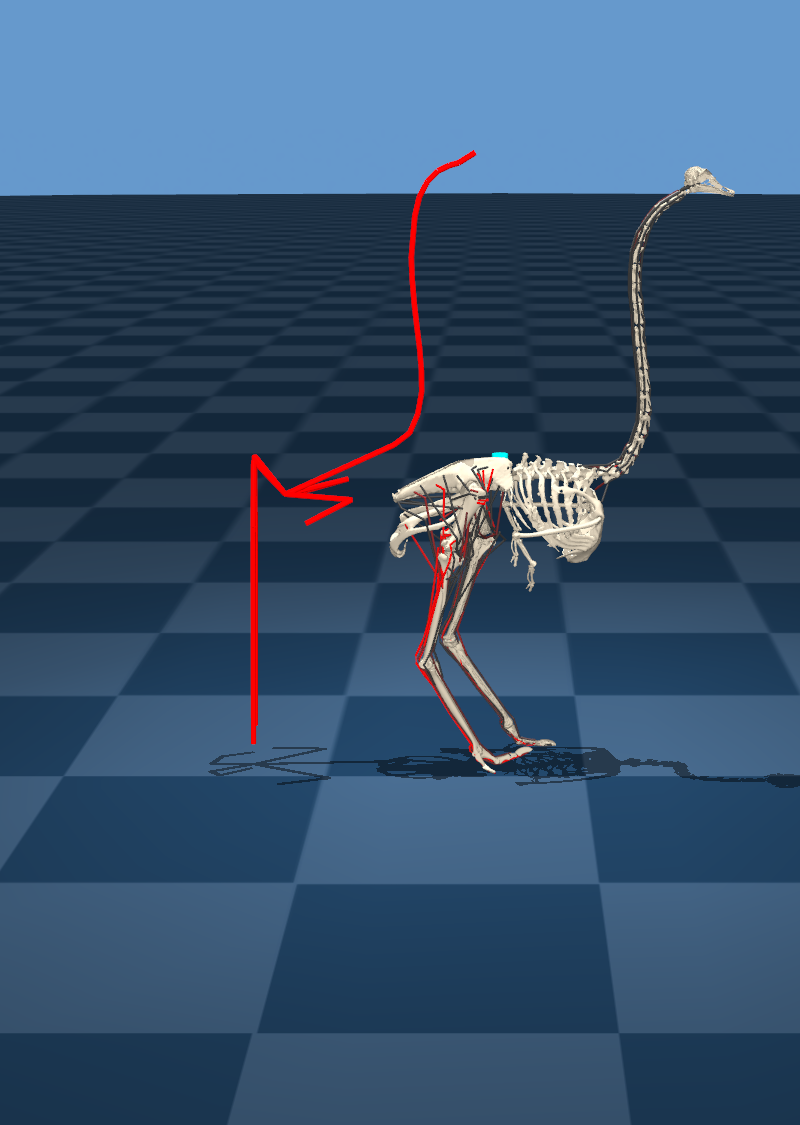}
    \includegraphics[trim=160 180 30 80, clip,width=0.192\columnwidth]{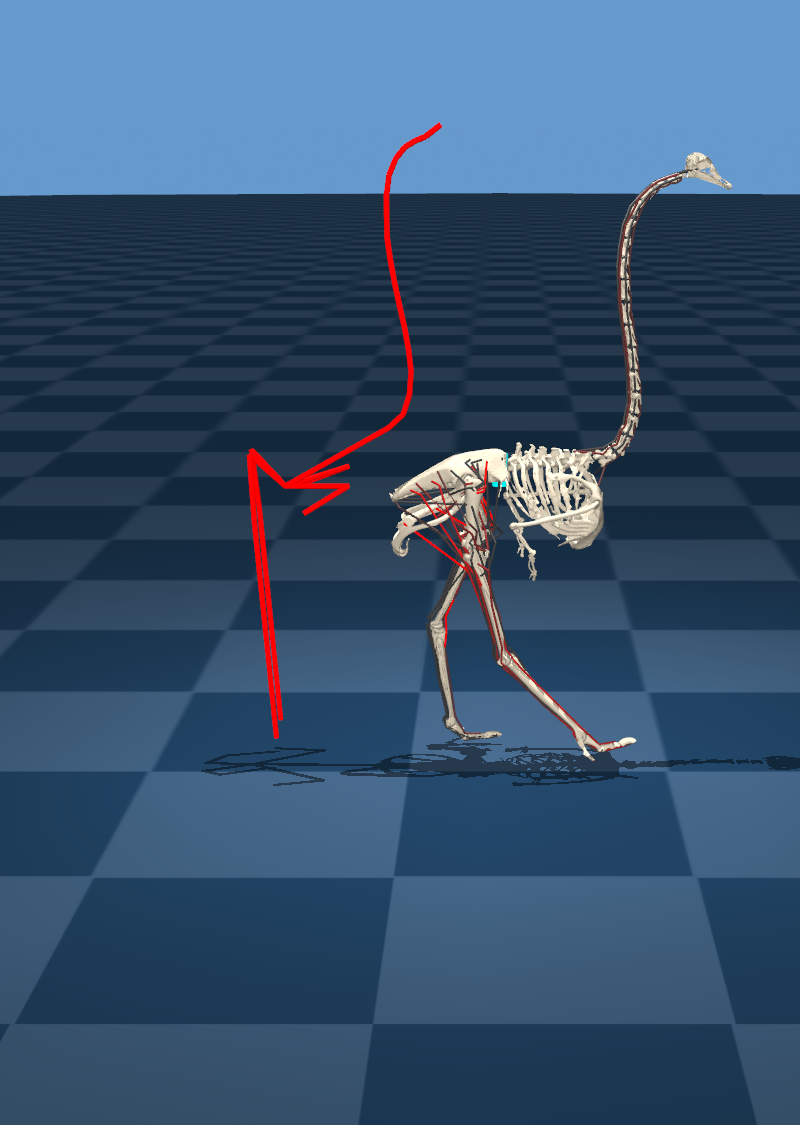}

    \captionof{figure}{
    \ys{Locomotion sequences of Humanoid (top three rows) and Ostrich (bottom two rows).
    The policies are trained under the \textit{Velocity + Pose + Energy} configuration, where the goal includes a target velocity, target pose, and target energy expenditure, resulting in a broader variety of gait patterns.
    Humanoid: high-energy running with flexed knees (first row), mid-energy toe-walking (second row), and low-energy natural gait (third row). Ostrich: high-energy running with flexed ankle (often perceived as knees) (fourth row) and mid-energy walking with extended ankles (fifth row).
    }
    }    
    \label{fig:goal_config_pose}
\end{figure}

\begin{figure}[t]
    \centering
    
    \reflectbox{\includegraphics[trim=40 70 40 15, clip, width=0.092\textwidth]{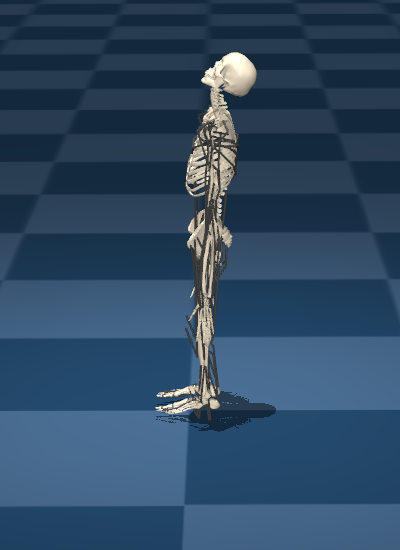}}
    \reflectbox{\includegraphics[trim=40 60 40 25, clip, width=0.092\textwidth]{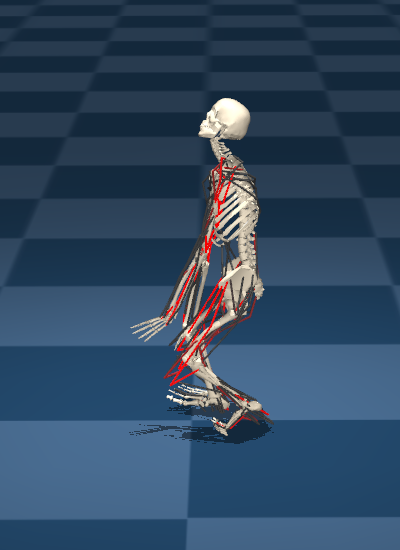}}
    \reflectbox{\includegraphics[trim=40 60 40 25, clip, width=0.092\textwidth]{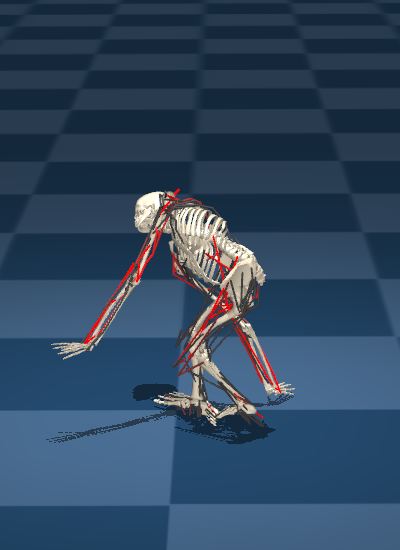}}
    \reflectbox{\includegraphics[trim=40 60 40 25, clip, width=0.092\textwidth]{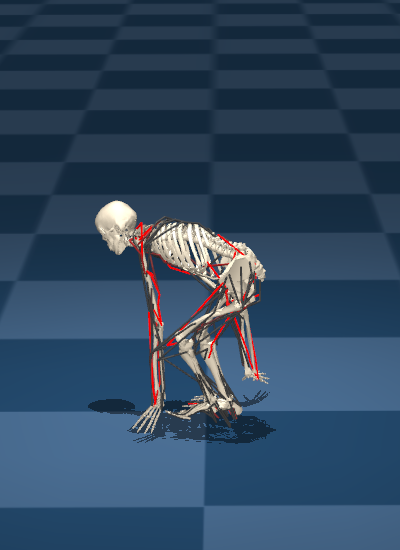}}
    \reflectbox{\includegraphics[trim=40 60 40 25, clip, width=0.092\textwidth]{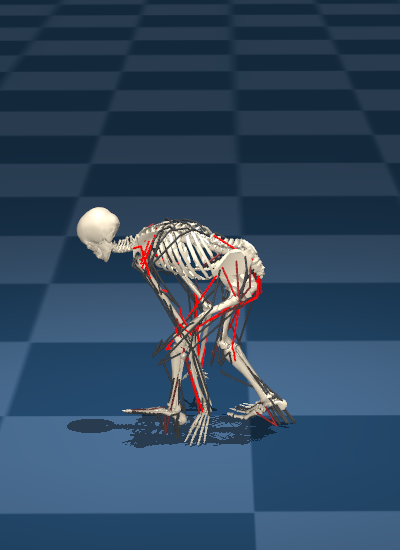}}
    \reflectbox{\includegraphics[trim=40 60 40 25, clip, width=0.092\textwidth]{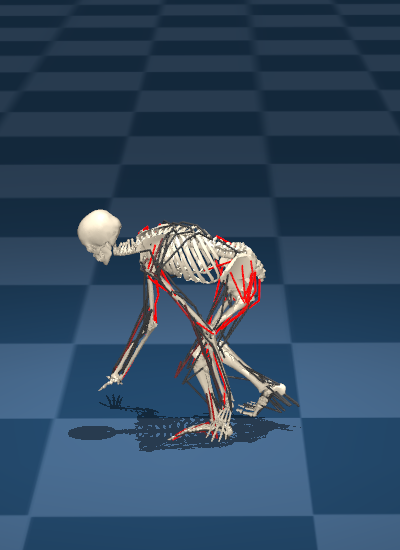}}
    \reflectbox{\includegraphics[trim=40 60 40 25, clip, width=0.092\textwidth]{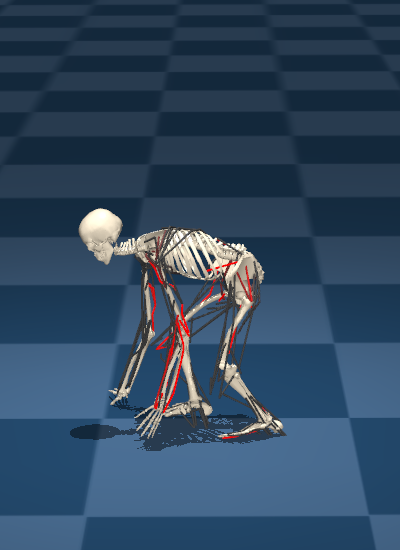}}
    \reflectbox{\includegraphics[trim=40 60 40 25, clip, width=0.092\textwidth]{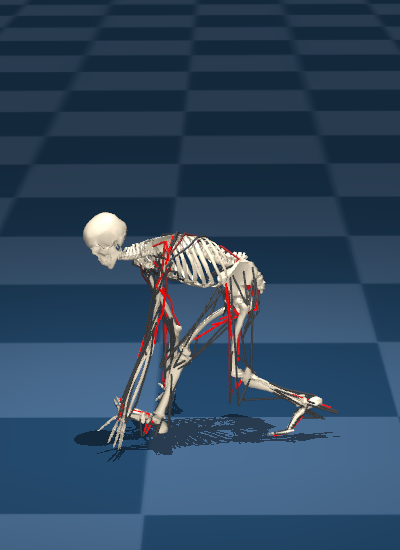}}
    \reflectbox{\includegraphics[trim=40 60 40 25, clip, width=0.092\textwidth]{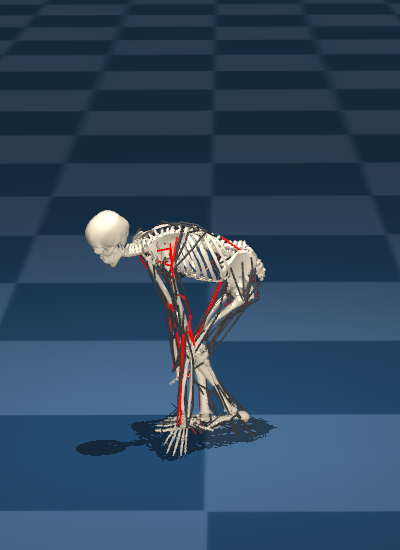}}
    \reflectbox{\includegraphics[trim=40 60 40 25, clip, width=0.092\textwidth]{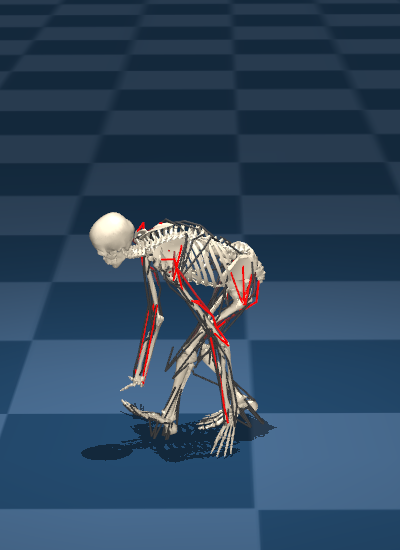}}
    
    \caption{Locomotion sequence of Chimanoid,
    \ys{trained under the \textit{Velocity Only} configuration.}
    Note that a trot-like quadrupedal gait emerges despite both the target pose used during training and the initial simulation pose being set to the default standing pose.
    }    
    \label{fig:locomotion_chimanoid}
\end{figure}

We demonstrate the capability of the learned FreeMusco model to generate diverse locomotion behaviors under various goal configurations \ys{(Figure~\ref{fig:locomotion_human_ostrich}--\ref{fig:locomotion_chimanoid})}.
\ys{Across characters, the resulting motions reflect the given goal inputs: 
locomotion direction and speed modulation for target velocity, 
facing direction adjustment for target facing direction, pose changes for target pose, and energy regulation for target energy level.
Importantly, all goal inputs can be modified at runtime, and the character smoothly and stably transitions between behaviors in response.
Although the models are trained with different goal configurations, they exhibit consistent locomotion under equivalent evaluation settings. For example, the policy trained with the \textit{Velocity + Pose + Energy} configuration produces a consistent gait pattern similar to the \textit{Velocity Only} configuration when evaluated with the default standing pose and minimum energy as targets.}

\begin{figure*}[h]
    \centering

    \includegraphics[trim=50 50 50 30, clip, width=0.096\textwidth]{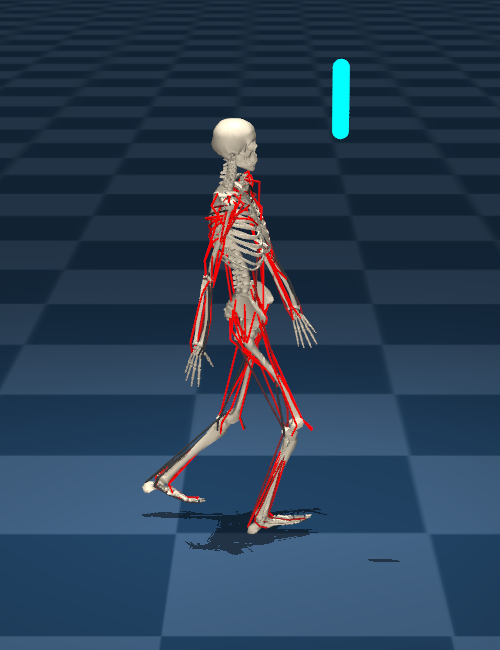}
    \includegraphics[trim=50 50 50 30, clip,width=0.096\textwidth]{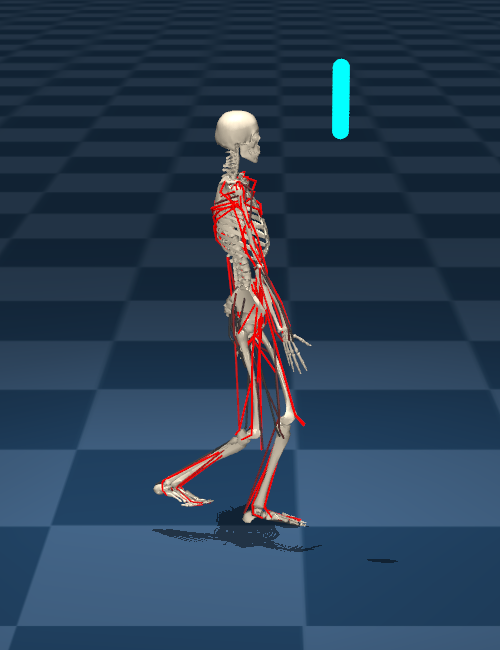}
    \includegraphics[trim=50 50 50 30, clip,width=0.096\textwidth]{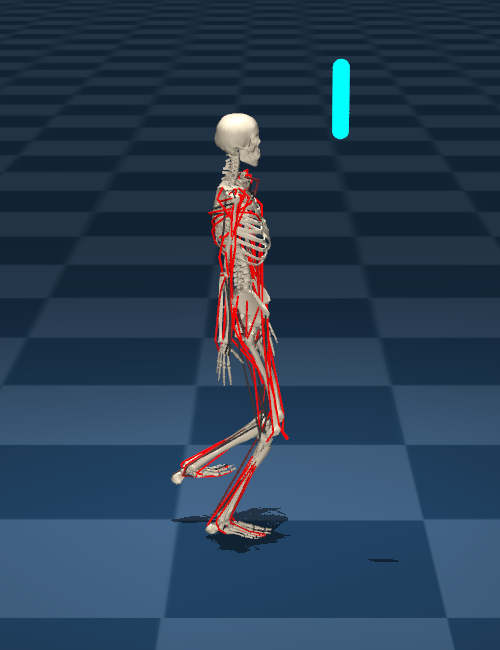}
    \includegraphics[trim=50 50 50 30, clip,width=0.096\textwidth]{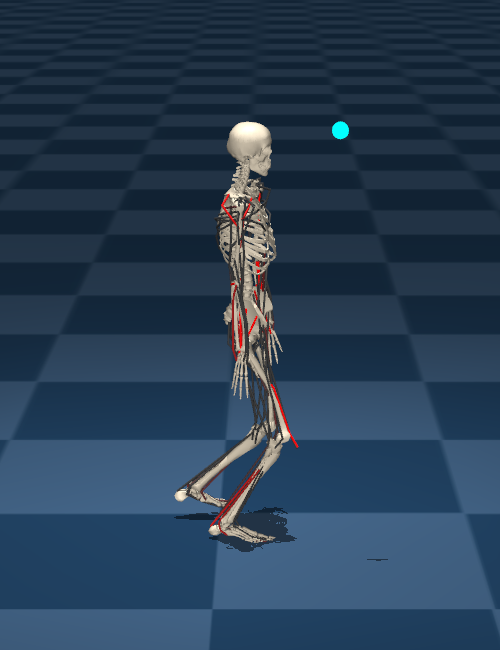}
    \includegraphics[trim=50 50 50 30, clip,width=0.096\textwidth]{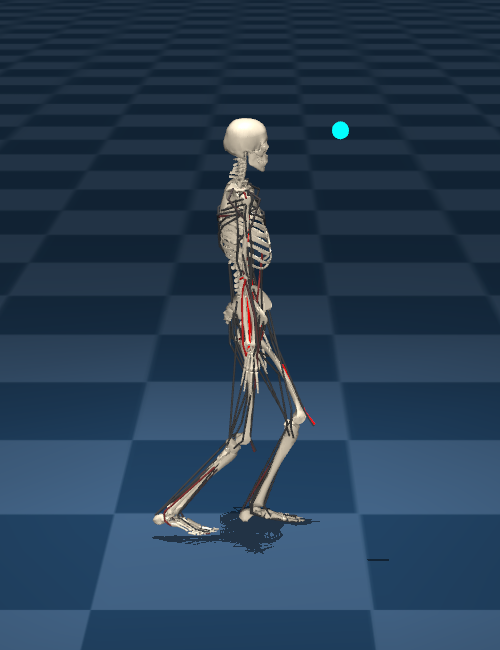}
    \includegraphics[trim=50 50 50 30, clip,width=0.096\textwidth]{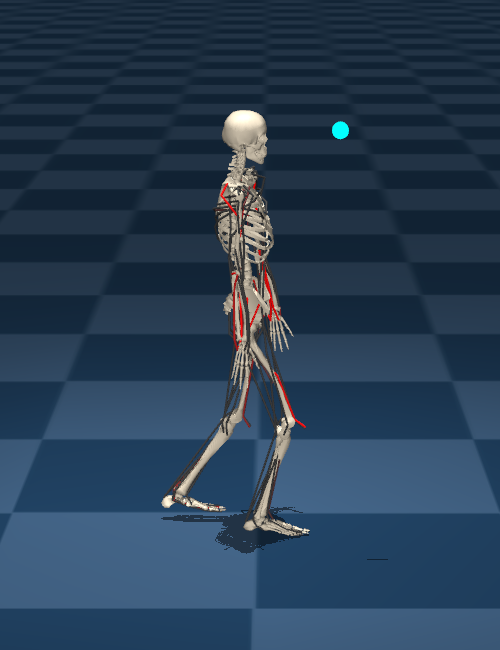}
    \includegraphics[trim=50 50 50 30, clip,width=0.096\textwidth]{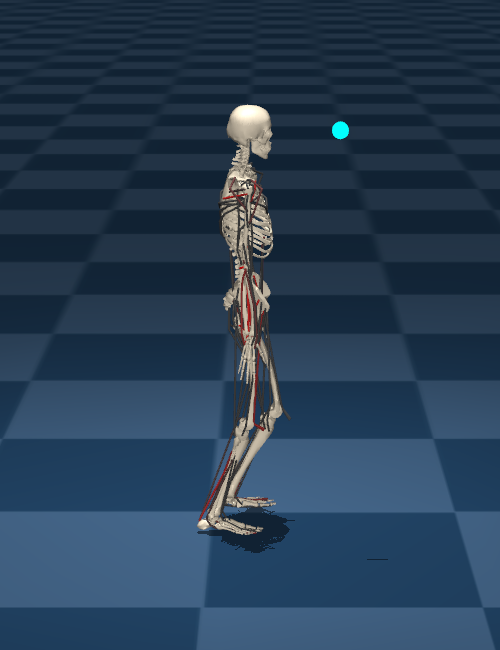}
    \includegraphics[trim=50 50 50 30, clip,width=0.096\textwidth]{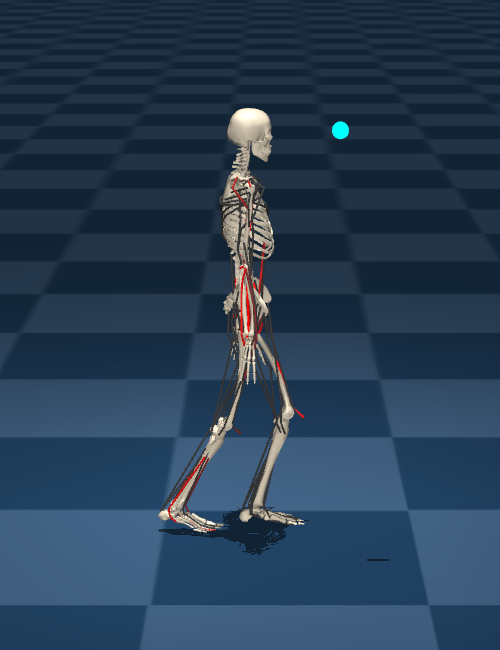}
    \includegraphics[trim=50 50 50 30, clip,width=0.096\textwidth]{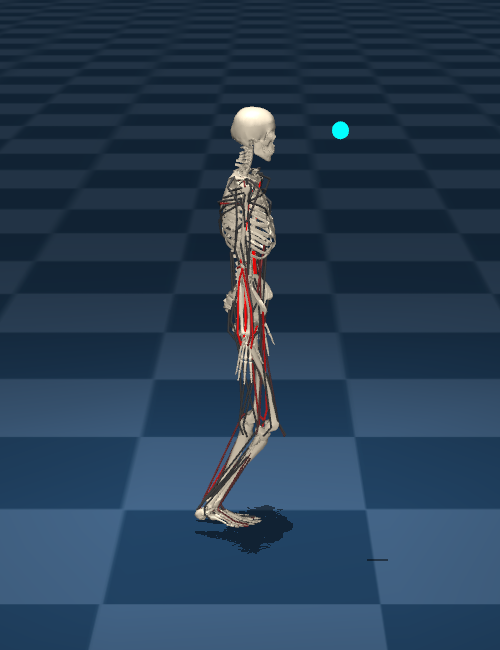}
    \includegraphics[trim=50 50 50 30, clip,width=0.096\textwidth]{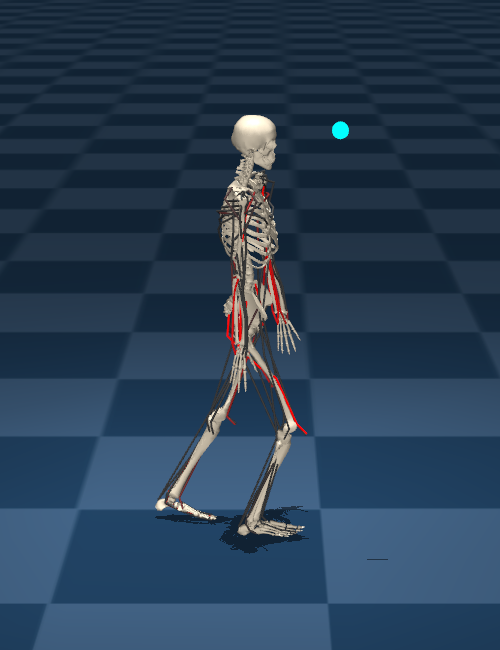}

    \includegraphics[trim=50 50 50 30, clip,width=0.096\textwidth]{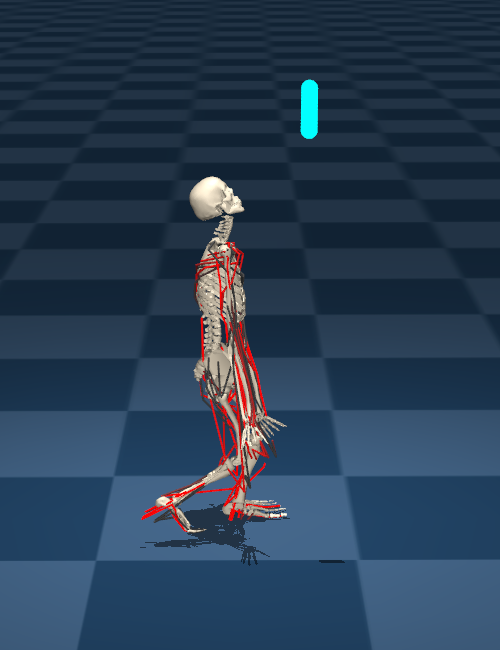}
    \includegraphics[trim=50 50 50 30, clip,width=0.096\textwidth]{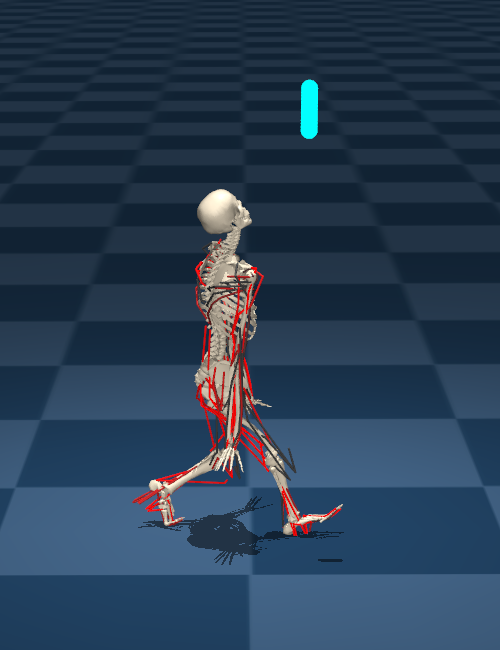}
    \includegraphics[trim=50 50 50 30, clip,width=0.096\textwidth]{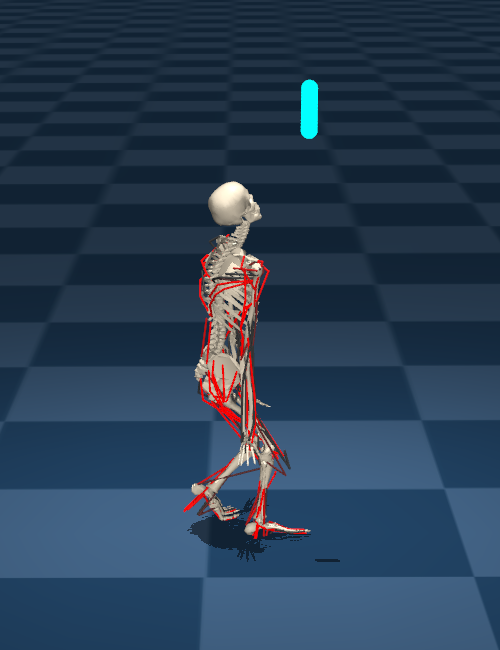}
    \includegraphics[trim=50 50 50 30, clip,width=0.096\textwidth]{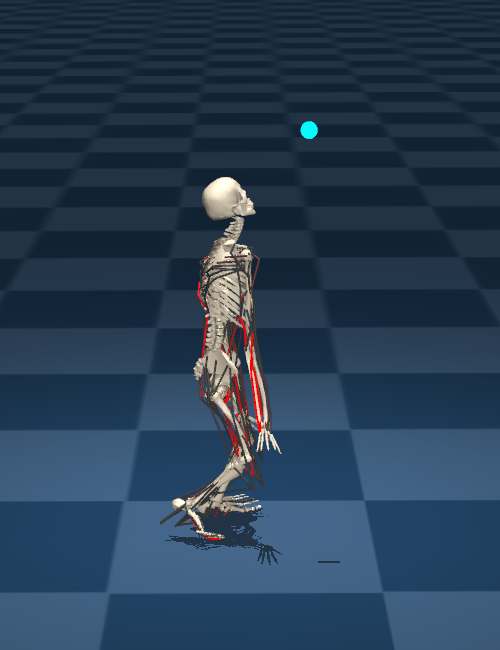}
    \includegraphics[trim=50 50 50 30, clip,width=0.096\textwidth]{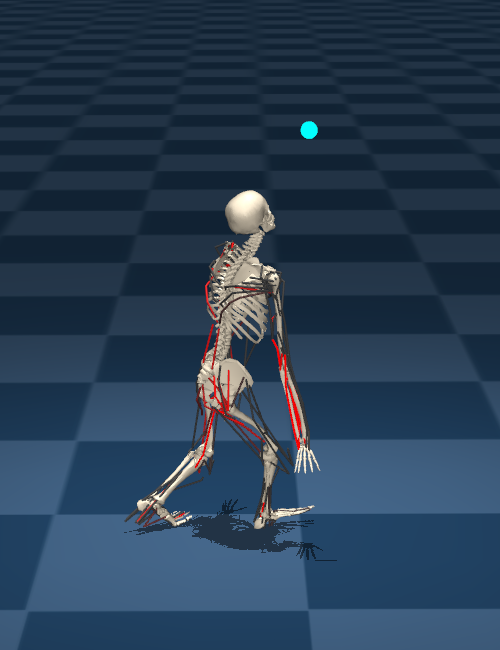}
    \includegraphics[trim=50 50 50 30, clip,width=0.096\textwidth]{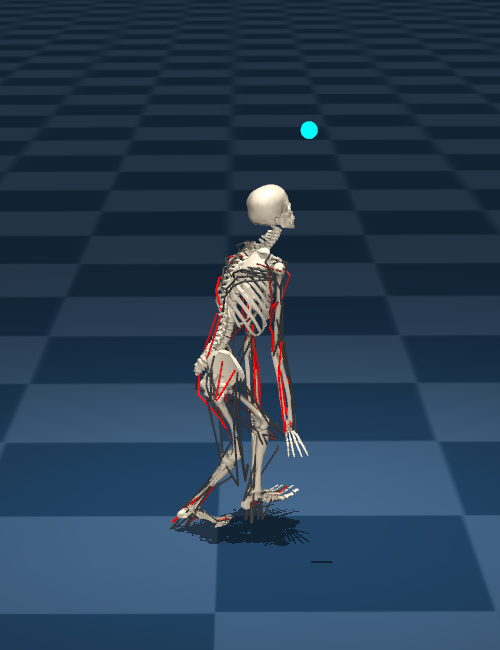}
    \includegraphics[trim=50 50 50 30, clip,width=0.096\textwidth]{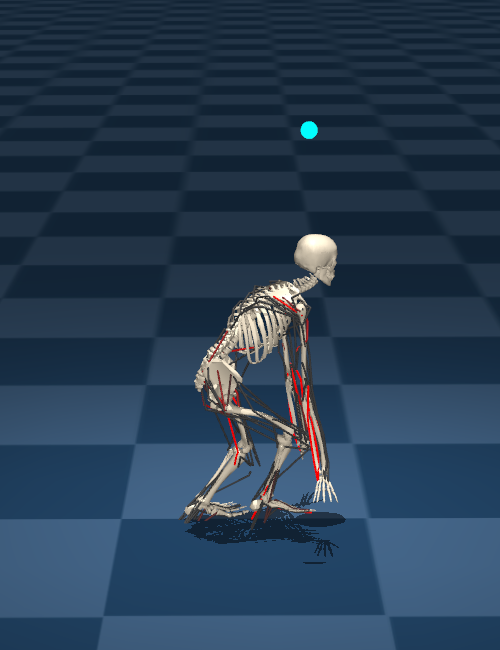}
    \includegraphics[trim=50 50 50 30, clip,width=0.096\textwidth]{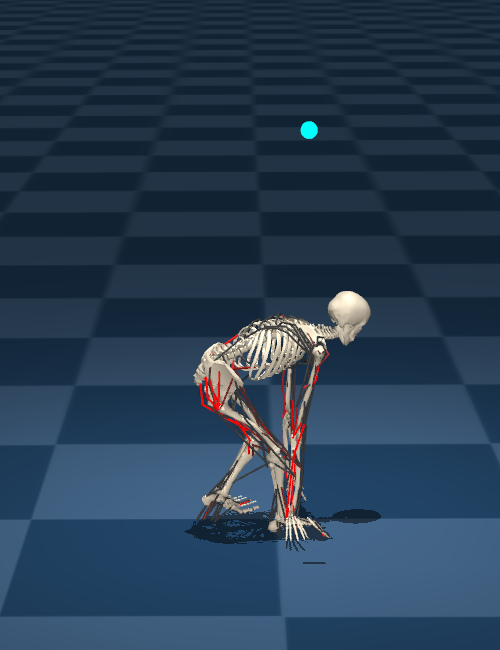}
    \includegraphics[trim=50 50 50 30, clip,width=0.096\textwidth]{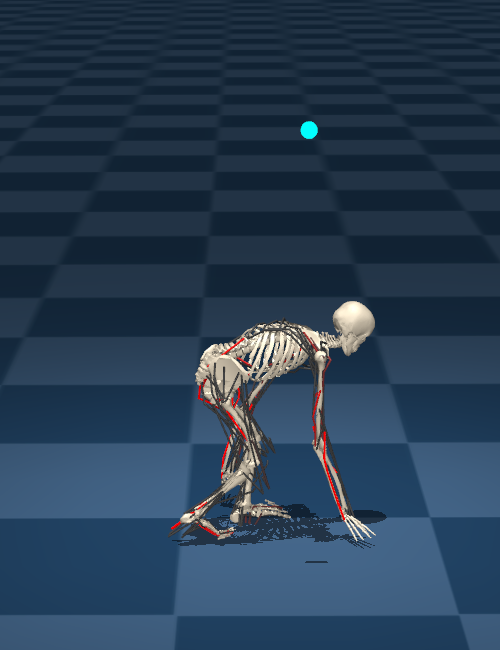}
    \includegraphics[trim=50 50 50 30, clip,width=0.096\textwidth]{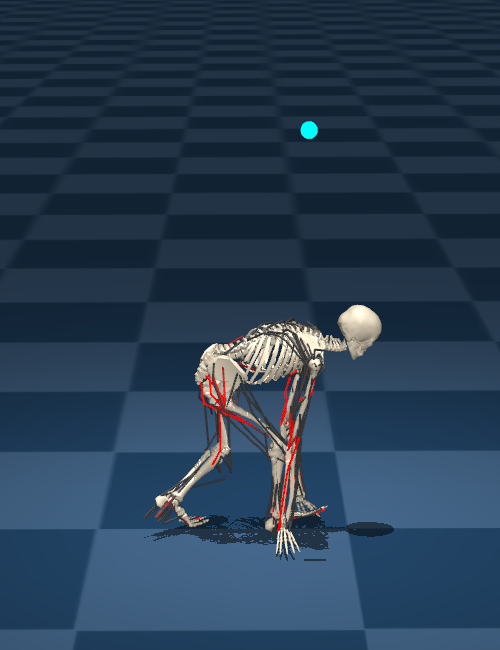}
    
    \caption{
    \ys{Locomotion sequences of Humanoid (top) and Chimanoid (bottom) trained under the \textit{Velocity + Energy} configuration, with target energy gradually decreased at runtime.
    For the Humanoid, lower target energy produces slower but still bipedal locomotion. For the Chimanoid, starting from a bipedal pose, high energy supports bipedal walking, but reduced energy naturally induces a shift to quadrupedal locomotion.}
    }    
    \label{fig:humnaoid_chimanoid_energy}
\end{figure*}

{\color{ys}
\subsection{Emergent Gait Strategies Across Morphologies}
\label{sec:emergent_gait}

A key result highlighting the morphology-adaptive nature of our framework is observed in Chimanoid—a humanoid variant with elongated arms (1.2×) and shortened legs (0.7×), resembling a chimpanzee-like morphology. Although trained with a goal configuration set to the default standing pose, Chimanoid does not remain bipedal (Figure~\ref{fig:locomotion_chimanoid}). Instead, a trot-like quadrupedal gait emerges without any motion priors, swinging diagonally opposite limbs in synchrony and making ground contact with its arms.

Energy modulation further reveals morphology-dependent gait strategies (Figure~\ref{fig:humnaoid_chimanoid_energy}). For the Humanoid, decreasing the target energy results in slower but still bipedal locomotion, while increasing the energy induces more dynamic motions with larger arm and leg swings. For the Chimanoid, when initialized in a bipedal pose, it can walk bipedally under high-energy conditions but naturally shifts to quadrupedal walking as the energy target decreases. When the energy is raised again, the character does not revert to a bipedal gait but instead reinforces a stronger quadrupedal pattern, indicating that quadrupedal locomotion is inherently more efficient for its morphology.

These findings collectively show that energy-efficient locomotion strategies depend on morphology, and that our framework can autonomously discover morphology-aligned gaits without requiring motion references.
}

\subsection{Unconditional Random Sampling from Latent Space}

\begin{figure}
    \centering

    \includegraphics[trim=500 00 750 250, clip, width=0.325\linewidth]{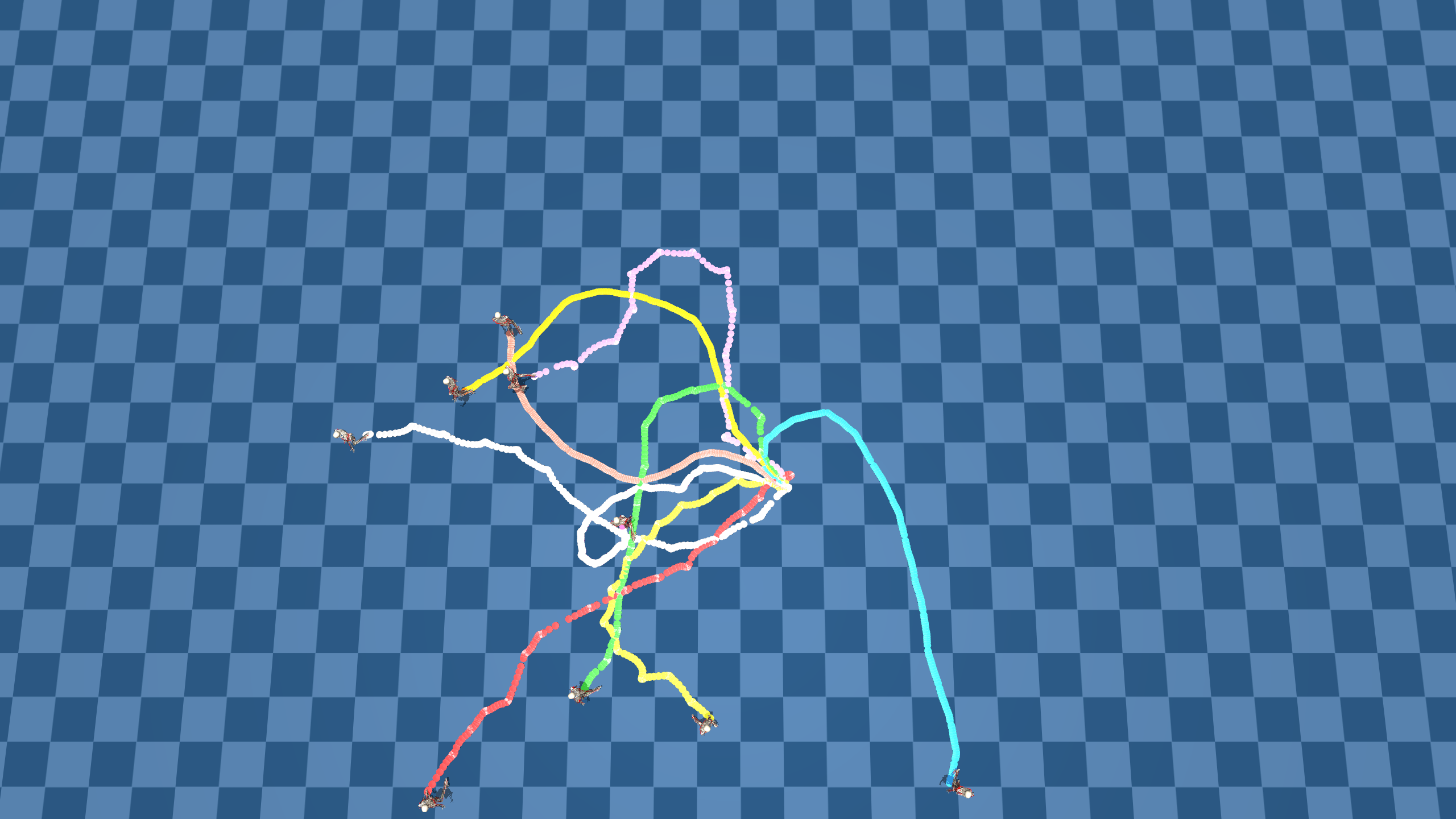}
    \includegraphics[trim=600 00 650 250, clip, width=0.325\linewidth]{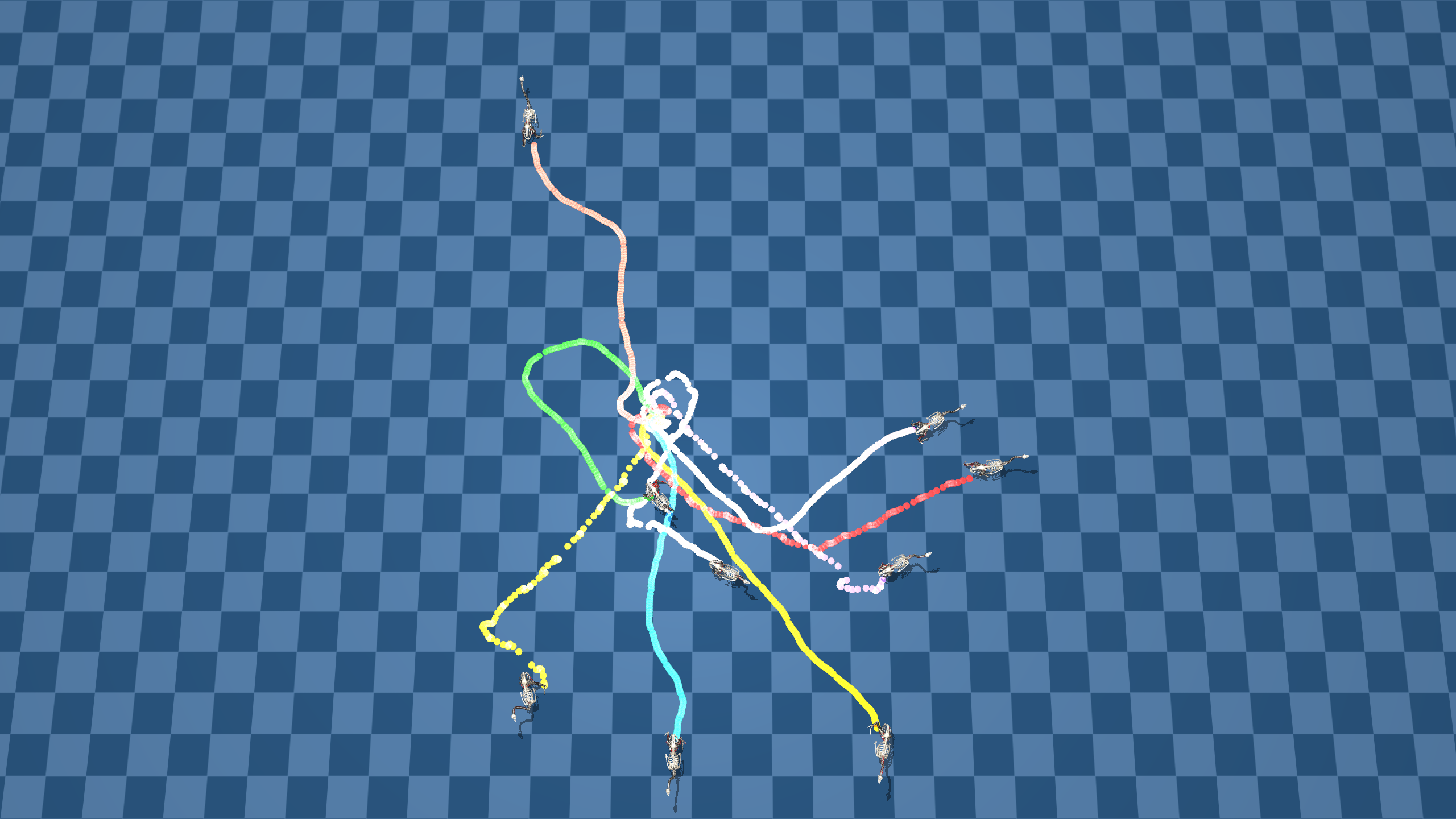}
    \includegraphics[trim=600 00 650 250, clip, width=0.325\linewidth]{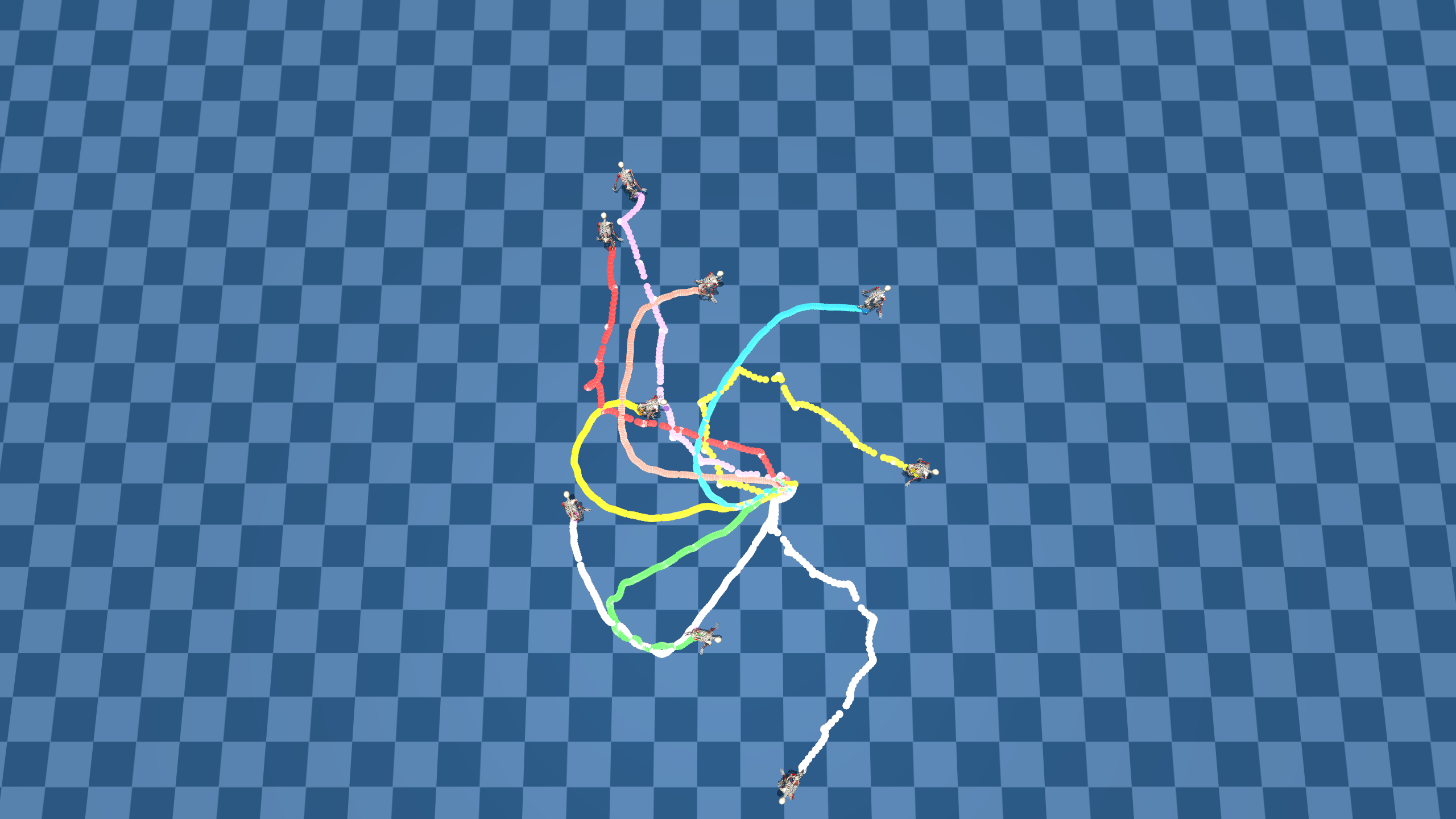}

    \captionof{figure}{Trajectories \ys{of Humanoid (left), Ostrich (center), and Chimanoid (right)} generated from different latent vectors sampled at each timestep, starting from the same initial state.}
    \label{fig:latent_sampling_paths}    
\end{figure}

To qualitatively assess the diversity and structure of the learned latent space, we conduct an unconditional sampling experiment.
At every timestep, we randomly sample a latent vector \( z \) from the prior distribution used during training, and decode it through the low-level policy without any task-specific conditioning.

This experiment reveals that the learned space encodes a wide range of
plausible locomotion behaviors.
The resulting motions exhibit distinct gaits, directions, and stepping patterns across different morphologies, despite the absence of any explicit goal input. Notably, the motions remain biomechanically valid and stable, indicating that the latent space captures structured motor behaviors rather than arbitrary muscle activations.
Figure~\ref{fig:latent_sampling_paths} visualizes trajectories from a shared initial state under different sampled latents, further illustrating the diversity encoded in the latent space.

\subsection{High-Level Control}

\begin{figure}
    \centering
    \subfloat[Goal Navigation]{   
        \includegraphics[trim=350 450 400 0, clip, width=0.5\linewidth]{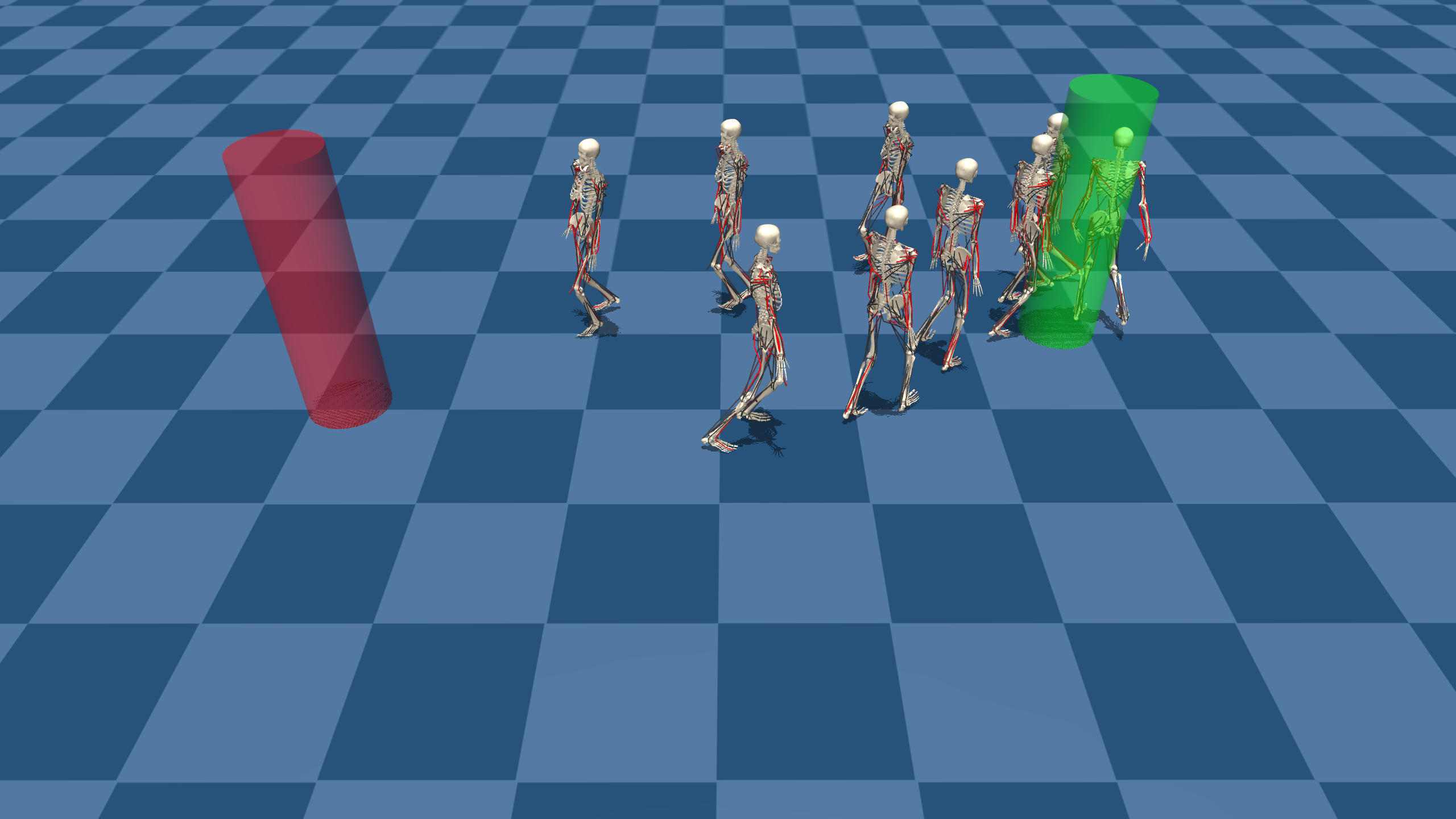}
    }%
    \subfloat[Path Following]{         
        \includegraphics[trim=400 250 350 200, clip, width=0.5\linewidth]{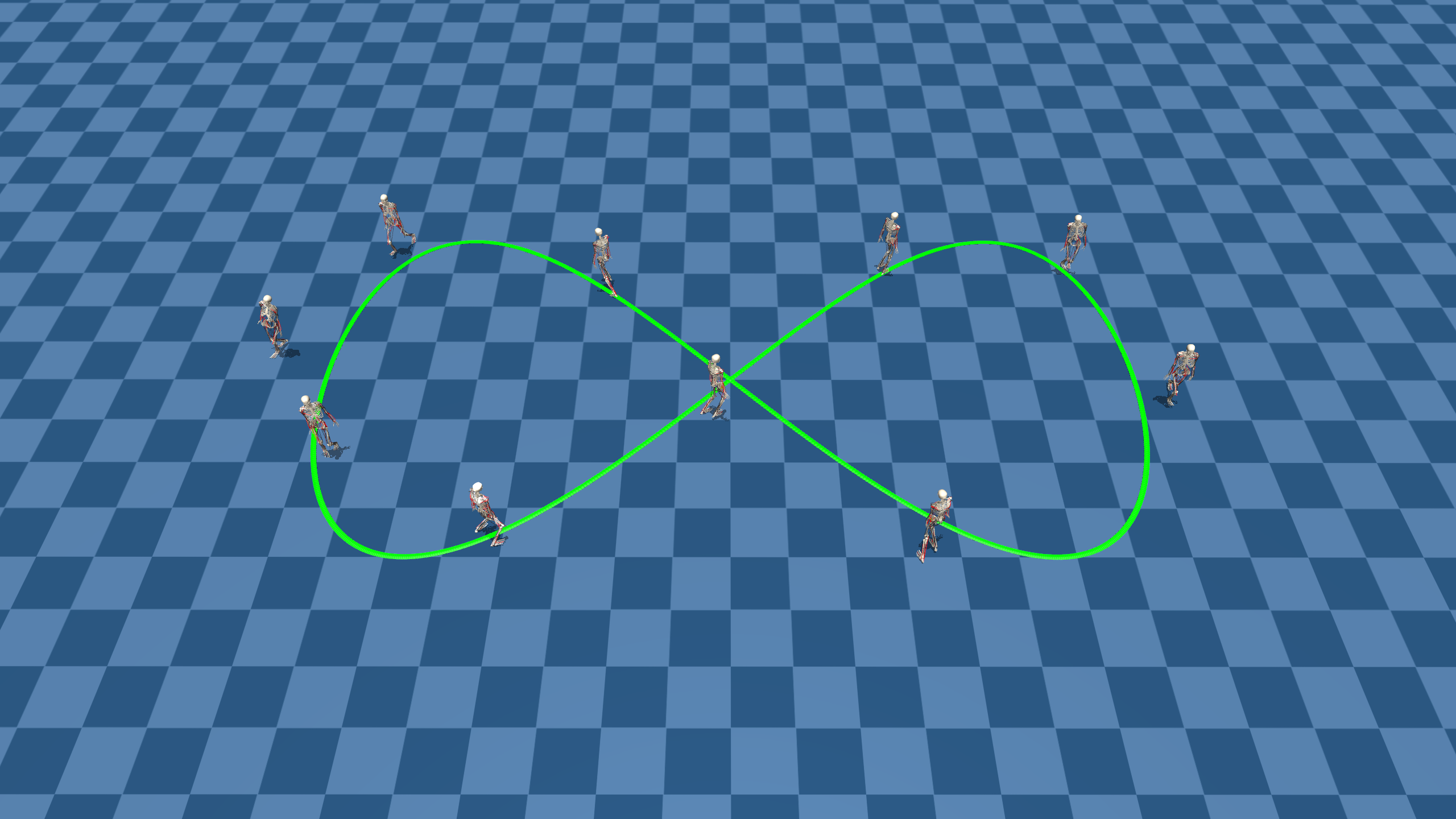}
    }%
    \caption{Downstream tasks for high-level control.}
    \label{fig:tasks}    
\end{figure}

To demonstrate the applicability of the learned latent space, we train high-level control policies for two downstream tasks: point-goal navigation and path following (Figure~\ref{fig:tasks}). In both tasks, the high-level policy replaces the posterior encoder and receives a task-specific goal as input along with the character’s current state. It outputs a residual latent vector, which is added to the prior encoder output to produce the final control input for the low-level policy (decoder).

Since the physical environment used for high-level tasks is identical to that used during latent space training, the same world model remains valid and can be reused. This allows the latent space to support both model-based and model-free high-level training strategies. In practice, we observed minimal difference in performance between the two approaches. For simplicity, the main results presented here are generated using the model-free strategy, where the high-level policy is trained via PPO to output residual latent vectors that guide the low-level controller in achieving the task objectives. Each task is implemented with task states and reward formulations following prior work~\cite{AMP,won_physics-based_2022}.

Qualitative results show that the high-level controller produces coherent, goal-directed behaviors across different morphologies. In the navigation task, characters consistently reach the target across different morphologies, demonstrating goal-directed behavior without task-specific supervision. In the path following task, the policy enables smooth tracking with heading and stepping \ys{continuously} adapted to the \ys{ribbon-shaped} path. These results demonstrate that the learned latent space supports diverse locomotion and serves as an effective interface for high-level task control.

\subsection{\ys{Effect of Temporally Averaged Loss Formulation}}

\begin{table}[h]
  \centering
  \small
  \caption{\ys{Quantitative comparisons between our proposed temporally averaged loss formulation versus per-step loss formulation. The values in the table correspond to the mean of several measurements obtained over one cycle at the nominal walking speed (1.2m/s). ROM denotes the joint range of motion, computed for each cycle as the difference between the maximum and minimum angles, and then averaged over multiple cycles.}}
  \label{tab:effect_loss}
\begin{tabular}{llcc}
\toprule
 & Metric & TempAvg (Ours) & Per-step \\
\midrule
\multirow{3}{*}{$L_{pose}$} & Hip ROM (Sagittal plane) & 33.3° & 23.7° \\
 & Shoulder ROM (Sagittal plane) & 12.8° & 5.3° \\
 & Pelvis ROM & & \\
 & \quad Sagittal plane & 7.6° & 4.8° \\
 & \quad Lateral plane & 7.6° & 3.0° \\
\midrule
\multirow{2}{*}{$L_{up}$} & Pelvis ROM & & \\
 & \quad Sagittal plane & 7.6° & 3.4° \\
 & \quad Lateral plane & 7.6° & 2.3° \\
\midrule
\multirow{2}{*}{$L_{vel}$} & Pelvis Velocity (Forward) & 1.16 m/s & 0.48 m/s \\
 & Pelvis ROM & & \\
 & \quad Sagittal plane & 7.6° & 3.9° \\
 & \quad Lateral plane & 7.6° & 4.8° \\
\bottomrule
\end{tabular}
\end{table}

\begin{figure}
    \centering
    \includegraphics[width=0.159\columnwidth]{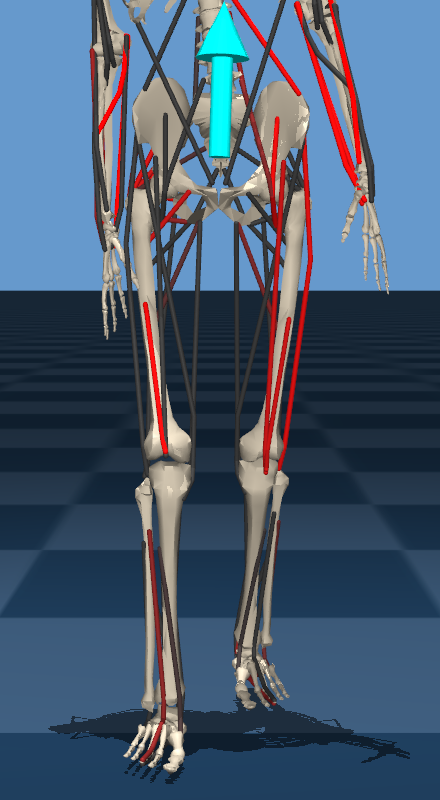}
    \includegraphics[width=0.159\columnwidth]{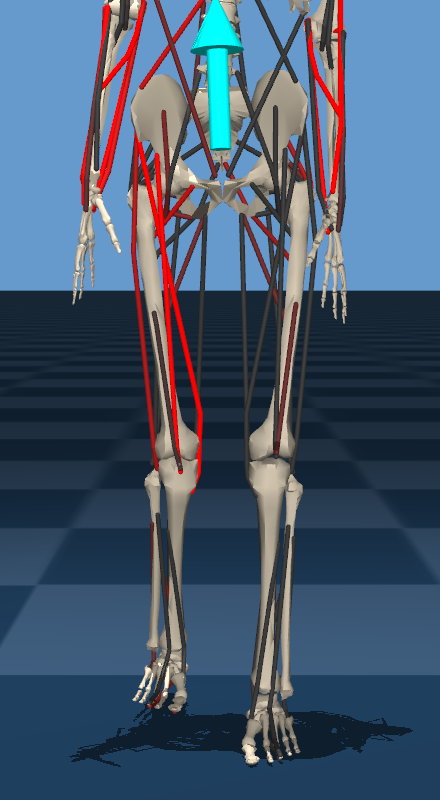}
    \includegraphics[width=0.159\columnwidth]{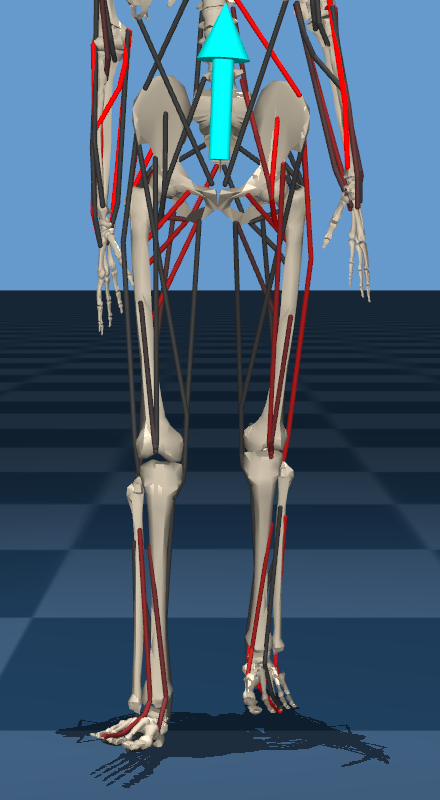}
    \includegraphics[width=0.159\columnwidth]{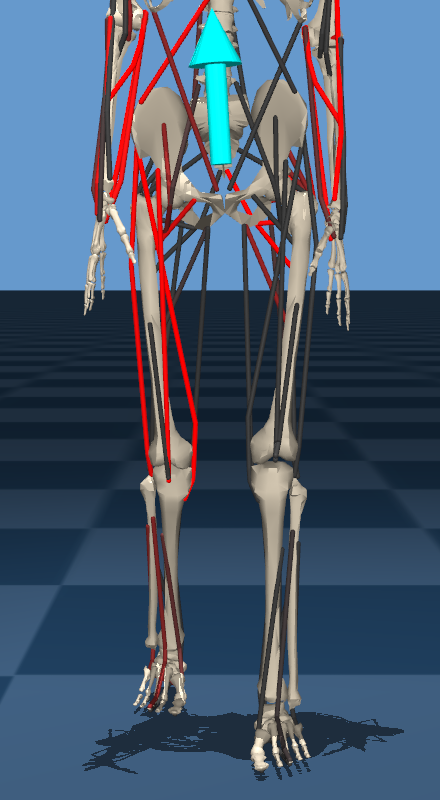}
    \includegraphics[width=0.159\columnwidth]{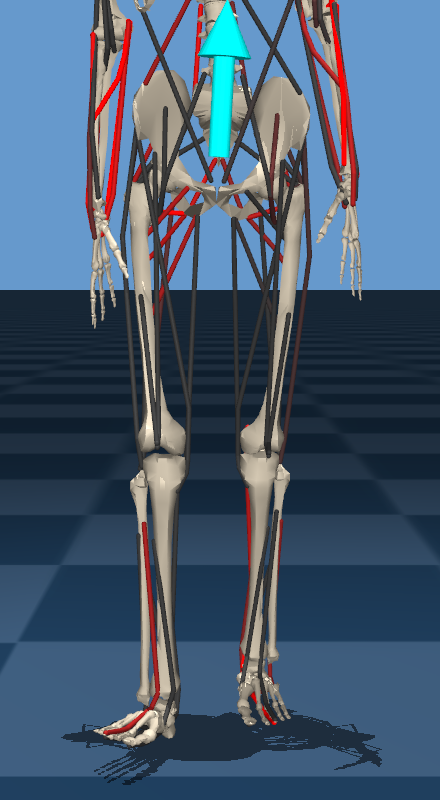}
    \includegraphics[width=0.159\columnwidth]{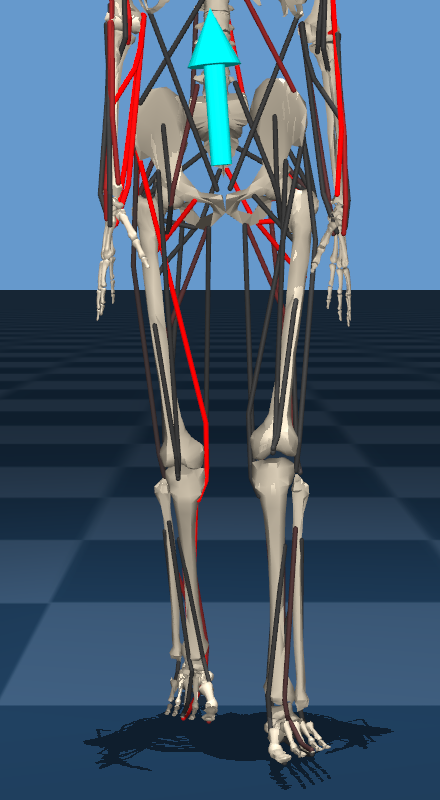}

    \includegraphics[width=0.159\columnwidth]{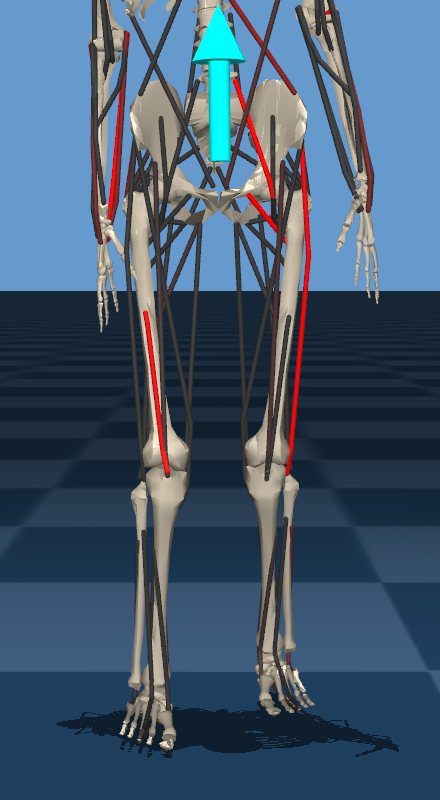}
    \includegraphics[width=0.159\columnwidth]{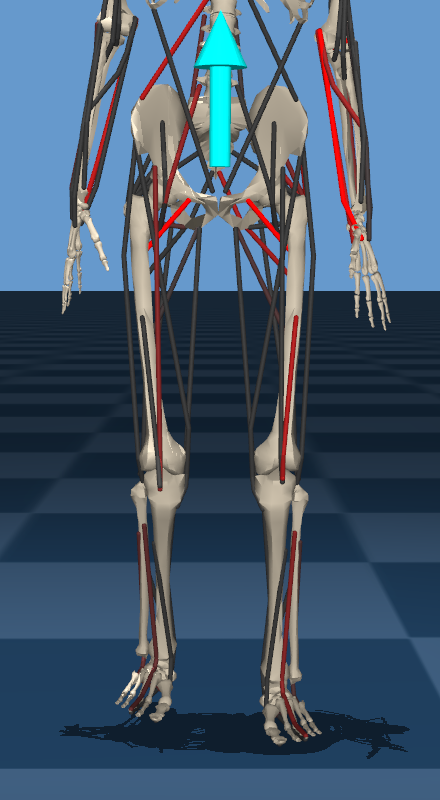}
    \includegraphics[width=0.159\columnwidth]{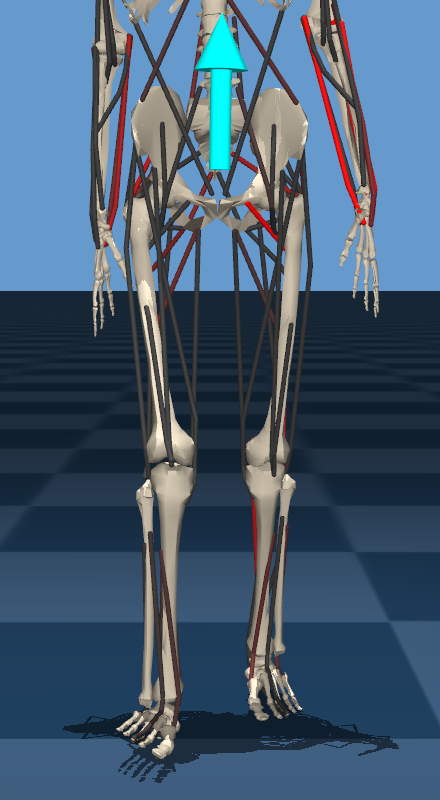}
    \includegraphics[width=0.159\columnwidth]{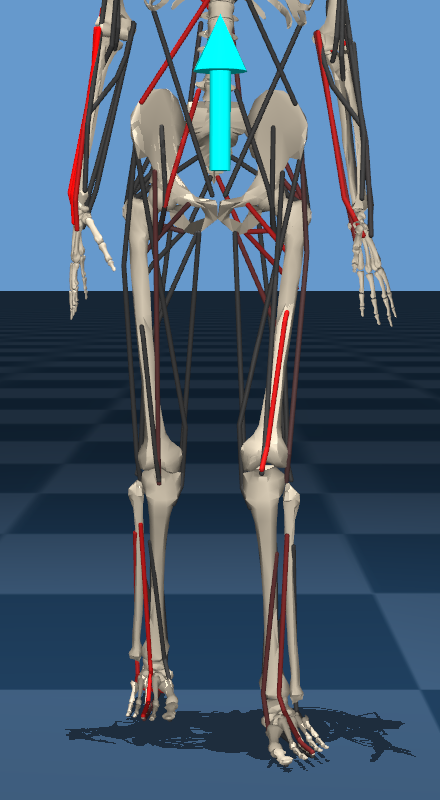}
    \includegraphics[width=0.159\columnwidth]{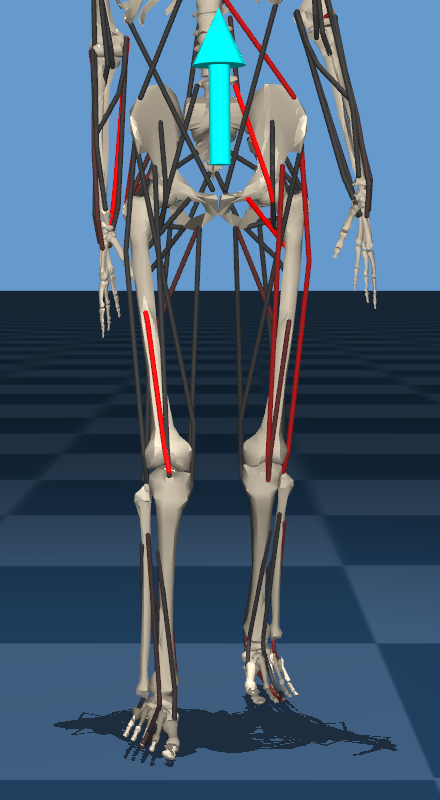}
    \includegraphics[width=0.159\columnwidth]{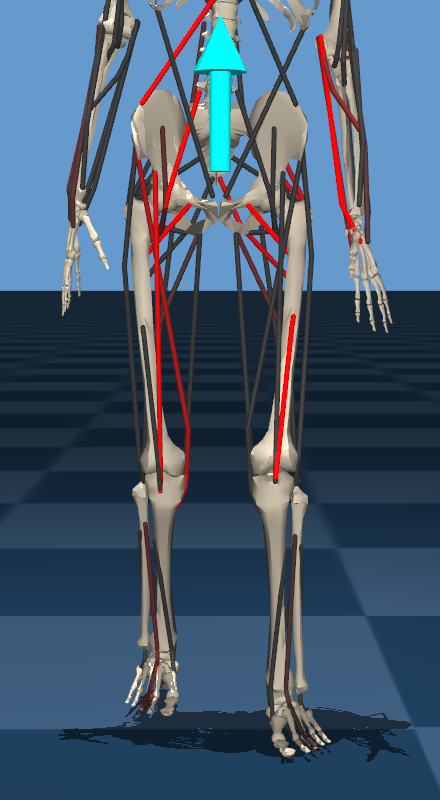}

    \captionof{figure}{
    \ys{Effect of \( L_\text{up} \) on pelvic motion. Our temporally averaged formulation (top) produces more natural pelvic rotation during the leg swing phase, whereas the per-step formulation (bottom) suppresses this rotation.}
    }    
    \label{fig:ablation_updir}    
\end{figure}

\ys{We demonstrate the effect of our proposed temporally averaged loss formulation (Equation~\eqref{eq:avg_loss}) on locomotion quality and dynamics, comparing it against the per-step formulation (Equation~\eqref{eq:step_loss}) commonly used in imitation learning (Table~\ref{tab:effect_loss}).}

\ys{For \( L_\text{pose} \), per-step loss enforces rigid frame-level pose matching, leading to limited tolerance for natural variation. The resulting gait is characterized by reduced ROM—hip (23.7° vs. 33.3°) and shoulder (5.3° vs. 12.8°)—and diminished pelvic oscillations (4.8° sagittal, 3.0° lateral vs. 7.6° in both planes). In the accompanying video, this manifests as a slightly crouched gait with reduced step length and limited pelvic rotation.}

\ys{For \( L_\text{up} \), the per-step formulation strongly constrains pelvic dynamics, yielding only 2.3° lateral and 3.4° sagittal ROM compared to 7.6° in both planes under the temporally averaged formulation. 
Qualitatively, pelvic rotation appears nearly rigid, with natural oscillatory components largely absent (Figure~\ref{fig:ablation_updir}).}

\ys{For \( L_\text{vel} \), the per-step formulation fails to maintain the target walking speed, achieving only 0.48 m/s versus 1.16 m/s under our method, and concurrently dampens pelvic motion (3.9° sagittal, 4.8° lateral vs. 7.6° in both planes). 
This produces a short-stepped gait with suppressed pelvic rotation.} 

These observations
suggest that strictly enforcing frame-level accuracy can hinder the emergence of biomechanically plausible dynamics. In contrast, our temporally averaged loss encourages smoother motion trajectories and allows natural oscillation patterns to emerge over time.

\subsection{\ys{Comparison with Torque-Actuated Humanoid}}

\begin{figure}
    \centering
    \reflectbox{\includegraphics[trim=50 30 50 30, clip, width=0.092\textwidth]{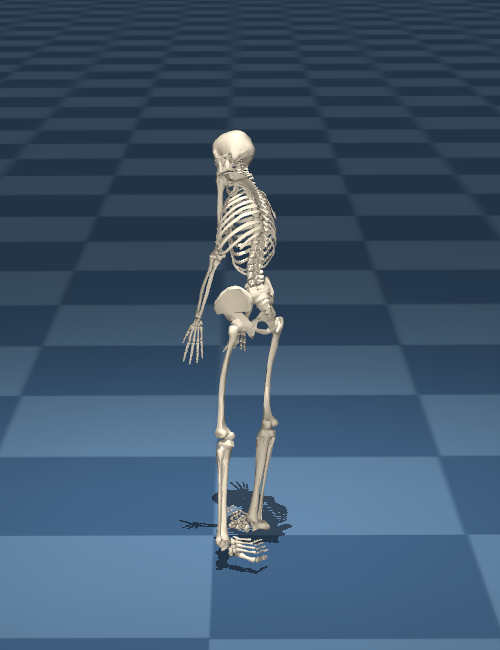}}
    \reflectbox{\includegraphics[trim=50 30 50 30, clip, width=0.092\textwidth]{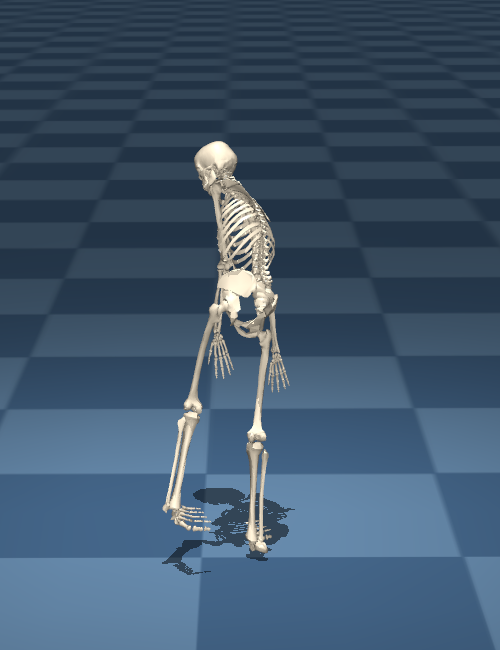}}
    \reflectbox{\includegraphics[trim=50 30 50 30, clip, width=0.092\textwidth]{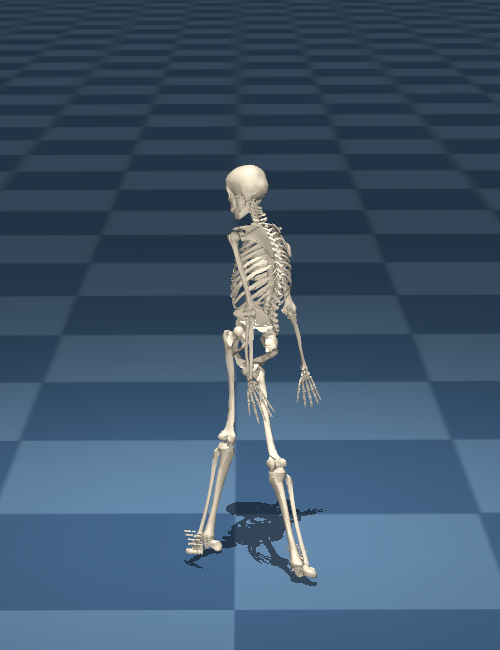}}
    \reflectbox{\includegraphics[trim=50 30 50 30, clip, width=0.092\textwidth]{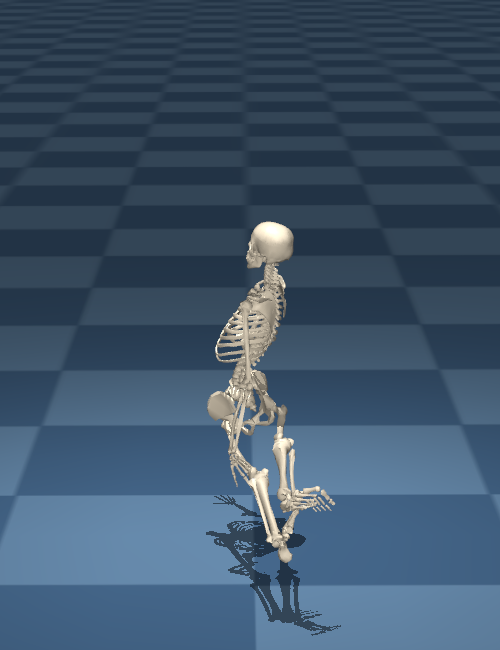}}
    \reflectbox{\includegraphics[trim=50 30 50 30, clip, width=0.092\textwidth]{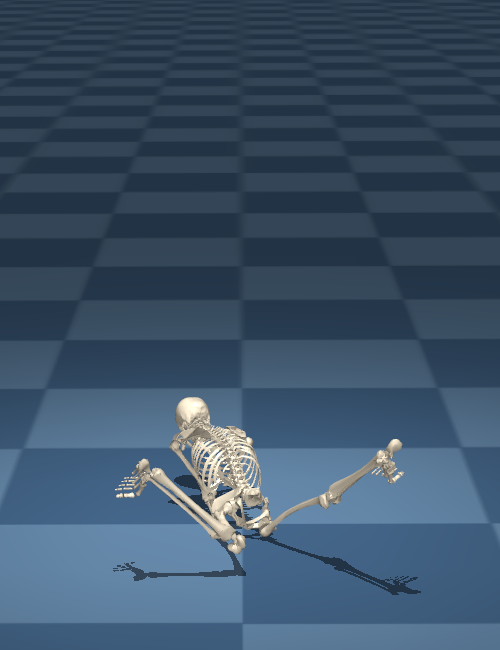}}

    \includegraphics[trim=50 30 50 30, clip, width=0.092\textwidth]{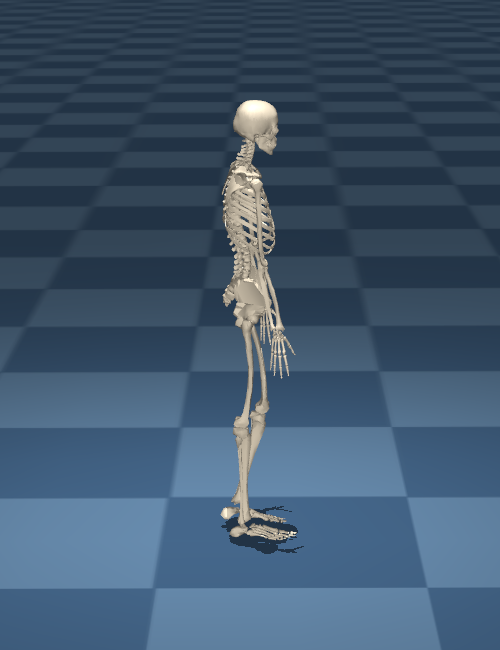}
    \includegraphics[trim=50 30 50 30, clip, width=0.092\textwidth]{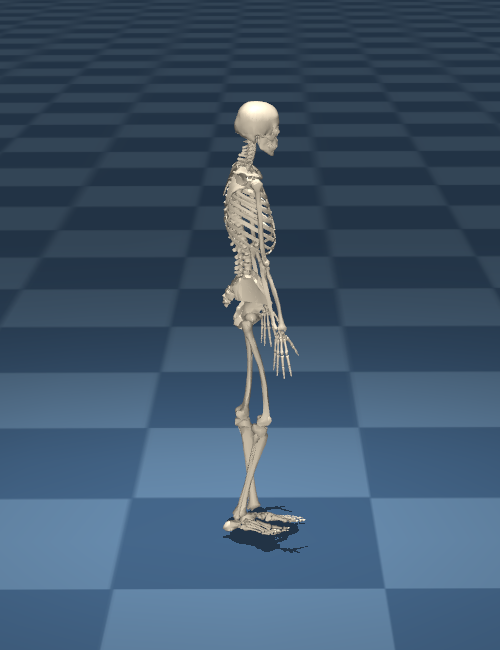}
    \includegraphics[trim=50 30 50 30, clip, width=0.092\textwidth]{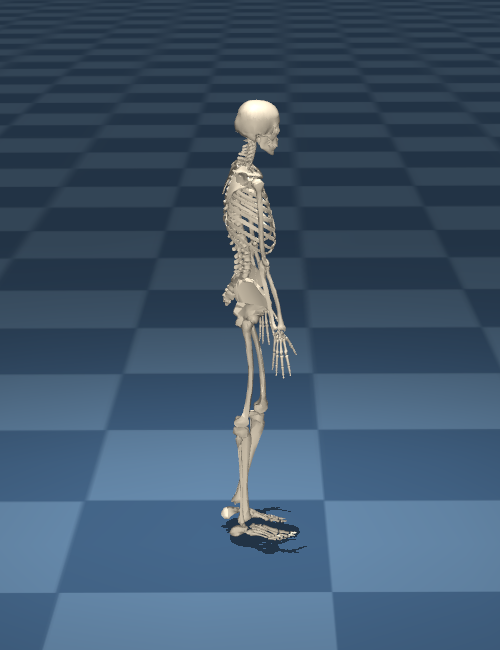}
    \includegraphics[trim=50 30 50 30, clip, width=0.092\textwidth]{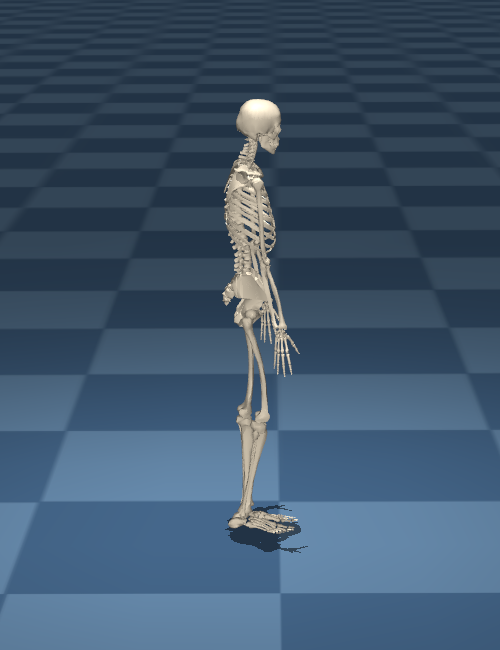}
    \includegraphics[trim=50 30 50 30, clip, width=0.092\textwidth]{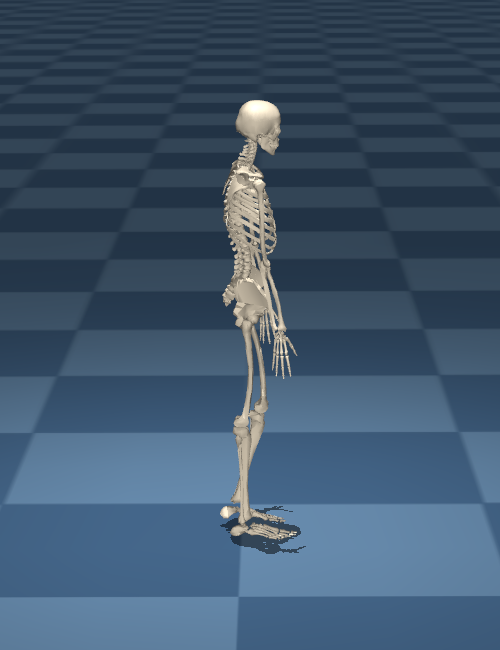}

    \includegraphics[trim=50 30 50 30, clip, width=0.092\textwidth]{images/final_baseloco/cropped_17.png}
    \includegraphics[trim=50 30 50 30, clip, width=0.092\textwidth]{images/final_baseloco/cropped_18.png}
    \includegraphics[trim=50 30 50 30, clip, width=0.092\textwidth]{images/final_baseloco/cropped_19.png}
    \includegraphics[trim=50 30 50 30, clip, width=0.092\textwidth]{images/final_baseloco/cropped_20.png}
    \includegraphics[trim=50 30 50 30, clip, width=0.092\textwidth]{images/final_baseloco/cropped_21.png}

    \caption{
    \ys{Torque-actuated humanoid fails to maintain balance without torque range tuning (top), and produces an unnatural short-stepped, high-frequency walking pattern even with tuning (middle), whereas our muscle-actuated humanoid exhibits natural locomotion (bottom).}
    }    
    \label{fig:ablation_torque}
\end{figure}

\ys{As shown in Section~\ref{sec:emergent_gait}, our framework enables different musculoskeletal characters to exhibit distinct and plausible locomotion patterns.
Here, we further highlight the role of musculoskeletal models through a comparison with a torque-actuated humanoid trained under the same loss formulation.}

\ys{Unlike our musculoskeletal agents, the torque-actuated humanoid failed to learn even basic balanced walking behavior (Figure~\ref{fig:ablation_torque}).
Even with manual tuning of joint torque limits, it only converged to a very short-stepped, high-frequency stepping pattern.
This contrast indicates that the musculoskeletal model not only facilitates stable and robust locomotion but also implicitly restricts the solution space to biomechanically plausible behaviors, serving as a strong prior when no motion references are available.}

\section{Conclusion and Discussion}

We introduced \textit{FreeMusco}, a motion-free framework that \ys{jointly} learns latent locomotion \ys{representations and} control policies for musculoskeletal characters without \ys{relying on} motion data.
\ys{By leveraging musculoskeletal models as strong priors and integrating a locomotion objective with temporally averaged loss terms,}
our method enables the emergence of diverse, morphology-adaptive, and energy-aware locomotion behaviors.
\ys{Randomization of target poses and energy levels during training further promotes behavioral diversity, allowing locomotion to be flexibly modulated in both form and intensity.}
The learned latent space supports high-level control, including goal navigation and path following, across human, non-human, and synthetic characters.
\ys{These} results demonstrate the feasibility and versatility of motion-free learning as a viable alternative to demonstration-based character control.

{\color{ys}
Using the default standing pose as the target was intended primarily as a minimal guide for learning locomotion, rather than as an optimal choice. Our results show that characters with different morphologies are able to develop distinct and morphology-appropriate gaits under the same target pose (Humanoid vs. Chimanoid).
We further show that varying the target pose influences the style of the resulting motion.
This suggests that some poses may yield more natural or more efficient walking patterns, yet identifying a universally optimal pose is inherently difficult due to tradeoffs in human gait—balancing energy efficiency, robustness, and injury avoidance. Exploring this design space more systematically represents an interesting direction for future work.

One concrete artifact highlighting this issue is the stiffness of the Humanoid’s arms.
This appears to stem from our naive target pose selection, where the elbows are placed in a fully extended configuration at the limit of their range of motion. As a result, the pose tracking loss, even under temporal averaging, tends to favor straightened arms throughout locomotion.
As shown in the accompanying video, this effect is mitigated when target poses with slightly bent elbows are used, suggesting that more careful target pose choice could improve upper-body realism.

Another limitation concerns turning behavior. In our current formulation, direction control is defined with respect to the root (pelvis) orientation, which sometimes produces unnatural turning motions.
Incorporating head-first orientation cues, as observed in natural animal and human locomotion, could lead to more coordinated and lifelike turning strategies within our framework.}
Future work \ys{also} includes extending the framework to richer motor skills such as manipulation or multi-contact actions, and exploring automatic discovery of goal structures to reduce manual design effort.

\begin{acks}
This work was supported by the National Research Foundation of
Korea (NRF) grant (RS-2023-00222776), and by Culture, Sports and Tourism R\&D Program through the Korea Creative Content Agency grant funded by the Ministry of Culture, Sports and Tourism in 2024 (RS-2024-00399136).
\end{acks}

\bibliographystyle{ACM-Reference-Format}
\bibliography{main}


\begin{thebibliography}{55}


\ifx \showCODEN    \undefined \def \showCODEN     #1{\unskip}     \fi
\ifx \showISBNx    \undefined \def \showISBNx     #1{\unskip}     \fi
\ifx \showISBNxiii \undefined \def \showISBNxiii  #1{\unskip}     \fi
\ifx \showISSN     \undefined \def \showISSN      #1{\unskip}     \fi
\ifx \showLCCN     \undefined \def \showLCCN      #1{\unskip}     \fi
\ifx \shownote     \undefined \def \shownote      #1{#1}          \fi
\ifx \showarticletitle \undefined \def \showarticletitle #1{#1}   \fi
\ifx \showURL      \undefined \def \showURL       {\relax}        \fi
\providecommand\bibfield[2]{#2}
\providecommand\bibinfo[2]{#2}
\providecommand\natexlab[1]{#1}
\providecommand\showeprint[2][]{arXiv:#2}

\bibitem[Bae et~al\mbox{.}(2025)]%
        {bae_versatile_2025}
\bibfield{author}{\bibinfo{person}{Jinseok Bae}, \bibinfo{person}{Jungdam Won}, \bibinfo{person}{Donggeun Lim}, \bibinfo{person}{Inwoo Hwang}, {and} \bibinfo{person}{Young~Min Kim}.} \bibinfo{year}{2025}\natexlab{}.
\newblock \showarticletitle{Versatile {Physics}-based {Character} {Control} with {Hybrid} {Latent} {Representation}}.
\newblock \bibinfo{journal}{\emph{Computer Graphics Forum}} \bibinfo{volume}{44}, \bibinfo{number}{2} (\bibinfo{year}{2025}), \bibinfo{pages}{e70018}.
\newblock
\showISSN{1467-8659}
\href{https://doi.org/10.1111/cgf.70018}{doi:\nolinkurl{10.1111/cgf.70018}}
\newblock
\shownote{\_eprint: https://onlinelibrary.wiley.com/doi/pdf/10.1111/cgf.70018}.


\bibitem[Bae et~al\mbox{.}(2023)]%
        {bae_pmp_2023}
\bibfield{author}{\bibinfo{person}{Jinseok Bae}, \bibinfo{person}{Jungdam Won}, \bibinfo{person}{Donggeun Lim}, \bibinfo{person}{Cheol-Hui Min}, {and} \bibinfo{person}{Young~Min Kim}.} \bibinfo{year}{2023}\natexlab{}.
\newblock \showarticletitle{{PMP}: {Learning} to {Physically} {Interact} with {Environments} using {Part}-wise {Motion} {Priors}}. In \bibinfo{booktitle}{\emph{{ACM} {SIGGRAPH} 2023 {Conference} {Proceedings}}} \emph{(\bibinfo{series}{{SIGGRAPH} '23})}. \bibinfo{publisher}{Association for Computing Machinery}, \bibinfo{address}{New York, NY, USA}, \bibinfo{pages}{1--10}.
\newblock
\showISBNx{9798400701597}


\bibitem[Berg et~al\mbox{.}(2023)]%
        {berg_sar_2023}
\bibfield{author}{\bibinfo{person}{Cameron Berg}, \bibinfo{person}{Vittorio Caggiano}, {and} \bibinfo{person}{Vikash Kumar}.} \bibinfo{year}{2023}\natexlab{}.
\newblock \showarticletitle{{SAR}: {Generalization} of {Physiological} {Dexterity} via {Synergistic} {Action} {Representation}}. In \bibinfo{booktitle}{\emph{Robotics: {Science} and {Systems} {XIX}}}. \bibinfo{publisher}{Robotics: Science and Systems Foundation}.
\newblock
\showISBNx{978-0-9923747-9-2}
\href{https://doi.org/10.15607/RSS.2023.XIX.007}{doi:\nolinkurl{10.15607/RSS.2023.XIX.007}}


\bibitem[Bergamin et~al\mbox{.}(2019)]%
        {bergamin_drecon_2019}
\bibfield{author}{\bibinfo{person}{Kevin Bergamin}, \bibinfo{person}{Simon Clavet}, \bibinfo{person}{Daniel Holden}, {and} \bibinfo{person}{James~Richard Forbes}.} \bibinfo{year}{2019}\natexlab{}.
\newblock \showarticletitle{{DReCon}: data-driven responsive control of physics-based characters}.
\newblock \bibinfo{journal}{\emph{ACM Transactions on Graphics (TOG)}}  \bibinfo{volume}{38} (\bibinfo{date}{Nov.} \bibinfo{year}{2019}), \bibinfo{pages}{206:1--206:11}.
\newblock
\showISSN{0730-0301}


\bibitem[Coros et~al\mbox{.}(2010)]%
        {coros2010generalized}
\bibfield{author}{\bibinfo{person}{Stelian Coros}, \bibinfo{person}{Philippe Beaudoin}, {and} \bibinfo{person}{Michiel Van~de Panne}.} \bibinfo{year}{2010}\natexlab{}.
\newblock \showarticletitle{Generalized biped walking control}.
\newblock \bibinfo{journal}{\emph{ACM Transactions On Graphics (TOG)}} \bibinfo{volume}{29}, \bibinfo{number}{4} (\bibinfo{year}{2010}), \bibinfo{pages}{1--9}.
\newblock


\bibitem[Dou et~al\mbox{.}(2023)]%
        {dou_case_2023}
\bibfield{author}{\bibinfo{person}{Zhiyang Dou}, \bibinfo{person}{Xuelin Chen}, \bibinfo{person}{Qingnan Fan}, \bibinfo{person}{Taku Komura}, {and} \bibinfo{person}{Wenping Wang}.} \bibinfo{year}{2023}\natexlab{}.
\newblock \showarticletitle{C·{ASE}: {Learning} {seConditional} {Adversarial} {Skill} {Embeddings} for {Physics}-based {Characters}}. In \bibinfo{booktitle}{\emph{{SIGGRAPH} {Asia} 2023 {Conference} {Papers}}} \emph{(\bibinfo{series}{{SA} '23})}. \bibinfo{pages}{1--11}.
\newblock
\showISBNx{9798400703157}


\bibitem[Feng et~al\mbox{.}(2023)]%
        {feng_musclevae_2023}
\bibfield{author}{\bibinfo{person}{Yusen Feng}, \bibinfo{person}{Xiyan Xu}, {and} \bibinfo{person}{Libin Liu}.} \bibinfo{year}{2023}\natexlab{}.
\newblock \showarticletitle{MuscleVAE: Model-Based Controllers of Muscle-Actuated Characters}. In \bibinfo{booktitle}{\emph{SIGGRAPH Asia 2023 Conference Papers}} (Sydney, NSW, Australia) \emph{(\bibinfo{series}{SA '23})}. \bibinfo{publisher}{Association for Computing Machinery}, \bibinfo{address}{New York, NY, USA}, Article \bibinfo{articleno}{3}, \bibinfo{numpages}{11}~pages.
\newblock
\showISBNx{9798400703157}
\href{https://doi.org/10.1145/3610548.3618137}{doi:\nolinkurl{10.1145/3610548.3618137}}


\bibitem[Fussell et~al\mbox{.}(2021)]%
        {fussel_supertrack_2021}
\bibfield{author}{\bibinfo{person}{Levi Fussell}, \bibinfo{person}{Kevin Bergamin}, {and} \bibinfo{person}{Daniel Holden}.} \bibinfo{year}{2021}\natexlab{}.
\newblock \showarticletitle{SuperTrack: motion tracking for physically simulated characters using supervised learning}.
\newblock \bibinfo{journal}{\emph{ACM Trans. Graph.}} \bibinfo{volume}{40}, \bibinfo{number}{6}, Article \bibinfo{articleno}{197} (\bibinfo{date}{Dec.} \bibinfo{year}{2021}), \bibinfo{numpages}{13}~pages.
\newblock
\showISSN{0730-0301}
\href{https://doi.org/10.1145/3478513.3480527}{doi:\nolinkurl{10.1145/3478513.3480527}}


\bibitem[Geijtenbeek et~al\mbox{.}(2013)]%
        {thomas_flexiblemuscle_2013}
\bibfield{author}{\bibinfo{person}{Thomas Geijtenbeek}, \bibinfo{person}{Michiel van~de Panne}, {and} \bibinfo{person}{A.~Frank van~der Stappen}.} \bibinfo{year}{2013}\natexlab{}.
\newblock \showarticletitle{Flexible Muscle-Based Locomotion for Bipedal Creatures}.
\newblock \bibinfo{journal}{\emph{ACM Transactions on Graphics}} \bibinfo{volume}{32}, \bibinfo{number}{6} (\bibinfo{year}{2013}).
\newblock


\bibitem[Heess et~al\mbox{.}(2016)]%
        {heess_learning_2016}
\bibfield{author}{\bibinfo{person}{Nicolas Heess}, \bibinfo{person}{Greg Wayne}, \bibinfo{person}{Yuval Tassa}, \bibinfo{person}{Timothy Lillicrap}, \bibinfo{person}{Martin Riedmiller}, {and} \bibinfo{person}{David Silver}.} \bibinfo{year}{2016}\natexlab{}.
\newblock \showarticletitle{Learning and {Transfer} of {Modulated} {Locomotor} {Controllers}}.
\newblock \bibinfo{journal}{\emph{arXiv:1610.05182 [cs]}} (\bibinfo{date}{Oct.} \bibinfo{year}{2016}).
\newblock
\urldef\tempurl%
\url{http://arxiv.org/abs/1610.05182}
\showURL{%
\tempurl}
\newblock
\shownote{arXiv: 1610.05182}.


\bibitem[Hodgins et~al\mbox{.}(1995)]%
        {hodgins_animating_1995}
\bibfield{author}{\bibinfo{person}{Jessica~K. Hodgins}, \bibinfo{person}{Wayne~L. Wooten}, \bibinfo{person}{David~C. Brogan}, {and} \bibinfo{person}{James~F. O'Brien}.} \bibinfo{year}{1995}\natexlab{}.
\newblock \showarticletitle{Animating human athletics}. In \bibinfo{booktitle}{\emph{Proceedings of the 22nd annual conference on {Computer} graphics and interactive techniques}} \emph{(\bibinfo{series}{{SIGGRAPH} '95})}. \bibinfo{publisher}{Association for Computing Machinery}, \bibinfo{address}{New York, NY, USA}, \bibinfo{pages}{71--78}.
\newblock
\showISBNx{978-0-89791-701-8}


\bibitem[Ikkala and H{\"a}m{\"a}l{\"a}inen(2022)]%
        {ikkala_converting_2022}
\bibfield{author}{\bibinfo{person}{Aleksi Ikkala} {and} \bibinfo{person}{Perttu H{\"a}m{\"a}l{\"a}inen}.} \bibinfo{year}{2022}\natexlab{}.
\newblock \showarticletitle{Converting Biomechanical Models from OpenSim to MuJoCo}. In \bibinfo{booktitle}{\emph{Converging Clinical and Engineering Research on Neurorehabilitation IV}}, \bibfield{editor}{\bibinfo{person}{Diego Torricelli}, \bibinfo{person}{Metin Akay}, {and} \bibinfo{person}{Jose~L. Pons}} (Eds.). \bibinfo{publisher}{Springer International Publishing}, \bibinfo{address}{Cham}, \bibinfo{pages}{277--281}.
\newblock
\showISBNx{978-3-030-70316-5}


\bibitem[Kim et~al\mbox{.}(2025)]%
        {kim_physicsfc_2025}
\bibfield{author}{\bibinfo{person}{Minsu Kim}, \bibinfo{person}{Eunho Jung}, {and} \bibinfo{person}{Yoonsang Lee}.} \bibinfo{year}{2025}\natexlab{}.
\newblock \showarticletitle{{PhysicsFC}: {Learning} {User}-{Controlled} {Skills} for a {Physics}-{Based} {Football} {Player} {Controller}}.
\newblock \bibinfo{journal}{\emph{ACM Trans. Graph.}} \bibinfo{volume}{44}, \bibinfo{number}{4} (\bibinfo{date}{July} \bibinfo{year}{2025}), \bibinfo{pages}{130:1--130:21}.
\newblock
\showISSN{0730-0301}
\href{https://doi.org/10.1145/3731425}{doi:\nolinkurl{10.1145/3731425}}


\bibitem[Kim and Lee(2023)]%
        {kim_learning_2023}
\bibfield{author}{\bibinfo{person}{Minkwan Kim} {and} \bibinfo{person}{Yoonsang Lee}.} \bibinfo{year}{2023}\natexlab{}.
\newblock \showarticletitle{Learning {Human}-like {Locomotion} {Based} on {Biological} {Actuation} and {Rewards}}. In \bibinfo{booktitle}{\emph{{ACM} {SIGGRAPH} 2023 {Posters}}} \emph{(\bibinfo{series}{{SIGGRAPH} '23})}. \bibinfo{publisher}{Association for Computing Machinery}, \bibinfo{address}{New York, NY, USA}, \bibinfo{pages}{1--2}.
\newblock
\showISBNx{979-8-4007-0152-8}
\href{https://doi.org/10.1145/3588028.3603646}{doi:\nolinkurl{10.1145/3588028.3603646}}


\bibitem[Kwon et~al\mbox{.}(2023)]%
        {kwon_adaptive_2023}
\bibfield{author}{\bibinfo{person}{Taesoo Kwon}, \bibinfo{person}{Taehong Gu}, \bibinfo{person}{Jaewon Ahn}, {and} \bibinfo{person}{Yoonsang Lee}.} \bibinfo{year}{2023}\natexlab{}.
\newblock \showarticletitle{Adaptive {Tracking} of a {Single}-{Rigid}-{Body} {Character} in {Various} {Environments}}. In \bibinfo{booktitle}{\emph{{SIGGRAPH} {Asia} 2023 {Conference} {Papers}}} \emph{(\bibinfo{series}{{SA} '23})}. \bibinfo{address}{New York, NY, USA}, \bibinfo{pages}{1--11}.
\newblock
\showISBNx{9798400703157}


\bibitem[Kwon et~al\mbox{.}(2020)]%
        {kwon2020fast}
\bibfield{author}{\bibinfo{person}{Taesoo Kwon}, \bibinfo{person}{Yoonsang Lee}, {and} \bibinfo{person}{Michiel Van De~Panne}.} \bibinfo{year}{2020}\natexlab{}.
\newblock \showarticletitle{Fast and flexible multilegged locomotion using learned centroidal dynamics}.
\newblock \bibinfo{journal}{\emph{ACM Transactions on Graphics (TOG)}} \bibinfo{volume}{39}, \bibinfo{number}{4} (\bibinfo{year}{2020}), \bibinfo{pages}{46--1}.
\newblock


\bibitem[La~Barbera et~al\mbox{.}(2021)]%
        {barbera_ostrichrl_2021}
\bibfield{author}{\bibinfo{person}{Vittorio La~Barbera}, \bibinfo{person}{Fabio Pardo}, \bibinfo{person}{Yuval Tassa}, \bibinfo{person}{Monica Daley}, \bibinfo{person}{Christopher Richards}, \bibinfo{person}{Petar Kormushev}, {and} \bibinfo{person}{John Hutchinson}.} \bibinfo{year}{2021}\natexlab{}.
\newblock \showarticletitle{{OstrichRL}: A Musculoskeletal Ostrich Simulation to Study Bio-mechanical Locomotion}. In \bibinfo{booktitle}{\emph{NeurIPS 2021 Deep Reinforcement Learning Workshop, 35th Conference on Neural Information Processing Systems (NeurIPS 2021)}}.
\newblock
\urldef\tempurl%
\url{https://kormushev.com/papers/Pardo_NeurIPS-2021.pdf}
\showURL{%
\tempurl}


\bibitem[Lee et~al\mbox{.}(2022)]%
        {chimeras22}
\bibfield{author}{\bibinfo{person}{Seyoung Lee}, \bibinfo{person}{Jiye Lee}, {and} \bibinfo{person}{Jehee Lee}.} \bibinfo{year}{2022}\natexlab{}.
\newblock \showarticletitle{Learning {Virtual} {Chimeras} by {Dynamic} {Motion} {Reassembly}}.
\newblock \bibinfo{journal}{\emph{ACM Transactions on Graphics}} \bibinfo{volume}{41}, \bibinfo{number}{6} (\bibinfo{year}{2022}), \bibinfo{pages}{182:1--182:13}.
\newblock
\showISSN{0730-0301}
\href{https://doi.org/10.1145/3550454.3555489}{doi:\nolinkurl{10.1145/3550454.3555489}}


\bibitem[Lee et~al\mbox{.}(2021)]%
        {Lee:2021:Parameterized}
\bibfield{author}{\bibinfo{person}{Seyoung Lee}, \bibinfo{person}{Sunmin Lee}, \bibinfo{person}{Yongwoo Lee}, {and} \bibinfo{person}{Jehee Lee}.} \bibinfo{year}{2021}\natexlab{}.
\newblock \showarticletitle{Learning a family of motor skills from a single motion clip}.
\newblock \bibinfo{journal}{\emph{ACM Transactions on Graphics}} \bibinfo{volume}{40}, \bibinfo{number}{4} (\bibinfo{date}{July} \bibinfo{year}{2021}), \bibinfo{pages}{93:1--93:13}.
\newblock
\showISSN{0730-0301}


\bibitem[Lee et~al\mbox{.}(2019)]%
        {lee_scalablemuscle_2019}
\bibfield{author}{\bibinfo{person}{Seunghwan Lee}, \bibinfo{person}{Moonseok Park}, \bibinfo{person}{Kyoungmin Lee}, {and} \bibinfo{person}{Jehee Lee}.} \bibinfo{year}{2019}\natexlab{}.
\newblock \showarticletitle{Scalable muscle-actuated human simulation and control}.
\newblock \bibinfo{journal}{\emph{ACM Trans. Graph.}} \bibinfo{volume}{38}, \bibinfo{number}{4}, Article \bibinfo{articleno}{73} (\bibinfo{date}{jul} \bibinfo{year}{2019}), \bibinfo{numpages}{13}~pages.
\newblock
\showISSN{0730-0301}
\href{https://doi.org/10.1145/3306346.3322972}{doi:\nolinkurl{10.1145/3306346.3322972}}


\bibitem[Lee et~al\mbox{.}(2010)]%
        {lee_data-driven_2010}
\bibfield{author}{\bibinfo{person}{Yoonsang Lee}, \bibinfo{person}{Sungeun Kim}, {and} \bibinfo{person}{Jehee Lee}.} \bibinfo{year}{2010}\natexlab{}.
\newblock \showarticletitle{Data-driven biped control}.
\newblock \bibinfo{journal}{\emph{ACM Trans. Graph.}} \bibinfo{volume}{29}, \bibinfo{number}{4} (\bibinfo{year}{2010}), \bibinfo{pages}{1--8}.
\newblock
\href{https://doi.org/10.1145/1778765.1781155}{doi:\nolinkurl{10.1145/1778765.1781155}}


\bibitem[Lee et~al\mbox{.}(2014)]%
        {lee_locomotion_2014}
\bibfield{author}{\bibinfo{person}{Yoonsang Lee}, \bibinfo{person}{Moon~Seok Park}, \bibinfo{person}{Taesoo Kwon}, {and} \bibinfo{person}{Jehee Lee}.} \bibinfo{year}{2014}\natexlab{}.
\newblock \showarticletitle{Locomotion {Control} for {Many}-muscle {Humanoids}}.
\newblock \bibinfo{journal}{\emph{ACM Trans. Graph.}} \bibinfo{volume}{33}, \bibinfo{number}{6} (\bibinfo{year}{2014}), \bibinfo{pages}{218:1--218:11}.
\newblock
\showISSN{0730-0301}
\href{https://doi.org/10.1145/2661229.2661233}{doi:\nolinkurl{10.1145/2661229.2661233}}


\bibitem[Liu et~al\mbox{.}(2012)]%
        {liu_terrain_2012}
\bibfield{author}{\bibinfo{person}{Libin Liu}, \bibinfo{person}{KangKang Yin}, \bibinfo{person}{Michiel van~de Panne}, {and} \bibinfo{person}{Baining Guo}.} \bibinfo{year}{2012}\natexlab{}.
\newblock \showarticletitle{Terrain runner: control, parameterization, composition, and planning for highly dynamic motions}.
\newblock \bibinfo{journal}{\emph{ACM Trans. Graph.}} \bibinfo{volume}{31}, \bibinfo{number}{6} (\bibinfo{date}{Nov.} \bibinfo{year}{2012}), \bibinfo{pages}{154:1--154:10}.
\newblock
\showISSN{0730-0301}
\href{https://doi.org/10.1145/2366145.2366173}{doi:\nolinkurl{10.1145/2366145.2366173}}


\bibitem[Merel et~al\mbox{.}(2019)]%
        {DBLP:conf/iclr/MerelHGAPWTH19}
\bibfield{author}{\bibinfo{person}{Josh Merel}, \bibinfo{person}{Leonard Hasenclever}, \bibinfo{person}{Alexandre Galashov}, \bibinfo{person}{Arun Ahuja}, \bibinfo{person}{Vu Pham}, \bibinfo{person}{Greg Wayne}, \bibinfo{person}{Yee~Whye Teh}, {and} \bibinfo{person}{Nicolas Heess}.} \bibinfo{year}{2019}\natexlab{}.
\newblock \showarticletitle{Neural Probabilistic Motor Primitives for Humanoid Control}. In \bibinfo{booktitle}{\emph{7th International Conference on Learning Representations, {ICLR} 2019, New Orleans, LA, USA, May 6-9, 2019}}.
\newblock


\bibitem[Park et~al\mbox{.}(2025)]%
        {park_magnet_2025}
\bibfield{author}{\bibinfo{person}{Jungnam Park}, \bibinfo{person}{Euikyun Jung}, \bibinfo{person}{Jehee Lee}, {and} \bibinfo{person}{Jungdam Won}.} \bibinfo{year}{2025}\natexlab{}.
\newblock \showarticletitle{{MAGNET}: {Muscle} {Activation} {Generation} {Networks} for {Diverse} {Human} {Movement}}. In \bibinfo{booktitle}{\emph{Proceedings of the {Special} {Interest} {Group} on {Computer} {Graphics} and {Interactive} {Techniques} {Conference} {Conference} {Papers}}} \emph{(\bibinfo{series}{{SIGGRAPH} {Conference} {Papers} '25})}. \bibinfo{publisher}{Association for Computing Machinery}, \bibinfo{address}{New York, NY, USA}, \bibinfo{pages}{1--11}.
\newblock
\showISBNx{979-8-4007-1540-2}
\href{https://doi.org/10.1145/3721238.3730617}{doi:\nolinkurl{10.1145/3721238.3730617}}


\bibitem[Park et~al\mbox{.}(2022)]%
        {park_generative_2022}
\bibfield{author}{\bibinfo{person}{Jungnam Park}, \bibinfo{person}{Sehee Min}, \bibinfo{person}{Phil~Sik Chang}, \bibinfo{person}{Jaedong Lee}, \bibinfo{person}{Moon~Seok Park}, {and} \bibinfo{person}{Jehee Lee}.} \bibinfo{year}{2022}\natexlab{}.
\newblock \showarticletitle{Generative {GaitNet}}. In \bibinfo{booktitle}{\emph{{ACM} {SIGGRAPH} 2022 {Conference} {Proceedings}}} \emph{(\bibinfo{series}{{SIGGRAPH} '22})}. \bibinfo{publisher}{Association for Computing Machinery}, \bibinfo{address}{New York, NY, USA}, \bibinfo{pages}{1--9}.
\newblock
\showISBNx{978-1-4503-9337-9}
\href{https://doi.org/10.1145/3528233.3530717}{doi:\nolinkurl{10.1145/3528233.3530717}}


\bibitem[Park et~al\mbox{.}(2019)]%
        {Park2019}
\bibfield{author}{\bibinfo{person}{Soohwan Park}, \bibinfo{person}{Hoseok Ryu}, \bibinfo{person}{Seyoung Lee}, \bibinfo{person}{Sunmin Lee}, {and} \bibinfo{person}{Jehee Lee}.} \bibinfo{year}{2019}\natexlab{}.
\newblock \showarticletitle{Learning predict-and-simulate policies from unorganized human motion data}.
\newblock \bibinfo{journal}{\emph{ACM Transactions on Graphics}} \bibinfo{volume}{38}, \bibinfo{number}{6} (\bibinfo{year}{2019}), \bibinfo{pages}{205:1--205:11}.
\newblock
\showISSN{0730-0301}


\bibitem[Peng et~al\mbox{.}(2018)]%
        {peng2018deepmimic}
\bibfield{author}{\bibinfo{person}{Xue~Bin Peng}, \bibinfo{person}{Pieter Abbeel}, \bibinfo{person}{Sergey Levine}, {and} \bibinfo{person}{Michiel van~de Panne}.} \bibinfo{year}{2018}\natexlab{}.
\newblock \showarticletitle{Deepmimic: Example-guided deep reinforcement learning of physics-based character skills}.
\newblock \bibinfo{journal}{\emph{ACM Transactions on Graphics (TOG)}} \bibinfo{volume}{37}, \bibinfo{number}{4} (\bibinfo{year}{2018}), \bibinfo{pages}{1--14}.
\newblock


\bibitem[Peng et~al\mbox{.}(2015)]%
        {peng2015dynamic}
\bibfield{author}{\bibinfo{person}{Xue~Bin Peng}, \bibinfo{person}{Glen Berseth}, {and} \bibinfo{person}{Michiel Van~de Panne}.} \bibinfo{year}{2015}\natexlab{}.
\newblock \showarticletitle{Dynamic terrain traversal skills using reinforcement learning}.
\newblock \bibinfo{journal}{\emph{ACM Transactions on Graphics (TOG)}} \bibinfo{volume}{34}, \bibinfo{number}{4} (\bibinfo{year}{2015}), \bibinfo{pages}{1--11}.
\newblock


\bibitem[Peng et~al\mbox{.}(2016)]%
        {peng2016terrain}
\bibfield{author}{\bibinfo{person}{Xue~Bin Peng}, \bibinfo{person}{Glen Berseth}, {and} \bibinfo{person}{Michiel Van~de Panne}.} \bibinfo{year}{2016}\natexlab{}.
\newblock \showarticletitle{Terrain-adaptive locomotion skills using deep reinforcement learning}.
\newblock \bibinfo{journal}{\emph{ACM Transactions on Graphics (TOG)}} \bibinfo{volume}{35}, \bibinfo{number}{4} (\bibinfo{year}{2016}), \bibinfo{pages}{1--12}.
\newblock


\bibitem[Peng et~al\mbox{.}(2022)]%
        {ASE}
\bibfield{author}{\bibinfo{person}{Xue~Bin Peng}, \bibinfo{person}{Yunrong Guo}, \bibinfo{person}{Lina Halper}, \bibinfo{person}{Sergey Levine}, {and} \bibinfo{person}{Sanja Fidler}.} \bibinfo{year}{2022}\natexlab{}.
\newblock \showarticletitle{ASE: large-scale reusable adversarial skill embeddings for physically simulated characters}.
\newblock \bibinfo{journal}{\emph{ACM Transactions on Graphics}} \bibinfo{volume}{41}, \bibinfo{number}{4} (\bibinfo{date}{July} \bibinfo{year}{2022}), \bibinfo{pages}{1–17}.
\newblock


\bibitem[Peng et~al\mbox{.}(2021)]%
        {AMP}
\bibfield{author}{\bibinfo{person}{Xue~Bin Peng}, \bibinfo{person}{Ze Ma}, \bibinfo{person}{Pieter Abbeel}, \bibinfo{person}{Sergey Levine}, {and} \bibinfo{person}{Angjoo Kanazawa}.} \bibinfo{year}{2021}\natexlab{}.
\newblock \showarticletitle{AMP: adversarial motion priors for stylized physics-based character control}.
\newblock \bibinfo{journal}{\emph{ACM Transactions on Graphics}} \bibinfo{volume}{40}, \bibinfo{number}{4}, Article \bibinfo{articleno}{144} (\bibinfo{date}{jul} \bibinfo{year}{2021}), \bibinfo{numpages}{20}~pages.
\newblock
\showISSN{0730-0301}


\bibitem[Schumacher et~al\mbox{.}(2025)]%
        {schumacher_emergence_2025}
\bibfield{author}{\bibinfo{person}{Pierre Schumacher}, \bibinfo{person}{Thomas Geijtenbeek}, \bibinfo{person}{Vittorio Caggiano}, \bibinfo{person}{Vikash Kumar}, \bibinfo{person}{Syn Schmitt}, \bibinfo{person}{Georg Martius}, {and} \bibinfo{person}{Daniel F.~B. Haeufle}.} \bibinfo{year}{2025}\natexlab{}.
\newblock \showarticletitle{Emergence of natural and robust bipedal walking by learning from biologically plausible objectives}.
\newblock \bibinfo{journal}{\emph{iScience}} \bibinfo{volume}{28}, \bibinfo{number}{4} (\bibinfo{date}{April} \bibinfo{year}{2025}), \bibinfo{pages}{112203}.
\newblock
\showISSN{2589-0042}
\href{https://doi.org/10.1016/j.isci.2025.112203}{doi:\nolinkurl{10.1016/j.isci.2025.112203}}


\bibitem[Schumacher et~al\mbox{.}(2023)]%
        {schumacher2023:deprl}
\bibfield{author}{\bibinfo{person}{Pierre Schumacher}, \bibinfo{person}{Daniel~F.B. Haeufle}, \bibinfo{person}{Dieter B{\"u}chler}, \bibinfo{person}{Syn Schmitt}, {and} \bibinfo{person}{Georg Martius}.} \bibinfo{year}{2023}\natexlab{}.
\newblock \showarticletitle{DEP-RL: Embodied Exploration for Reinforcement Learning in Overactuated and Musculoskeletal Systems}. In \bibinfo{booktitle}{\emph{Proceedings of the Eleventh International Conference on Learning Representations (ICLR)}}.
\newblock
\urldef\tempurl%
\url{https://openreview.net/forum?id=C-xa_D3oTj6}
\showURL{%
\tempurl}


\bibitem[Serifi et~al\mbox{.}(2024)]%
        {serifi_robot_2024}
\bibfield{author}{\bibinfo{person}{Agon Serifi}, \bibinfo{person}{Ruben Grandia}, \bibinfo{person}{Espen Knoop}, \bibinfo{person}{Markus Gross}, {and} \bibinfo{person}{Moritz Bächer}.} \bibinfo{year}{2024}\natexlab{}.
\newblock \showarticletitle{Robot {Motion} {Diffusion} {Model}: {Motion} {Generation} for {Robotic} {Characters}}. In \bibinfo{booktitle}{\emph{{SIGGRAPH} {Asia} 2024 {Conference} {Papers}}} \emph{(\bibinfo{series}{{SA} '24})}. \bibinfo{publisher}{Association for Computing Machinery}, \bibinfo{address}{New York, NY, USA}, \bibinfo{pages}{1--9}.
\newblock
\showISBNx{9798400711312}
\href{https://doi.org/10.1145/3680528.3687626}{doi:\nolinkurl{10.1145/3680528.3687626}}


\bibitem[Tessler et~al\mbox{.}(2024)]%
        {tessler_maskedmimic_2024}
\bibfield{author}{\bibinfo{person}{Chen Tessler}, \bibinfo{person}{Yunrong Guo}, \bibinfo{person}{Ofir Nabati}, \bibinfo{person}{Gal Chechik}, {and} \bibinfo{person}{Xue~Bin Peng}.} \bibinfo{year}{2024}\natexlab{}.
\newblock \showarticletitle{{MaskedMimic}: {Unified} {Physics}-{Based} {Character} {Control} {Through} {Masked} {Motion} {Inpainting}}.
\newblock \bibinfo{journal}{\emph{ACM Trans. Graph.}} \bibinfo{volume}{43}, \bibinfo{number}{6} (\bibinfo{date}{Nov.} \bibinfo{year}{2024}), \bibinfo{pages}{209:1--209:21}.
\newblock
\showISSN{0730-0301}
\href{https://doi.org/10.1145/3687951}{doi:\nolinkurl{10.1145/3687951}}


\bibitem[Tessler et~al\mbox{.}(2023)]%
        {CALM}
\bibfield{author}{\bibinfo{person}{Chen Tessler}, \bibinfo{person}{Yoni Kasten}, \bibinfo{person}{Yunrong Guo}, \bibinfo{person}{Shie Mannor}, \bibinfo{person}{Gal Chechik}, {and} \bibinfo{person}{Xue~Bin Peng}.} \bibinfo{year}{2023}\natexlab{}.
\newblock \showarticletitle{CALM: Conditional Adversarial Latent Models for Directable Virtual Characters}. In \bibinfo{booktitle}{\emph{ACM SIGGRAPH 2023 Conference Proceedings}} \emph{(\bibinfo{series}{SIGGRAPH '23})}.
\newblock


\bibitem[Todorov et~al\mbox{.}(2012)]%
        {todorov_mujoco_2012}
\bibfield{author}{\bibinfo{person}{Emanuel Todorov}, \bibinfo{person}{Tom Erez}, {and} \bibinfo{person}{Yuval Tassa}.} \bibinfo{year}{2012}\natexlab{}.
\newblock \showarticletitle{MuJoCo: A physics engine for model-based control}. In \bibinfo{booktitle}{\emph{2012 IEEE/RSJ International Conference on Intelligent Robots and Systems}}. \bibinfo{pages}{5026--5033}.
\newblock
\href{https://doi.org/10.1109/IROS.2012.6386109}{doi:\nolinkurl{10.1109/IROS.2012.6386109}}


\bibitem[Truong et~al\mbox{.}(2024)]%
        {truong_pdp_2024}
\bibfield{author}{\bibinfo{person}{Takara~Everest Truong}, \bibinfo{person}{Michael Piseno}, \bibinfo{person}{Zhaoming Xie}, {and} \bibinfo{person}{Karen Liu}.} \bibinfo{year}{2024}\natexlab{}.
\newblock \showarticletitle{{PDP}: {Physics}-{Based} {Character} {Animation} via {Diffusion} {Policy}}. In \bibinfo{booktitle}{\emph{{SIGGRAPH} {Asia} 2024 {Conference} {Papers}}} \emph{(\bibinfo{series}{{SA} '24})}. \bibinfo{publisher}{Association for Computing Machinery}, \bibinfo{address}{New York, NY, USA}, \bibinfo{pages}{1--10}.
\newblock
\showISBNx{9798400711312}
\href{https://doi.org/10.1145/3680528.3687683}{doi:\nolinkurl{10.1145/3680528.3687683}}


\bibitem[Wang et~al\mbox{.}(2024)]%
        {wang_strategy_2024}
\bibfield{author}{\bibinfo{person}{Jiashun Wang}, \bibinfo{person}{Jessica Hodgins}, {and} \bibinfo{person}{Jungdam Won}.} \bibinfo{year}{2024}\natexlab{}.
\newblock \showarticletitle{Strategy and {Skill} {Learning} for {Physics}-based {Table} {Tennis} {Animation}}. In \bibinfo{booktitle}{\emph{{ACM} {SIGGRAPH} 2024 {Conference} {Papers}}} \emph{(\bibinfo{series}{{SIGGRAPH} '24})}. \bibinfo{publisher}{Association for Computing Machinery}, \bibinfo{address}{New York, NY, USA}, \bibinfo{pages}{1--11}.
\newblock
\showISBNx{9798400705250}
\href{https://doi.org/10.1145/3641519.3657437}{doi:\nolinkurl{10.1145/3641519.3657437}}


\bibitem[Wang et~al\mbox{.}(2009)]%
        {wang_optimizing_2009}
\bibfield{author}{\bibinfo{person}{Jack~M. Wang}, \bibinfo{person}{David~J. Fleet}, {and} \bibinfo{person}{Aaron Hertzmann}.} \bibinfo{year}{2009}\natexlab{}.
\newblock \showarticletitle{Optimizing walking controllers}. In \bibinfo{booktitle}{\emph{{ACM} {SIGGRAPH} {Asia} 2009 papers}} \emph{(\bibinfo{series}{{SIGGRAPH} {Asia} '09})}. \bibinfo{publisher}{ACM}, \bibinfo{address}{New York, NY, USA}, \bibinfo{pages}{168:1--168:8}.
\newblock
\showISBNx{978-1-60558-858-2}


\bibitem[Wang et~al\mbox{.}(2010)]%
        {wang_optimizing_2010}
\bibfield{author}{\bibinfo{person}{Jack~M. Wang}, \bibinfo{person}{David~J. Fleet}, {and} \bibinfo{person}{Aaron Hertzmann}.} \bibinfo{year}{2010}\natexlab{}.
\newblock \showarticletitle{Optimizing walking controllers for uncertain inputs and environments}.
\newblock \bibinfo{journal}{\emph{ACM Trans. Graph.}} \bibinfo{volume}{29}, \bibinfo{number}{4} (\bibinfo{year}{2010}), \bibinfo{pages}{1--8}.
\newblock
\href{https://doi.org/10.1145/1778765.1778810}{doi:\nolinkurl{10.1145/1778765.1778810}}


\bibitem[Wang et~al\mbox{.}(2012)]%
        {wang_optimizing_2012}
\bibfield{author}{\bibinfo{person}{Jack~M. Wang}, \bibinfo{person}{Samuel~R. Hamner}, \bibinfo{person}{Scott~L. Delp}, {and} \bibinfo{person}{Vladlen Koltun}.} \bibinfo{year}{2012}\natexlab{}.
\newblock \showarticletitle{Optimizing locomotion controllers using biologically-based actuators and objectives}.
\newblock \bibinfo{journal}{\emph{ACM Trans. Graph.}} \bibinfo{volume}{31}, \bibinfo{number}{4}, Article \bibinfo{articleno}{25} (\bibinfo{date}{July} \bibinfo{year}{2012}), \bibinfo{numpages}{11}~pages.
\newblock
\showISSN{0730-0301}
\href{https://doi.org/10.1145/2185520.2185521}{doi:\nolinkurl{10.1145/2185520.2185521}}


\bibitem[Wang et~al\mbox{.}(2020)]%
        {wang2020unicon}
\bibfield{author}{\bibinfo{person}{Tingwu Wang}, \bibinfo{person}{Yunrong Guo}, \bibinfo{person}{Maria Shugrina}, {and} \bibinfo{person}{Sanja Fidler}.} \bibinfo{year}{2020}\natexlab{}.
\newblock \bibinfo{title}{UniCon: Universal Neural Controller For Physics-based Character Motion}.
\newblock
\showeprint[arxiv]{2011.15119}~[cs.GR]


\bibitem[Won et~al\mbox{.}(2020)]%
        {won_scalable_2020}
\bibfield{author}{\bibinfo{person}{Jungdam Won}, \bibinfo{person}{Deepak Gopinath}, {and} \bibinfo{person}{Jessica Hodgins}.} \bibinfo{year}{2020}\natexlab{}.
\newblock \showarticletitle{A scalable approach to control diverse behaviors for physically simulated characters}.
\newblock \bibinfo{journal}{\emph{ACM Transactions on Graphics}} \bibinfo{volume}{39}, \bibinfo{number}{4} (\bibinfo{date}{July} \bibinfo{year}{2020}), \bibinfo{pages}{33:33:1--33:33:12}.
\newblock
\showISSN{0730-0301}
\href{https://doi.org/10.1145/3386569.3392381}{doi:\nolinkurl{10.1145/3386569.3392381}}


\bibitem[Won et~al\mbox{.}(2022)]%
        {won_physics-based_2022}
\bibfield{author}{\bibinfo{person}{Jungdam Won}, \bibinfo{person}{Deepak Gopinath}, {and} \bibinfo{person}{Jessica Hodgins}.} \bibinfo{year}{2022}\natexlab{}.
\newblock \showarticletitle{Physics-based character controllers using conditional {VAEs}}.
\newblock \bibinfo{journal}{\emph{ACM Transactions on Graphics}} \bibinfo{volume}{41}, \bibinfo{number}{4} (\bibinfo{date}{July} \bibinfo{year}{2022}), \bibinfo{pages}{96:1--96:12}.
\newblock
\showISSN{0730-0301}


\bibitem[Xie et~al\mbox{.}(2020)]%
        {allsteps20}
\bibfield{author}{\bibinfo{person}{Zhaoming Xie}, \bibinfo{person}{Hung~Yu Ling}, \bibinfo{person}{Nam~Hee Kim}, {and} \bibinfo{person}{Michiel van~de Panne}.} \bibinfo{year}{2020}\natexlab{}.
\newblock \showarticletitle{ALLSTEPS: Curriculum-driven Learning of Stepping Stone Skills}.
\newblock \bibinfo{journal}{\emph{Computer Graphics Forum}} \bibinfo{volume}{39}, \bibinfo{number}{8} (\bibinfo{year}{2020}), \bibinfo{pages}{213--224}.
\newblock
\href{https://doi.org/10.1111/cgf.14115}{doi:\nolinkurl{10.1111/cgf.14115}}
\showeprint{https://onlinelibrary.wiley.com/doi/pdf/10.1111/cgf.14115}


\bibitem[Xu et~al\mbox{.}(2023)]%
        {xu_composite_2023}
\bibfield{author}{\bibinfo{person}{Pei Xu}, \bibinfo{person}{Xiumin Shang}, \bibinfo{person}{Victor Zordan}, {and} \bibinfo{person}{Ioannis Karamouzas}.} \bibinfo{year}{2023}\natexlab{}.
\newblock \showarticletitle{Composite {Motion} {Learning} with {Task} {Control}}.
\newblock \bibinfo{journal}{\emph{ACM Transactions on Graphics}} \bibinfo{volume}{42}, \bibinfo{number}{4} (\bibinfo{date}{July} \bibinfo{year}{2023}), \bibinfo{pages}{93:1--93:16}.
\newblock
\showISSN{0730-0301}


\bibitem[Yao et~al\mbox{.}(2022)]%
        {yao_controlvae_2022}
\bibfield{author}{\bibinfo{person}{Heyuan Yao}, \bibinfo{person}{Zhenhua Song}, \bibinfo{person}{Baoquan Chen}, {and} \bibinfo{person}{Libin Liu}.} \bibinfo{year}{2022}\natexlab{}.
\newblock \showarticletitle{{ControlVAE}: {Model}-{Based} {Learning} of {Generative} {Controllers} for {Physics}-{Based} {Characters}}.
\newblock \bibinfo{journal}{\emph{ACM Transactions on Graphics}} \bibinfo{volume}{41}, \bibinfo{number}{6} (\bibinfo{date}{Nov.} \bibinfo{year}{2022}), \bibinfo{pages}{183:1--183:16}.
\newblock
\showISSN{0730-0301}


\bibitem[Yao et~al\mbox{.}(2024)]%
        {yao_moconvq_2024}
\bibfield{author}{\bibinfo{person}{Heyuan Yao}, \bibinfo{person}{Zhenhua Song}, \bibinfo{person}{Yuyang Zhou}, \bibinfo{person}{Tenglong Ao}, \bibinfo{person}{Baoquan Chen}, {and} \bibinfo{person}{Libin Liu}.} \bibinfo{year}{2024}\natexlab{}.
\newblock \showarticletitle{{MoConVQ}: {Unified} {Physics}-{Based} {Motion} {Control} via {Scalable} {Discrete} {Representations}}.
\newblock \bibinfo{journal}{\emph{ACM Transactions on Graphics}} \bibinfo{volume}{43}, \bibinfo{number}{4} (\bibinfo{date}{July} \bibinfo{year}{2024}), \bibinfo{pages}{144:1--144:21}.
\newblock
\showISSN{0730-0301}


\bibitem[Yin et~al\mbox{.}(2007)]%
        {yin2007simbicon}
\bibfield{author}{\bibinfo{person}{KangKang Yin}, \bibinfo{person}{Kevin Loken}, {and} \bibinfo{person}{Michiel Van~de Panne}.} \bibinfo{year}{2007}\natexlab{}.
\newblock \showarticletitle{Simbicon: Simple biped locomotion control}.
\newblock \bibinfo{journal}{\emph{ACM Transactions on Graphics (TOG)}} \bibinfo{volume}{26}, \bibinfo{number}{3} (\bibinfo{year}{2007}), \bibinfo{pages}{105--es}.
\newblock


\bibitem[Yu et~al\mbox{.}(2018)]%
        {yu_learningsymmetric_2018}
\bibfield{author}{\bibinfo{person}{Wenhao Yu}, \bibinfo{person}{Greg Turk}, {and} \bibinfo{person}{C.~Karen Liu}.} \bibinfo{year}{2018}\natexlab{}.
\newblock \showarticletitle{Learning {Symmetric} and {Low}-energy {Locomotion}}.
\newblock \bibinfo{journal}{\emph{ACM Trans. Graph.}} \bibinfo{volume}{37}, \bibinfo{number}{4} (\bibinfo{date}{July} \bibinfo{year}{2018}), \bibinfo{pages}{144:1--144:12}.
\newblock
\showISSN{0730-0301}
\href{https://doi.org/10.1145/3197517.3201397}{doi:\nolinkurl{10.1145/3197517.3201397}}


\bibitem[Zajac(1989)]%
        {zajac_muscle_1989}
\bibfield{author}{\bibinfo{person}{F~E Zajac}.} \bibinfo{year}{1989}\natexlab{}.
\newblock \showarticletitle{Muscle and tendon: properties, models, scaling, and application to biomechanics and motor control}.
\newblock \bibinfo{journal}{\emph{Critical Reviews in Biomedical Engineering}} \bibinfo{volume}{17}, \bibinfo{number}{4} (\bibinfo{year}{1989}), \bibinfo{pages}{359--411}.
\newblock
\showISSN{0278-940X}
\urldef\tempurl%
\url{http://www.ncbi.nlm.nih.gov/pubmed/2676342}
\showURL{%
\tempurl}


\bibitem[Zhang et~al\mbox{.}(2023)]%
        {zhang_learning_2023}
\bibfield{author}{\bibinfo{person}{Haotian Zhang}, \bibinfo{person}{Ye Yuan}, \bibinfo{person}{Viktor Makoviychuk}, \bibinfo{person}{Yunrong Guo}, \bibinfo{person}{Sanja Fidler}, \bibinfo{person}{Xue~Bin Peng}, {and} \bibinfo{person}{Kayvon Fatahalian}.} \bibinfo{year}{2023}\natexlab{}.
\newblock \showarticletitle{Learning {Physically} {Simulated} {Tennis} {Skills} from {Broadcast} {Videos}}.
\newblock \bibinfo{journal}{\emph{ACM Transactions on Graphics}} \bibinfo{volume}{42}, \bibinfo{number}{4} (\bibinfo{date}{July} \bibinfo{year}{2023}), \bibinfo{pages}{95:1--95:14}.
\newblock
\showISSN{0730-0301}
\href{https://doi.org/10.1145/3592408}{doi:\nolinkurl{10.1145/3592408}}


\bibitem[Zhu et~al\mbox{.}(2023)]%
        {zhu_neural_2023}
\bibfield{author}{\bibinfo{person}{Qingxu Zhu}, \bibinfo{person}{He Zhang}, \bibinfo{person}{Mengting Lan}, {and} \bibinfo{person}{Lei Han}.} \bibinfo{year}{2023}\natexlab{}.
\newblock \showarticletitle{Neural {Categorical} {Priors} for {Physics}-{Based} {Character} {Control}}.
\newblock \bibinfo{journal}{\emph{ACM Transactions on Graphics}} \bibinfo{volume}{42}, \bibinfo{number}{6} (\bibinfo{date}{Dec.} \bibinfo{year}{2023}), \bibinfo{pages}{178:1--178:16}.
\newblock
\showISSN{0730-0301}


\end{thebibliography}


\begin{thebibliography}{11}

\ifx \showCODEN    \undefined \def \showCODEN     #1{\unskip}     \fi
\ifx \showISBNx    \undefined \def \showISBNx     #1{\unskip}     \fi
\ifx \showISBNxiii \undefined \def \showISBNxiii  #1{\unskip}     \fi
\ifx \showISSN     \undefined \def \showISSN      #1{\unskip}     \fi
\ifx \showLCCN     \undefined \def \showLCCN      #1{\unskip}     \fi
\ifx \shownote     \undefined \def \shownote      #1{#1}          \fi
\ifx \showarticletitle \undefined \def \showarticletitle #1{#1}   \fi
\ifx \showURL      \undefined \def \showURL       {\relax}        \fi

\providecommand\bibfield[2]{#2}
\providecommand\bibinfo[2]{#2}
\providecommand\natexlab[1]{#1}
\providecommand\showeprint[2][]{arXiv:#2}

\bibitem[Feng et~al\mbox{.}(2023)]%
        {feng_musclevae_2023}
\bibfield{author}{\bibinfo{person}{Yusen Feng}, \bibinfo{person}{Xiyan Xu}, {and} \bibinfo{person}{Libin Liu}.} \bibinfo{year}{2023}\natexlab{}.
\newblock \showarticletitle{MuscleVAE: Model-Based Controllers of Muscle-Actuated Characters}. In \bibinfo{booktitle}{\emph{SIGGRAPH Asia 2023 Conference Papers}} (Sydney, NSW, Australia) \emph{(\bibinfo{series}{SA '23})}. \bibinfo{publisher}{Association for Computing Machinery}, \bibinfo{address}{New York, NY, USA}, Article \bibinfo{articleno}{3}, \bibinfo{numpages}{11}~pages.
\newblock
\showISBNx{9798400703157}
\href{https://doi.org/10.1145/3610548.3618137}{doi:\nolinkurl{10.1145/3610548.3618137}}


\bibitem[Fussell et~al\mbox{.}(2021)]%
        {fussel_supertrack_2021}
\bibfield{author}{\bibinfo{person}{Levi Fussell}, \bibinfo{person}{Kevin Bergamin}, {and} \bibinfo{person}{Daniel Holden}.} \bibinfo{year}{2021}\natexlab{}.
\newblock \showarticletitle{SuperTrack: motion tracking for physically simulated characters using supervised learning}.
\newblock \bibinfo{journal}{\emph{ACM Trans. Graph.}} \bibinfo{volume}{40}, \bibinfo{number}{6}, Article \bibinfo{articleno}{197} (\bibinfo{date}{Dec.} \bibinfo{year}{2021}), \bibinfo{numpages}{13}~pages.
\newblock
\showISSN{0730-0301}
\href{https://doi.org/10.1145/3478513.3480527}{doi:\nolinkurl{10.1145/3478513.3480527}}


\bibitem[Ikkala and H{\"a}m{\"a}l{\"a}inen(2022)]%
        {ikkala_converting_2022}
\bibfield{author}{\bibinfo{person}{Aleksi Ikkala} {and} \bibinfo{person}{Perttu H{\"a}m{\"a}l{\"a}inen}.} \bibinfo{year}{2022}\natexlab{}.
\newblock \showarticletitle{Converting Biomechanical Models from OpenSim to MuJoCo}. In \bibinfo{booktitle}{\emph{Converging Clinical and Engineering Research on Neurorehabilitation IV}}, \bibfield{editor}{\bibinfo{person}{Diego Torricelli}, \bibinfo{person}{Metin Akay}, {and} \bibinfo{person}{Jose~L. Pons}} (Eds.). \bibinfo{publisher}{Springer International Publishing}, \bibinfo{address}{Cham}, \bibinfo{pages}{277--281}.
\newblock
\showISBNx{978-3-030-70316-5}


\bibitem[La~Barbera et~al\mbox{.}(2021)]%
        {barbera_ostrichrl_2021}
\bibfield{author}{\bibinfo{person}{Vittorio La~Barbera}, \bibinfo{person}{Fabio Pardo}, \bibinfo{person}{Yuval Tassa}, \bibinfo{person}{Monica Daley}, \bibinfo{person}{Christopher Richards}, \bibinfo{person}{Petar Kormushev}, {and} \bibinfo{person}{John Hutchinson}.} \bibinfo{year}{2021}\natexlab{}.
\newblock \showarticletitle{{OstrichRL}: A Musculoskeletal Ostrich Simulation to Study Bio-mechanical Locomotion}. In \bibinfo{booktitle}{\emph{NeurIPS 2021 Deep Reinforcement Learning Workshop, 35th Conference on Neural Information Processing Systems (NeurIPS 2021)}}.
\newblock
\urldef\tempurl%
\url{https://kormushev.com/papers/Pardo_NeurIPS-2021.pdf}
\showURL{%
\tempurl}


\bibitem[Lee et~al\mbox{.}(2019)]%
        {lee_scalablemuscle_2019}
\bibfield{author}{\bibinfo{person}{Seunghwan Lee}, \bibinfo{person}{Moonseok Park}, \bibinfo{person}{Kyoungmin Lee}, {and} \bibinfo{person}{Jehee Lee}.} \bibinfo{year}{2019}\natexlab{}.
\newblock \showarticletitle{Scalable muscle-actuated human simulation and control}.
\newblock \bibinfo{journal}{\emph{ACM Trans. Graph.}} \bibinfo{volume}{38}, \bibinfo{number}{4}, Article \bibinfo{articleno}{73} (\bibinfo{date}{jul} \bibinfo{year}{2019}), \bibinfo{numpages}{13}~pages.
\newblock
\showISSN{0730-0301}
\href{https://doi.org/10.1145/3306346.3322972}{doi:\nolinkurl{10.1145/3306346.3322972}}


\bibitem[Lee et~al\mbox{.}(2014)]%
        {lee_locomotion_2014}
\bibfield{author}{\bibinfo{person}{Yoonsang Lee}, \bibinfo{person}{Moon~Seok Park}, \bibinfo{person}{Taesoo Kwon}, {and} \bibinfo{person}{Jehee Lee}.} \bibinfo{year}{2014}\natexlab{}.
\newblock \showarticletitle{Locomotion {Control} for {Many}-muscle {Humanoids}}.
\newblock \bibinfo{journal}{\emph{ACM Trans. Graph.}} \bibinfo{volume}{33}, \bibinfo{number}{6} (\bibinfo{year}{2014}), \bibinfo{pages}{218:1--218:11}.
\newblock
\showISSN{0730-0301}
\href{https://doi.org/10.1145/2661229.2661233}{doi:\nolinkurl{10.1145/2661229.2661233}}


\bibitem[Tessler et~al\mbox{.}(2023)]%
        {CALM}
\bibfield{author}{\bibinfo{person}{Chen Tessler}, \bibinfo{person}{Yoni Kasten}, \bibinfo{person}{Yunrong Guo}, \bibinfo{person}{Shie Mannor}, \bibinfo{person}{Gal Chechik}, {and} \bibinfo{person}{Xue~Bin Peng}.} \bibinfo{year}{2023}\natexlab{}.
\newblock \showarticletitle{CALM: Conditional Adversarial Latent Models for Directable Virtual Characters}. In \bibinfo{booktitle}{\emph{ACM SIGGRAPH 2023 Conference Proceedings}} \emph{(\bibinfo{series}{SIGGRAPH '23})}.
\newblock


\bibitem[Todorov et~al\mbox{.}(2012)]%
        {todorov_mujoco_2012}
\bibfield{author}{\bibinfo{person}{Emanuel Todorov}, \bibinfo{person}{Tom Erez}, {and} \bibinfo{person}{Yuval Tassa}.} \bibinfo{year}{2012}\natexlab{}.
\newblock \showarticletitle{MuJoCo: A physics engine for model-based control}. In \bibinfo{booktitle}{\emph{2012 IEEE/RSJ International Conference on Intelligent Robots and Systems}}. \bibinfo{pages}{5026--5033}.
\newblock
\href{https://doi.org/10.1109/IROS.2012.6386109}{doi:\nolinkurl{10.1109/IROS.2012.6386109}}


\bibitem[Wang et~al\mbox{.}(2012)]%
        {wang_optimizing_2012}
\bibfield{author}{\bibinfo{person}{Jack~M. Wang}, \bibinfo{person}{Samuel~R. Hamner}, \bibinfo{person}{Scott~L. Delp}, {and} \bibinfo{person}{Vladlen Koltun}.} \bibinfo{year}{2012}\natexlab{}.
\newblock \showarticletitle{Optimizing locomotion controllers using biologically-based actuators and objectives}.
\newblock \bibinfo{journal}{\emph{ACM Trans. Graph.}} \bibinfo{volume}{31}, \bibinfo{number}{4}, Article \bibinfo{articleno}{25} (\bibinfo{date}{July} \bibinfo{year}{2012}), \bibinfo{numpages}{11}~pages.
\newblock
\showISSN{0730-0301}
\href{https://doi.org/10.1145/2185520.2185521}{doi:\nolinkurl{10.1145/2185520.2185521}}


\bibitem[Yao et~al\mbox{.}(2022)]%
        {yao_controlvae_2022}
\bibfield{author}{\bibinfo{person}{Heyuan Yao}, \bibinfo{person}{Zhenhua Song}, \bibinfo{person}{Baoquan Chen}, {and} \bibinfo{person}{Libin Liu}.} \bibinfo{year}{2022}\natexlab{}.
\newblock \showarticletitle{{ControlVAE}: {Model}-{Based} {Learning} of {Generative} {Controllers} for {Physics}-{Based} {Characters}}.
\newblock \bibinfo{journal}{\emph{ACM Transactions on Graphics}} \bibinfo{volume}{41}, \bibinfo{number}{6} (\bibinfo{date}{Nov.} \bibinfo{year}{2022}), \bibinfo{pages}{183:1--183:16}.
\newblock
\showISSN{0730-0301}


\bibitem[Zajac(1989)]%
        {zajac_muscle_1989}
\bibfield{author}{\bibinfo{person}{F~E Zajac}.} \bibinfo{year}{1989}\natexlab{}.
\newblock \showarticletitle{Muscle and tendon: properties, models, scaling, and application to biomechanics and motor control}.
\newblock \bibinfo{journal}{\emph{Critical Reviews in Biomedical Engineering}} \bibinfo{volume}{17}, \bibinfo{number}{4} (\bibinfo{year}{1989}), \bibinfo{pages}{359--411}.
\newblock
\showISSN{0278-940X}
\urldef\tempurl%
\url{http://www.ncbi.nlm.nih.gov/pubmed/2676342}
\showURL{%
\tempurl}


\end{thebibliography}

\clearpage
\appendix
\begin{strip}
\raggedright 
{\sffamily\mdseries\fontsize{16}{24}\selectfont  Supplemental Material for Motion-Free Learning of Latent Control for Morphology-Adaptive Locomotion in Musculoskeletal Characters}
\end{strip}

\section{Muscle Modeling in MuJoCo}

In musculoskeletal simulation, a muscle-tendon actuator is typically modeled as a Hill-based muscle consisting of three components: the active and passive elements of the muscle fiber, and the tendon~\cite{zajac_muscle_1989}.
Our system simulates musculoskeletal movement using the Hill-type muscle model implemented in the MuJoCo engine~\cite{todorov_mujoco_2012}, which assumes inelastic tendons and represents each muscle actuator as a simplified Hill-type structure—composed of a muscle fiber and a fixed-length tendon.
This model is defined as follows:
\begin{equation}
F_{\text{muscle}} = - \left( act \cdot F_L(L) \cdot F_V(V) + F_P(L) \right),
\label{eq:muscle_force}
\end{equation}
where:
\begin{itemize}
    \item $L$ is the current muscle fiber length,
    \item $V$ is the contraction velocity,
    \item $F_L(L)$ is the \textit{force-length relationship}, a static gain function representing the maximum force producible at a given length,
    \item $F_V(V)$ is the \textit{force-velocity relationship}, a dynamic gain function varying with contraction speed,
    \item $F_P(L)$ is the \textit{passive force}, modeling the elastic resistance when the muscle is stretched.
\end{itemize}

$act \in [0, 1]$ denotes the muscle activation level, which serves as the control signal in our system. MuJoCo assumes an inelastic tendon, so the total muscle-tendon length is computed as the sum of the fiber length and the fixed tendon length. The negative sign in Equation~\eqref{eq:muscle_force} reflects the physical nature of muscles, which always exert pulling forces between their attachment points.

Muscle routing is constructed based on line segments, where each muscle connects between body links via multiple attachment points that are predefined as offsets on each link. This routing dynamically changes with the character’s posture, determining the muscle length $L$ and contraction velocity $V$ for all muscles.

The sum of the computed muscle forces is ultimately translated into joint torques. However, because the muscle paths vary with posture, these torques always fall within a more realistic torque region that reflects the posture-dependent variability of range of motion in real biological systems. This modeling approach not only constrains joint movements in a physiologically plausible way but also naturally incorporates muscle redundancy and force constraints observed in real musculoskeletal systems.

\section{Musculoskeletal Characters Details}

\begin{description}
    \item[Humanoid] 
    We use the musculoskeletal humanoid model with 120 muscles from \cite{lee_locomotion_2014}. It consists of 19 links, 39 DOFs, and a total body mass of 72\,kg. To support simulation in MuJoCo, we convert the original OpenSim model using the method from \cite{ikkala_converting_2022}. Capsule-shaped collision geometries are assigned to the foot and hand links to support stable contact handling in MuJoCo.

    \item[Ostrich] 
    To demonstrate generalizability to non-human animals, we employ an ostrich model equipped with 120 muscle actuators from \cite{barbera_ostrichrl_2021}.
    Unlike the Humanoid and Chimanoid, the original foot geometry is used as the collision mesh.
    
    \item[Chimanoid] 
    The Chimanoid is a fictional character designed by modifying the Humanoid model to exhibit chimpanzee-like proportions. Specifically, the arms are lengthened by 20\% and the legs shortened by 30\%. The masses of the affected limb segments are adjusted proportionally. To reflect the physical property that a chimpanzee’s center of mass is located higher and more forward than that of a human, the center of mass of the torso (thorax link) is shifted upward and forward by 20\,cm each. As with the Humanoid, capsule-shaped collision meshes are used for the feet and hands to ensure robust contact simulation.
\end{description}

\section{Latent Control Architecture Details}
    
\subsection{Representation of State, Action, Energy, and Goal}

\paragraph{State $\mathbf s_t$}
The character state $s_t$, used as input to the encoder, decoder, and world model, follows the implementation of prior work~\cite{fussel_supertrack_2021, yao_controlvae_2022}. It encodes the physical properties of each link in the character's body, expressed in the local frame of the pelvis (i.e., the root link). Specifically, for each link in the set of rigid body links, the state includes:
\begin{itemize}
    \item Position and orientation (6D representation),
    \item Linear and angular velocity,
    \item Height of each rigid link,
\end{itemize}
plus the upward direction vector of the root link.  
The total dimensionality is $16 \times N_\text{link} + 3$, where $N_\text{link}$ is the number of links.  
This formulation is consistently applied across all character models.

\paragraph{Action $\mathbf a_t$}
Although real musculoskeletal systems involve excitation dynamics where muscle activation gradually increases or decreases over time, we simplify the control input by directly using the muscle activation level $act \in [0,1]$ as the control signal. This value is used both for simulation in MuJoCo and for metabolic energy expenditure computation.

\paragraph{Metabolic Energy Expenditure $\mathbf e_t$}
We consider the metabolic energy expenditure in computing both the world model loss (Equation~\eqref{eq:L_world}) and the encoder/decoder loss (Equation~\eqref{eq:L_vae}), in order to promote the emergence of energy-efficient movement patterns and to enable the encoding of diverse energetic styles in the latent space. Following the formulation adopted in~\cite{wang_optimizing_2012}, we first compute the metabolic energy expenditure rate \(\dot{E}_i\) for each muscle \(i\) using the following decomposition:
\begin{equation}
    \dot{E}_i = \dot{A}_i + \dot{M}_i + \dot{S}_i + \dot{W}_i,
    \label{eq:energy_rate}
\end{equation}
where each term represents the rate of energy used for activation, maintenance, shortening, and mechanical work, respectively.  
The energy expenditure rates for all muscles form a vector \(\dot{\mathbf{E}} = [\dot{E}_1, \dots, \dot{E}_N]\), which is integrated over time to obtain the energy consumption vector \(\mathbf{e}_t\).

\paragraph{Goal $\mathbf g_t$}
The goal signal $\mathbf{g}_t$ serves as the conditioning input to the posterior encoder.
During training, we explore multiple configurations for defining $\mathbf{g}_t$, enabling the controller to learn from a variety of control objectives. Specifically, we adopt the following configurations, all centered around target velocity:

\begin{itemize}

    \item \textbf{Velocity Only}: Includes the target horizontal velocity and its difference from the current velocity.

    \item \textbf{Velocity + Direction}: Additionally includes the target facing direction and its difference from the current direction.

    \item \textbf{Velocity + Energy}: Additionally includes the desired total metabolic energy expenditure, along with its differences from the current values.

    \item \textbf{Velocity + Pose + Energy}: Additionally includes the target pose and desired total metabolic energy expenditure, along with their differences from the current values.

\end{itemize}

These variants enable the system to learn latent representations that support not only direction- and speed-aware control, but also form- and energy-modulated locomotion.

\subsection{Architecture}

The prior/posterior encoder, decoder, and world model are implemented in PyTorch, based on the architecture of ControlVAE~\cite{yao_controlvae_2022}, with several architectural modifications.
\ys{Specifically, both the prior and posterior encoders have two hidden layers with [512, 512] units.
A 64-dimensional latent vector $\mathbf z_t$ is then formed by summing their outputs}, which serves as the compact control representation of the character’s intended movement.
\ys{The decoder adopts a mixture-of-experts (MoE) structure, in which six expert networks independently produce outputs through three hidden layers of [512, 512, 512] units, and these outputs are then combined via a gating network with two hidden layers of size [64, 64] that outputs the weights for the weighted sum. The world model also uses three hidden layers of [512, 512, 512] units.}

Each hidden layer in the encoder and world model uses the ELU activation function. Following prior work~\cite{lee_scalablemuscle_2019}, the decoder uses a composite activation function, where the output is passed through a Tanh and then a ReLU function (i.e., $\mathrm{ReLU}(\tanh(\cdot))$), effectively constraining the muscle activations to a biologically plausible range.

Our implementation introduces three modifications compared to prior work:
(i) layer normalization is applied to all layers of the world model to improve training stability;
(ii) the world model is extended to predict not only state transitions but also metabolic energy expenditure at each timestep; and
(iii) we introduce a normalization step that projects the latent vector onto a hypersphere before feeding it into the decoder, encouraging structured and bounded latent representations that improve generalization to high-level task control~\cite{CALM}.

We use the RAdam optimizer with $\beta_1=0.9$, $\beta_2=0.999$. The learning rates are set to $1\mathrm{e}{-5}$ for the policy network and $2\mathrm{e}{-3}$ for the world model.

\section{Training Procedure Details}

Our latent policy model is trained through a three-stage procedure, which is repeated in each iteration to gradually improve the model.

\paragraph{Stage 1: Data Collection with Frozen Model}  
First, we collect simulation samples for training the world model. During this stage, all networks are frozen. At the beginning of the rollout, a random goal \( \mathbf{g}_t \) is sampled. Then, at every timestep, the current state \( \mathbf{s}_t \) is passed to both the prior and posterior encoders, and the current goal \( \mathbf{g}_t \) is additionally provided to the posterior encoder. The encoders produce latent codes, which are combined to form \( \mathbf{z}_t \), and this latent is decoded into the action \( \mathbf{a}_t \). With a 3\% probability at each timestep, the goal \( \mathbf{g}_t \) is resampled randomly to encourage exploration across diverse behavior patterns. The action is then applied to the physics simulator, which computes the resulting muscle forces and updates the character’s state accordingly.
The resulting transition tuple \( (\mathbf{s}_t, \mathbf{a}_t, \mathbf{s}_{t+1}) \), along with the goal \( \mathbf{g}_t \), is stored in a trajectory buffer.  
Additionally, the metabolic energy expenditure \( \mathbf{e}_t \) is computed based on muscle state variables, using Equation~\eqref{eq:energy_rate}, and the resulting values are integrated over time.  
The computed energy is also stored in the buffer.
These stored values serve as initial states ($\mathbf s_0$), activation trajectory ($\mathbf a_t$), and ground truth ($\mathbf s_{t+1}, \mathbf{e}_t$) for world model update (Stage 2), and as initial states ($\mathbf s_0$) and goal trajectory ($\mathbf g_t$) for encoder-decoder update (Stage 3). 
\ys{Following ControlVAE, the buffer is updated with 2,048 new samples per epoch and capped at 50,000 entries, using a FIFO (first-in-first-out) replacement policy.}

The goal \( \mathbf{g}_t \) can be composed of the following elements, each of which is randomly sampled from a predefined range when included:
\begin{itemize}
    \item \textbf{Target velocity magnitude} $\in [0.0,\ 4.25]$ m/s
    \item \textbf{Target movement direction} $\in [-\pi,\ \pi]$ radians
    \item \textbf{Target facing direction} $\in [-\pi,\ \pi]$ radians
    \item \textbf{Target pose}: sampled by perturbing key joint rotations around the default standing pose. 
Rather than specifying full-body postures, we limit the variation to a subset of joints that strongly influence gait style—specifically, the frontal-axis rotations of the hips, knees, and ankles. 
For each of these joints, the target rotation is randomly sampled within the anatomically valid range.
    \item \textbf{Target energy level} $\in [0.15,\ 0.21]$: representing the desired metabolic effort. 
    This corresponds to a scaled version of the total metabolic energy expenditure, with a scaling factor of 0.01 applied to match the range used during training. 
\end{itemize}

\paragraph{Stage 2: World Model Update}
Second, we update the network parameters of the world model. Specifically, we randomly sample a simulation trajectory from the buffer and use its initial state $\mathbf s_0$ as input. The world model is then rolled out for $T_W$ steps using the activation $\mathbf{a}_t$ stored in the same trajectory, and its parameters are optimized to minimize the following loss $L_{\text{world}}$, which measures the $\ell_1$ norm between the predicted and ground-truth simulation results:
\begin{align}
    L_{\text{world}} &= \sum_{t=0}^{T_{W}-1} \left( \left\lVert \overline{\mathbf s}_{t+1} - \mathbf s_{t+1} \right\rVert_{1,\mathbf W} + w_\text w \left\lVert \overline{\mathbf e}_t - \mathbf e_t \right\rVert_1 \right),
    \label{eq:L_world}
\end{align}
where $\overline{\mathbf s}_{t+1}$ and $\mathbf s_{t+1}$ are the ground-truth and predicted character states at timestep $t+1$, 
$\overline{\mathbf e}_t$ and $\mathbf e_t$ 
are the ground-truth and predicted metabolic energy expenditures at $t$, and \ys{$w_{\text{w}}$} is a weighting factor for the energy loss term, which we set to \ys{0.1.} 
The matrix $\mathbf{W}$ is a diagonal weight matrix that assigns a weight of 1 to position, orientation, and height dimensions, and a weight of 4 to linear and angular velocity dimensions.

Following prior work on predictive world models for physics-based simulation~\cite{fussel_supertrack_2021, yao_controlvae_2022, feng_musclevae_2023}, we predict changes in the character’s state rather than absolute states, and obtain the final result through timestep integration. Inspired by MuscleVAE~\cite{feng_musclevae_2023}, which trained the world model \ys{to predict fatigue at the body-part level}, we extend the world model to predict per-muscle metabolic energy expenditure, in addition to changes in rigid body states. The predicted energy values are used in the next stage to calculate the loss function for updating the encoder and decoder, serving as a regularizer to promote energy-aware movements and contribute to physical stability.

\paragraph{Stage 3: Encoder and Decoder Update}

With the world model frozen, we generate $T_P$-step trajectories using the current encoder-decoder model, and update its parameters by minimizing the loss $L_{\text{VAE}}$. Given an initial state $\mathbf{s}_0$ from a randomly sampled trajectory from the buffer, the encoder-decoder-world model pipeline produces a synthesized trajectory over $T_P$ steps, using the corresponding goal $\mathbf{g}_t$ stored in the same trajectory. The loss $L_{\text{VAE}}$ is defined as:
\begin{align}
L_\text{VAE} &= L_\text{objective} +  \beta \cdot L_\text{KL},
\label{eq:L_vae}
\end{align}
where
\begin{align}
    L_{\text{KL}} &= \sum_{t=0}^{T_{P}-1} \gamma^t \cdot D_{\text{KL}}\left(q(\mathbf z_t | \mathbf s_t, \mathbf g_t) \,\|\, p(\mathbf z_t | \mathbf s_t)\right).
\end{align}

The KL loss $L_{\text{KL}}$ is computed as the divergence between the posterior distribution $q(\mathbf z_t | \mathbf s_t, \mathbf g_t)$ and prior distribution $p(\mathbf z_t | \mathbf s_t)$, with $\beta$-scheduling applied as in prior work.
\ys{Since the posterior distribution is obtained by summing the prior and posterior encoder outputs, this term actually serves as a regularizer for the posterior encoder.}
A discount factor $\gamma$ is used over time, for which we use $\gamma = 0.99$.
A detailed description of our locomotion objective loss $L_{\text{objective}}$ is provided in Section~3.3 of the main paper.

These three stages are repeated for a total of 20,000 iterations to train the our model. The rollout horizon for the world model is set to $T_W = 8$, and the rollout horizon for the policy is set to $T_P = 32$, following the optimal settings suggested in \cite{fussel_supertrack_2021}, with adjustments made to suit our simulation environment.

\section{Locomotion Objective Loss Details}

\paragraph{Target height \( \overline{h}_t \) in \( L_\text{height} \)}  
The target height \( \overline{h}_t \) is set slightly below the character's root height in a nominal standing pose to allow for natural vertical fluctuations. Specifically, we use 0.8\,m for Humanoid, 0.75\,m for Ostrich, and 0.6\,m for Chimanoid.

\paragraph{Loss Weight Configuration}
We use the following scalar weights for each term in the total locomotion objective:  
\ys{\( w_v = 60 \), \( w_d = 6 \) for the control objective;  
\( w_h = 24 \), \( w_u = 3 \) for the balancing objective; and  
\( w_p = 0.06 \), \( w_e = 0.15 \) for the biomechanical objective.}
These values were chosen empirically to balance goal tracking, postural stability, and biomechanical plausibility. 

To slightly stabilize upper-body motion, a minor regularization term on torso orientation was applied to the Humanoid character.

\end{document}